\documentclass[10pt,journal,compsoc]{IEEEtran}
\ifCLASSOPTIONcompsoc
  \usepackage[nocompress]{cite}
\else
  \usepackage{cite}
\fi
\ifCLASSINFOpdf
\else
\fi

\usepackage{graphicx}
\usepackage{subfigure}
\usepackage{amsmath}
\usepackage{amssymb}
\usepackage{booktabs}
\usepackage{threeparttable}
\usepackage{color, xcolor}
\usepackage[pagebackref=true,breaklinks=true,colorlinks,bookmarks=false]{hyperref}
\usepackage{listings}
\usepackage{multirow}

\definecolor{codegreen}{rgb}{0,0.6,0}
\definecolor{codegray}{rgb}{0.5,0.5,0.5}
\definecolor{codepurple}{rgb}{0.58,0,0.82}
\definecolor{backcolour}{rgb}{0.95,0.95,0.92}
\lstdefinestyle{mystyle}{
    backgroundcolor=\color{backcolour},   
    commentstyle=\color{codegreen},
    keywordstyle=\color{magenta},
    numberstyle=\tiny\color{codegray},
    stringstyle=\color{codepurple},
    basicstyle=\ttfamily\footnotesize,
    breakatwhitespace=false,         
    breaklines=true,                 
    captionpos=b,                    
    keepspaces=true,                 
    numbers=left,                    
    numbersep=5pt,                  
    showspaces=false,                
    showstringspaces=false,
    showtabs=false,                  
    tabsize=2
}
\begin{document}
%
\title{High-Efficiency Neural Video Compression \\via Hierarchical Predictive Learning}
%
%
%
%

\author{Ming Lu, Zhihao Duan, Wuyang Cong, Dandan Ding, Fengqing Zhu, and Zhan Ma
\IEEEcompsocitemizethanks{\IEEEcompsocthanksitem M. Lu, W. Cong, and Z. Ma are with the School of Electronic Science and Engineering, Nanjing University, Nanjing, Jiangsu 210023, China (e-mail: minglu@nju.edu.cn; congwuyang@smail.nju.edu.cn; mazhan@nju.edu.cn).
\IEEEcompsocthanksitem Z. Duan and F. Zhu are with the Elmore Family School of Electrical and Computer Engineering, Purdue University, West Lafayette, IN 47907 USA (e-mail: duan90@purdue.edu, zhu0@purdue.edu).
\IEEEcompsocthanksitem D. Ding is with the School of Information Science and Technology, Hangzhou Normal University, Hangzhou, Zhejiang 311121, China (e-mail: dandanding@hznu.edu.cn).
}}

%
%

\markboth{Journal of \LaTeX\ Class Files,~Vol.~14, No.~8, August~2015}%
{Shell \MakeLowercase{\textit{et al.}}: Bare Demo of IEEEtran.cls for Computer Society Journals}
%



\IEEEtitleabstractindextext{%
\begin{abstract}
The enhanced Deep Hierarchical Video Compression—DHVC 2.0—has been introduced. This single-model neural video codec operates across a broad range of bitrates, delivering not only superior compression performance to representative methods but also impressive complexity efficiency, enabling real-time processing with a significantly smaller memory footprint on standard GPUs. These remarkable advancements stem from the use of hierarchical predictive coding. Each video frame is uniformly transformed into multiscale representations through hierarchical variational autoencoders. For a specific scale's feature representation of a frame, its corresponding latent residual variables are generated by referencing lower-scale spatial features from the same frame and then conditionally entropy-encoded using a probabilistic model whose parameters are predicted using same-scale temporal reference from previous frames and lower-scale spatial reference of the current frame. This feature-space processing operates from the lowest to the highest scale of each frame, completely eliminating the need for the complexity-intensive motion estimation and compensation techniques that have been standard in video codecs for decades.  The hierarchical approach facilitates parallel processing, accelerating both encoding and decoding, and supports transmission-friendly progressive decoding, making it particularly advantageous for networked video applications in the presence of packet loss. Source codes will be made available.
\end{abstract}

}

\maketitle

\IEEEdisplaynontitleabstractindextext

%
\IEEEpeerreviewmaketitle

\IEEEraisesectionheading{\section{Introduction}\label{sec:intro}}

%
%
%
%
\IEEEPARstart{R}{ecent} years, we have observed substantial progress in employing deep learning techniques to improve visual data compression~\cite{ma2019image,ding2021advances,quach2022survey}.  By utilizing data-driven learning in an end-to-end fashion, these methods significantly surpass traditional, rule-based approaches for compressing various types of data, including images~\cite{wallace1991jpeg,marcellin2000overview,lainema2012intra}, videos~\cite{wiegand2003overview, sullivan2012overview, bross2021overview}, point clouds~\cite{schwarz2018emerging,Cao_PCC_PIEEE}, among others.

\begin{figure}[t]
\centering
\includegraphics[width=\linewidth]{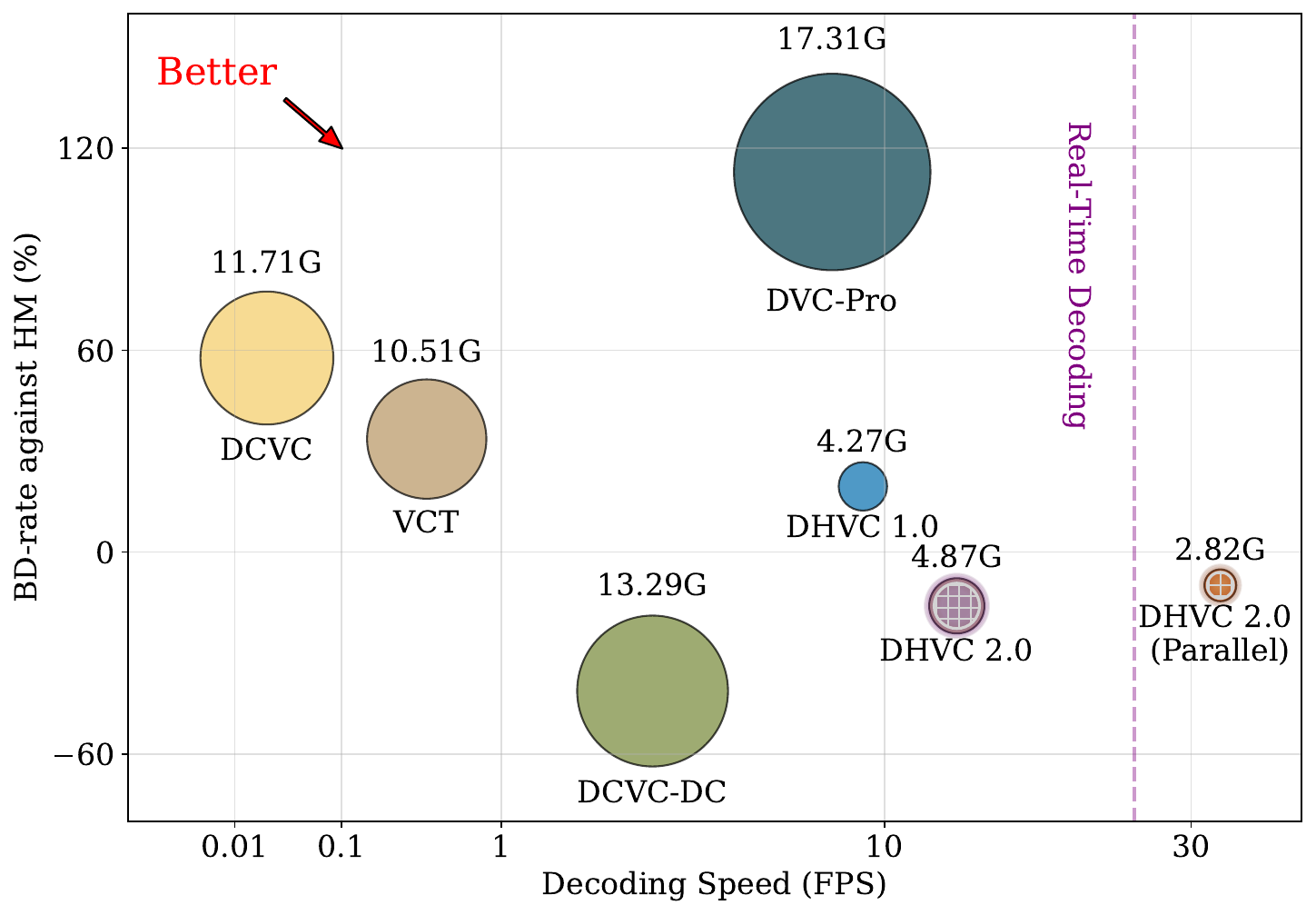}
\caption{{\bf Rate-distortion-complexity} of representative learned video codecs. The horizontal axis represents the decoding speed in frames per second (FPS), and the vertical axis shows the BD-rate~\cite{bjontegaard2001calculation} relative to HEVC reference model HM 16.26. The marker's size is proportional to the running memory of each method. The proposed DHVC 2.0 demonstrates a well-justified performance-complexity balance using a single model (marked with grid pattern). Particularly,  DHVC 2.0 (Parallel) is the only one running in real-time (i.e., over 25 FPS) with superior performance to HM16.26.}
\label{fig:r-d-c}
\end{figure}

\begin{figure*}[t]
\centering
\subfigure[]{\includegraphics[width=0.23\linewidth]{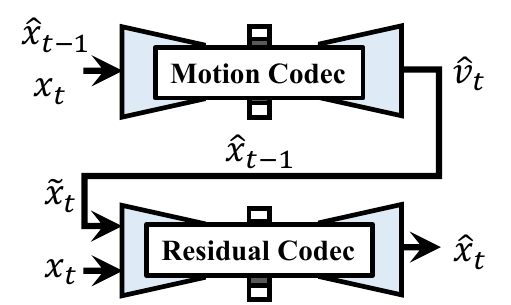}\label{sfig:motion_residual_coding}}
\hspace{0.5cm}
\subfigure[]{\includegraphics[width=0.21\linewidth]{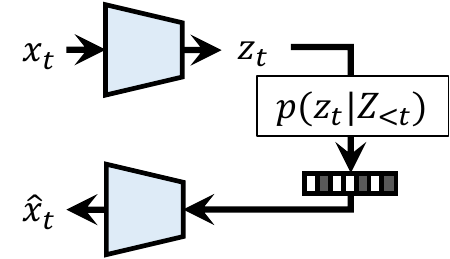}\label{sfig:single-scale_prob_predictive_coding}}
\hspace{0.5cm}
\subfigure[]{\includegraphics[width=0.4\linewidth]{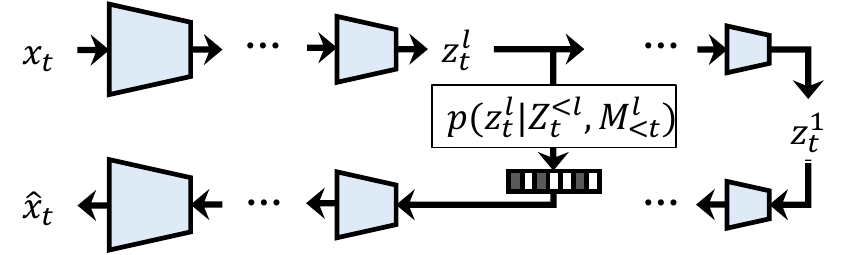}\label{sfig:hier_prob_pred_coding}}
\caption{{{\bf Inter-frame coding} using (a) hybrid motion \& residual coding~\cite{lu2019dvc,li2021deep}, (b) single-scale predictive coding~\cite{liu2020conditional,mentzer2022vct}, and (c) the proposed hierarchical predictive coding~\cite{lu2024deep}. $x$: pixel-space image; $v$: motion flow; $z$: feature-space latent variable; $Z, M$: accumulated reference features for conditional coding.}}
\end{figure*}

Despite the significant advancements, most learning-based methods have primarily focused on developing deep neural network (DNN) models to replace the rule-based tools used in widely adopted standardized codecs. In the context of learned video codecs, DNN models are designed within the same {\it hybrid coding} framework that has dominated for decades~\cite{sullivan2005video,habibi1974hybrid,roese1975combined}. These models leverage spatial correlations within individual frames for intra-frame compression and temporal correlations across consecutive frames for inter-frame compression. For inter-frame coding mode, the two-stage approach illustrated in Fig.~\ref{sfig:motion_residual_coding} is commonly employed: first, the motion representation (e.g., flow) is encoded, followed by the residual between the current and motion-compensated frame, either explicitly~\cite{lu2019dvc} or implicitly~\cite{liu2020neural}. 

However, this hybrid coding design typically requires separate models for intra-frame coding, inter-residual coding, motion coding, and motion estimation, which can be cumbersome and impractical. This is due to two main challenges: 1) the complexity and time-consuming nature of tuning numerous hyperparameters to train multiple models, and 2) the significant space-time complexity involved in both training and inference, which poses considerable difficulties for product-driven applications. Consequently, industry leaders are reluctant to adopt such a promising technology to replace existing video codec solutions like HEVC~\cite{sullivan2012overview}, particularly for real-time video services.

In contrast, this work investigates an alternative and promising approach to completely eliminate hybrid coding. This method, referred to as DHVC 2.0, represents a significant enhancement over our initial DHVC 1.0 (Deep Hierarchical Video Compression) presented in~\cite{lu2024deep}. The key contribution of DHVC lies in its innovative design of hierarchical predictive coding. Specifically, each frame in a video sequence is first transformed into feature-space multiscale representations using hierarchical variational autoencoders, as illustrated in Fig.~\ref{sfig:hier_prob_pred_coding}~\cite{lu2024deep,duan2023lossy,duan2023qarv}.  For the feature-space representation at a given scale of a specific frame, causal reference features from the lower scale of the same frame (e.g., spatially) and/or from the same scale in previous frames (e.g., temporally) are organized to enhance prediction within the conditional coding framework. To achieve this, simple yet efficient DNN models are developed to operate scale by scale, progressing from the lowest to the highest level of each frame, i.e., 1) {\it Spatial abstraction} to create more compact latent residual variables for compression, 2) {\it Entropy parameter estimation} using contexts predicted by spatial and temporal references to more accurately capture the probabilistic distribution of latent residual variables at each scale, and 3) {\it Spatiotemporal compensation} to more effectively restore multiscale decoded features for aggregation into temporal reference(s) or for pixel-space reconstruction.

The proposed hierarchical processing framework enables flexible selection of spatial and temporal features as appropriate conditional reference(s), either independently or in combination, for each scale. This approach integrates variable-rate intra- and inter-coding into a single model, significantly reducing the deployment cost during inference.

Additionally, parallel pipeline processing can be efficiently managed according to the hierarchical scales (see Fig.~\ref{fig:parallel}), greatly enhancing encoding and decoding speed. By fusing temporal reference in feature space for entropy parameter prediction and decoding-path compensation, we can effectively represent and incorporate temporal dependencies ~\cite{mentzer2022vct} without the need for traditional motion representations (e.g., flow estimation, motion coding) used in hybrid coding frameworks. This further decreases space-time complexity.

As a result, the proposed DHVC 2.0 showcases the following highlights based on extensive evaluations conducted with commonly used video sequences and testing conditions:
\begin{itemize}
    \item Competitive compression performance that demonstrates significant improvements over mainstream HEVC, as well as prominent learned video coding methods such as  DVC~\cite{lu2019dvc}, DCVC~\cite{li2021deep}, and VCT~\cite{mentzer2022vct}. It also shows promise in potentially surpassing the latest VVC~\cite{bross2021overview} and the state-of-the-art learned video codec - DCVC-DC~\cite{Li_2023_CVPR} (see Fig.~\ref{fig:rd_curve}); 
    \item Ultra-lightweight space-time complexity that enables real-time processing while significantly reducing memory usage on consumer-level GPUs\footnote{Our tests are performed on NVIDIA RTX 3090.}. Compared to DCVC-DC, our solution accelerates encoding and decoding speed by more than 10$\times$ and offers approximately a 4$\times$ reduction in memory footprint and KMACs per pixel (see Table~\ref{tab:effficiency}).
    \item Transmission-friendly progressive decoding that is inherently supported by the proposed hierarchical coding structure. For networked video applications with unstable connectivity (e.g., fluctuating bandwidth, congestion, packet loss, etc.), our solution can provide smooth streaming without service interruptions, even in the event of lost packets. This capability is particularly advantageous for internet video streaming providers (see Fig.~\ref{fig:loss_resiliance}).  
\end{itemize}

From version 1.0 to 2.0, substantial improvements have been made in DHVC:
\begin{itemize}
    \item Instead of accumulating latent residual variables across frames in DHVC 1.0, DHVC 2.0 has buffered and utilized multiscale decoded features capable of mining fine-grained temporal information for more accurate conditional probability estimation;
    \item  An auxiliary refinement has also been defined in DHVC 2.0, in which multiscale auxiliary features from reconstructed temporal reference frame(s) are extracted and fused with multiscale decoded features to enhance temporal reference features used for conditional entropy coding. In the meantime, this can mitigate the optimization collapse often encountered in hierarchical learning.
    \item A light neural network with depth-wise convolution and spatiotemporal attention has been adopted to replace simple convolutions for efficient processing. Furthermore, parallel encoding/decoding on multiple GPUs is exemplified, by which real-time processing and compelling compression performance\footnote{In contrast, real-time learned video codecs often came with notable performance drop, even significantly worse than HEVC reference model~\cite{MobileNVC}.} can both be guaranteed. 
\end{itemize}

Although still in its early stages, DHVC 2.0's groundbreaking advancements over existing solutions position it as the leading learned video codec from an application standpoint. We intend to make it publicly available shortly after the work is accepted, and we welcome all interested parties to contribute to its further improvement.

\begin{table}[t]
    \centering
    \caption{Notations}
    \label{tab:notation}
    \begin{tabular}{c|c}
    \hline  
      Abbreviation   &  Description \\
    \hline
      DNN & Deep Neural Network \\
      VAE   & Variational AutoEncoder \\
    \hline
    HEVC & High-Efficiency Video Coding~\cite{sullivan2012overview}\\
    VVC & Versatile Video Coding~\cite{bross2021overview}\\
    DVC & Deep Video Compression~\cite{lu2019dvc,lu2020end}\\
    DCVC & Deep Contextual Video Compression~\cite{li2021deep}\\
    DCVC-DC & DCVC with Diverse Contexts~\cite{Li_2023_CVPR}\\
    VCT & Video Compression Transformer~\cite{mentzer2022vct}\\
    DHVC & Deep Hierarchical Video Compression~\cite{lu2024deep}\\
    \hline
    BD-rate & Bj{\o}ntegaard Delta Rate~\cite{bjontegaard2001calculation}\\
    FPS & Frame per second\\
    \hline
    \end{tabular}
\end{table}

\section{Related Work} \label{sec:related_work}

The exploration of neural networks in visual data compression began in the late 1980s~\cite{luttrell1988image}. Recently, advancements in both hardware (e.g., GPUs) and software (e.g., PyTorch) have led to a surge in developing DNN models for visual data compression, particularly in video compression~\cite{ding2021advances}. DNNs can be applied to enhance specific coding tools, such as reference frame generation~\cite{ding2021neural}, adaptive filtering~\cite{ding2023neural}, and intra prediction~\cite{DNN-intra}, within traditional video codecs. Alternatively, DNN models can be designed to replace all coding modules, enabling end-to-end supervised learning aimed at optimizing the overall rate-distortion metric. This work focuses on the latter approach.

\subsection{Hybrid Coding}

Most learned video compression methods adopt the same hybrid coding framework \cite{habibi1974hybrid,roese1975combined} and then carefully design specific DNN models to represent intraframe texture, inter motion, inter residuals, and more.

In 2017, Chen {\it et al.} \cite{chen2017deepcoder} proposed DeepCoder to devise convolutional autoencoders for the representation of intra-predictive and residual signal (after intra- or inter-prediction). In DeepCoder, block-based motion estimation and compensation and H.264/AVC-compliant motion coding were used.

Later in 2019,  DVC was introduced~\cite{lu2019dvc}, in which pixel-domain explicit motion estimation was facilitated using a learnable network model (e.g., SpyNet \cite{ranjan2017optical}), and motion flow as well as the corresponding inter residual were compressed using variational autoencoders originated from~\cite{balle2018variational}. Follow-up enhancements included scale-space flow processing~\cite{agustsson2020scale}, implicit motion representation in either single scale~\cite{liu2020learned} or multiscale manners~\cite{liu2020neural}, feature-space motion and residual processing (under deformable convolutional framework)~\cite{hu2021fvc,hu2022fvc}, recurrent processing of motion and residual~\cite{yang2020learning} etc.

Discussions above dealt with motion and residual coding for interframe compression, where inter residual was generated through simple subtraction in either pixel or feature domain. Such residual coding was widely adopted in conventional video codecs for its tractable implementation in practice to remove redundancy, even though conditional coding could lead to lower Shannon entropy from a source coding perspective. Luckily, the emergence of DNN models has not only shown the powerful capacity to characterize underlying data but also offered much easier means to exploit data correlations in a conditional way.

As a result, Ladune \textit{et al.}~\cite{ladune2020optical} proposed CodecNet, which performed interframe encoding conditioned on temporally-warped reference (instead of deriving inter residual through simple subtractions). Then, Li \textit{et al.}~\cite{li2021deep} extended the idea by introducing DCVC, in which temporal reference(s) were aggregated and mined as contextual priors in the encoder, entropy coder and decoder for conditional coding. Subsequent explorations attempted to improve DCVC by implementing multiscale temporal contexts~\cite{sheng2022temporal,li2022hybrid,sheng2024vnvc}, multistage context modeling~\cite{Li_2023_CVPR,lu2022high}, feature modulation (to mitigate temporal quality degradation across frames)~\cite{li2024neural}, etc. Learnable network models were used to estimate motion flow and conduct motion compensation (e.g., warping) to produce feature-space temporal references in CodecNet, DCVC, and similar variants.

In DCVC and its followup works, the encoder basically leverages DNN models to derive {\it conditional residual} through more generalized linear superimposition and nonlinear activation (via stacked convolutions) instead of naive subtraction, by which more clustered zero-mean Gaussian distribution for underlying latent variable can be obtained for better compression (see illustrative subplots of feature-space latent variable in DVC and DCVC-DC of Fig.~\ref{fig:latent_visualization}).  This partially explains the performance gains of DCVC-alike solutions from earlier solutions such as DVC, etc.

As multiple, separate models were required to represent intraframe texture, interframe conditional residual, and motion in DCVC-alike solutions, they still belonged to the hybrid coding class. Facilitating hybrid coding with separate models is still impractical concerning cumbersome steps in training ~\cite{sheng2022temporal} and excessive space-time complexity  (see Table~\ref{tab:effficiency}).

\subsection{Unified Coding}  

Conditional entropy coding allowed unifying intraframe and interframe compression modes with a single model. This is because compressive latent variables are assumed to follow a probabilistic distribution such as Gaussian or Laplacian~\cite{balle2016density}, for which we just need to aggregate and embed spatial or spatiotemporal reference features as proper contextual conditions to predict entropy model parameters for faithfully capturing the distribution of underlying data in respective intra and inter coding modes.

To this end, Liu {\it et al.}~\cite{liu2020conditional} stacked convolutions to aggregate temporal and hyper features as the contextual condition to model the latent distribution of the current frame, while VCT~\cite{mentzer2022vct} adopted powerful Transformers to include spatial (autoregressive) and temporal features for the same purpose. Both methods exploited temporal correlations through simple feature-space computations (e.g., convolution or self-attention) instead of dedicated motion estimation and compensation used in the hybrid coding framework mentioned above.

However, as only single-scale reference features at a reduced resolution, e.g., $\frac{H}{16}\times\frac{W}{16}$ assuming the original input at a size of $H\times W$, were used in single-scale variational autoencoder (VAE)~\cite{liu2020conditional,mentzer2022vct}, their compression performance was still bounded. In contrast, our preliminary DHVC 1.0~\cite{lu2024deep}  proposed embedding multiscale reference features through hierarchical VAE, leading to much better compression performance, almost 8$\times$ encoding/decoding speedup,  $>8\times$ reduction of KMACs/pixel, and close to 4$\times$ decrease of running memory, when compared with VCT in Table~\ref{tab:effficiency}. 

The performance improvement of DHVC 1.0 from VCT mainly appreciates the multiscale features that can better characterize fine-grained spatiotemporal details to improve conditional entropy coding. Another novel piece is we encode {\it latent residual variables}. Note that scale-wise latent residual variables are derived through cross-scale abstraction using the lower-scale reference of the same frame, resulting in a more concentrated distribution that is earlier to compress. 

The hierarchical structure of DHVC 1.0 also allows us to just apply simple convolutions instead of complicated Transformers.  In VCT, to best characterize dependency between compressive latent variables across successive frames, the deployed Transformer model comes with sizeable parameters (e.g., 750M), which also leads to considerable consumption of running memory (e.g., $>10$G). In the meantime, a local autoregressive model is devised to better aggregate spatial features for referencing, which significantly increases the computational complexity (e.g., more than 3000 KMACs/pixel compared with $\approx 400$ KMACs/pixel required by DHVC 1.0).

This work, DHVC 2.0, inherits the novel hierarchical structure used in DHVC 1.0~\cite{lu2024deep} but substantially improves it with much better performance and lower complexity (see Table~\ref{tab:effficiency} and brief discussions in Sec.~\ref{sec:intro}).

\section{Predictive Coding: From Single-scale to Multiscale Probabilistic Modeling}

This section briefly reviews probabilistic predictive coding for inter-frame compression in learned video codecs.

\subsection{Single-scale Probabilistic Modeling} \label{sec:3-single_predictive_coding}

Assuming a video sequence $X = \{ x_1, ..., x_T\}$ containing $T$ frames, feature-space probabilistic predictive video coding~\cite{habibian2019video,liu2020conditional,mentzer2022vct} generally uses the (same) lossy image compression model to process each frame but applies proper conditional contexts to exploit temporal dependency. As a result, conventional motion estimation and compensation used for interframe compression are avoided~\cite{lu2019dvc,li2021deep}.

\begin{figure*}[htbp]
\centering
\subfigure[Joint~\cite{minnen2018joint}]{\includegraphics[width=0.24\linewidth]{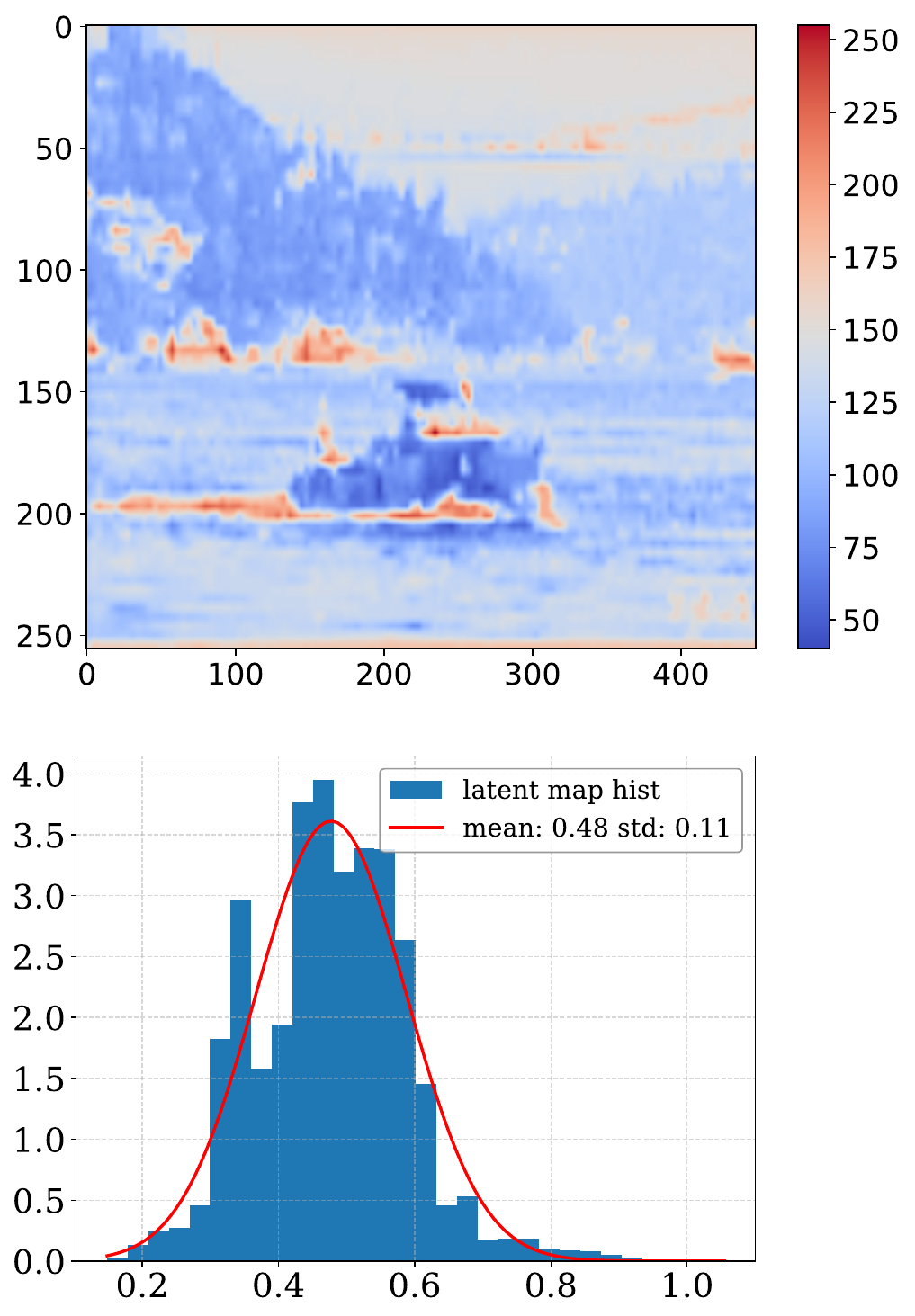}\label{subfig:joint_latent}}
\subfigure[DVC~\cite{lu2019dvc}]{\includegraphics[width=0.24\linewidth]{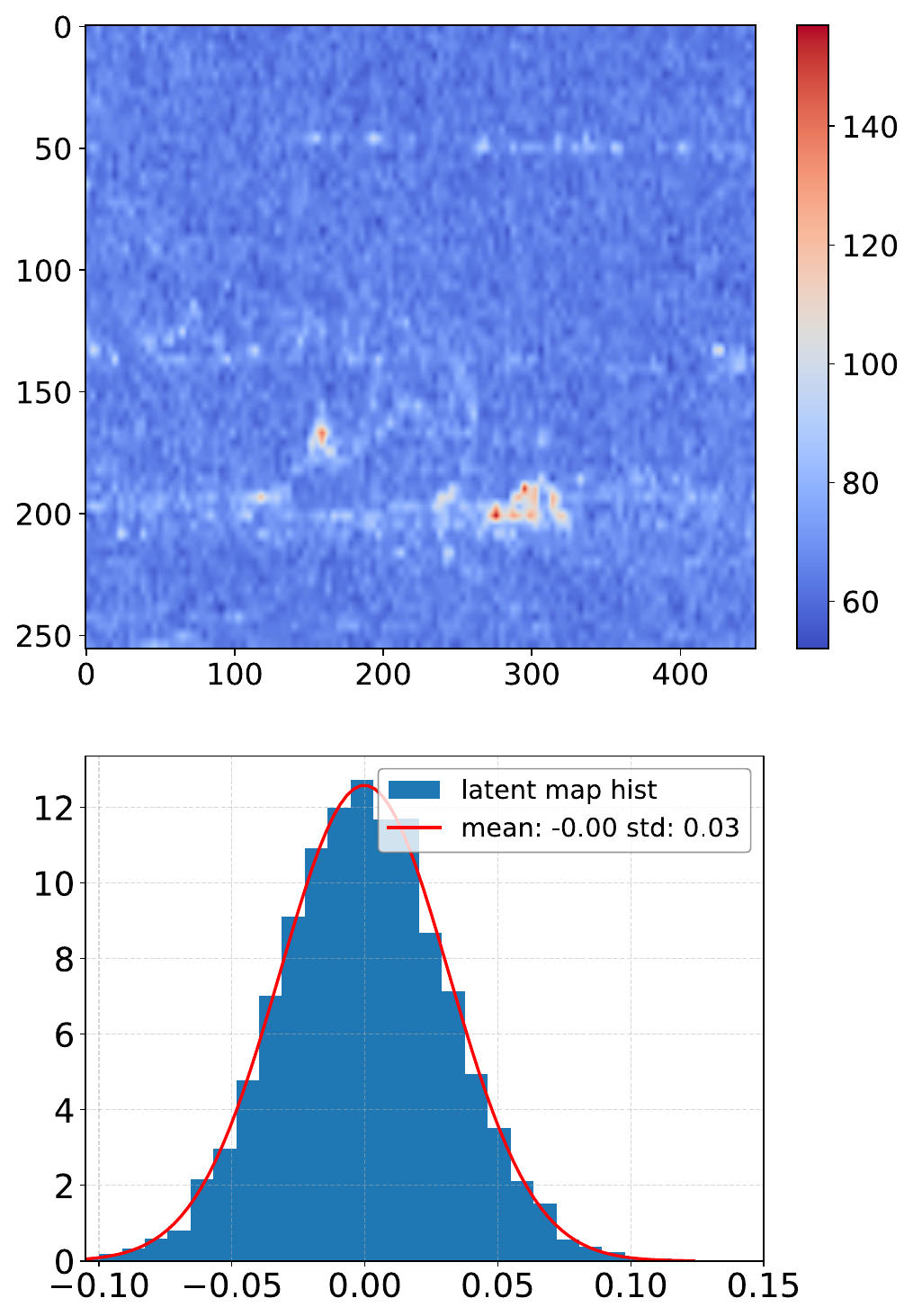}\label{subfig:dvc_latent}}
\subfigure[DCVC-DC~\cite{Li_2023_CVPR}]{\includegraphics[width=0.24\linewidth]{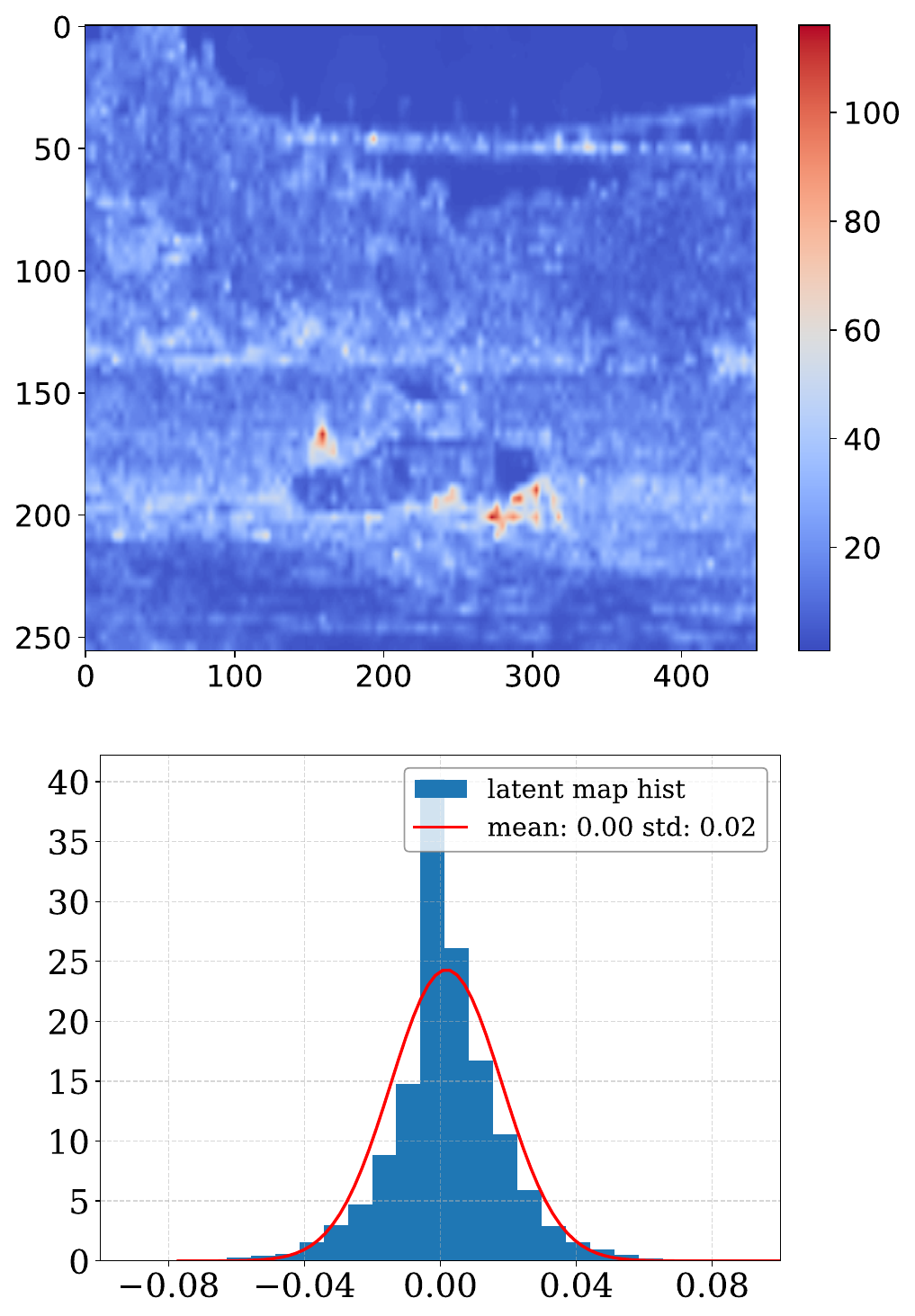}\label{subfig:dcvcdc_latent}}
\subfigure[VCT~\cite{mentzer2022vct}]{\includegraphics[width=0.24\linewidth]{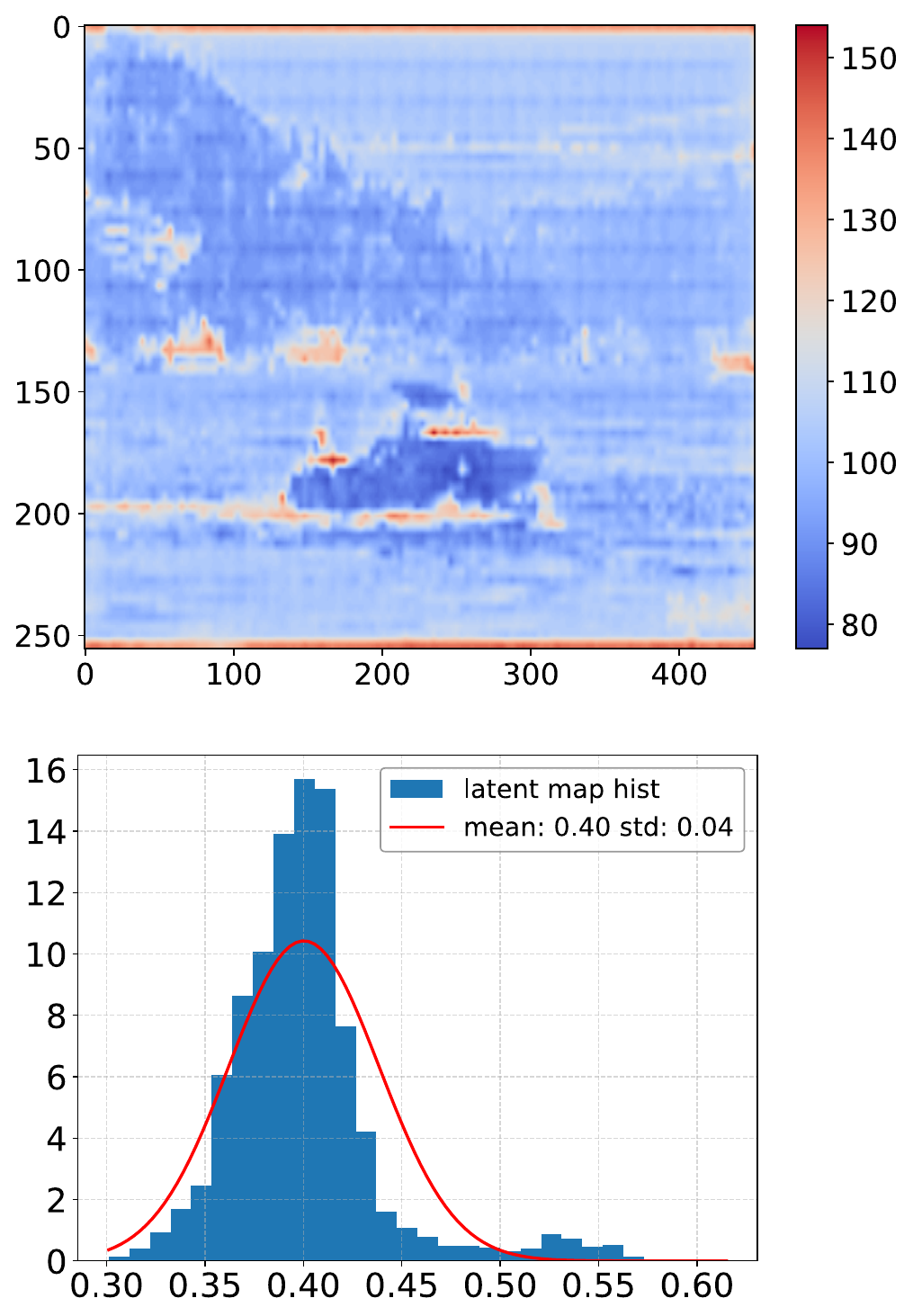}\label{subfig:vct_latent}} \\
\subfigure[DHVC]{\includegraphics[width=0.96\linewidth]{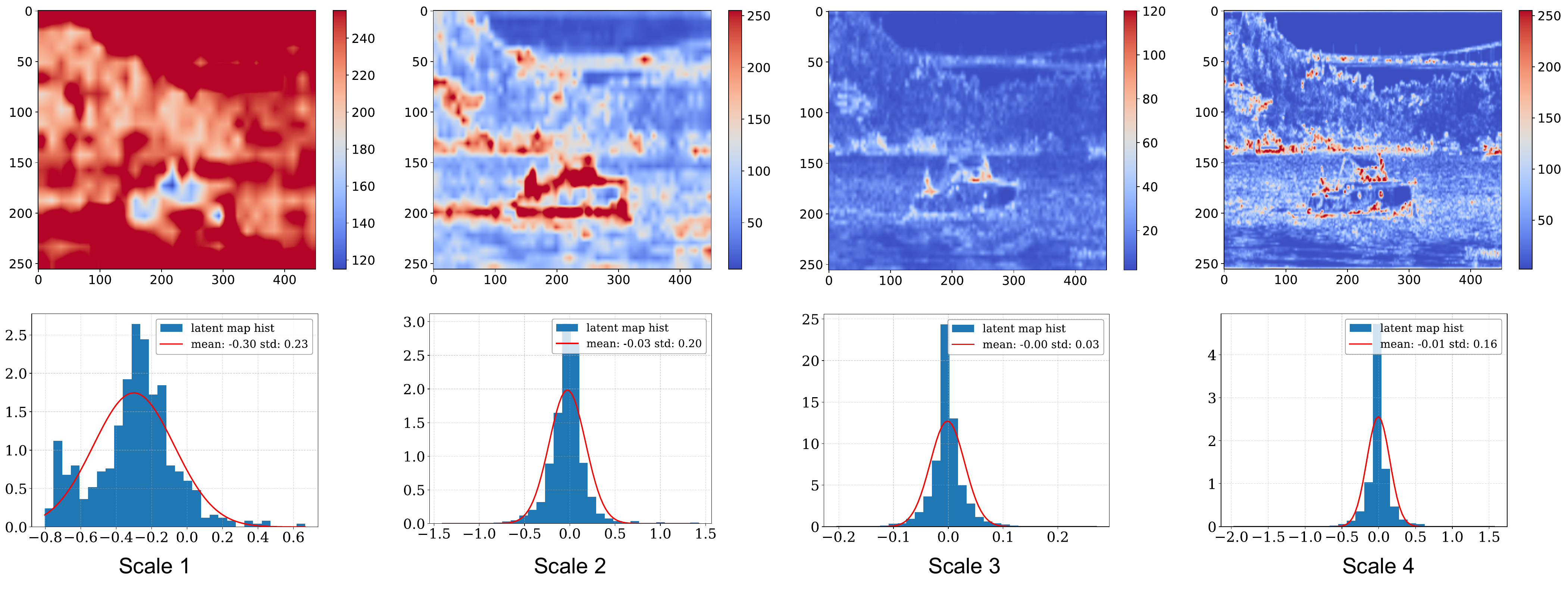}\label{subfig:hierarchical_latent}}
\caption{{\bf Visualizations of the distribution of latent variables} for representative learned compression methods. (a) image coder using Joint~\cite{minnen2018joint}, and video coders using respective (b)  DVC~\cite{lu2019dvc}, (c) DCVC-DC~\cite{Li_2023_CVPR}, (d) VCT~\cite{mentzer2022vct}, and (e) DHVC (hierarchical scale 1 to  4 from the leftmost to rightmost subplot).
Joint and VCT share a similar distribution without implementing a residual coding mechanism as others. DHVC performs scale-wise residual coding from scale 1 (the leftmost subplot) to scale 4 (the rightmost subplot). }
\label{fig:latent_visualization}
\end{figure*}

As proven in~\cite{balle2018variational}, lossy image coding can be implemented using a single-scale VAE. To clarify, the VAE models data (saying image $x_t$) by a joint distribution: 
\begin{equation}
    p(x_t, z_t) = p(x_t \mid z_t) \cdot p(z_t),
\end{equation}
where $z_t$ is latent variables transformed from $x_t$ using an approximate posterior $q(z_t \mid x_t)$ (i.e., the encoder). $p(z_t)$ describes the prior distribution, and a decoder $p(x_t \mid z_t)$ maps $z_t$ to pixel-space reconstruction. Letting $\hat{x}_t \sim p(x_t \mid z_t)$ denote the reconstruction, the VAE objective can be written as~\cite{balle2018variational, duan2023lossy}:
\begin{equation}
\label{eq:lossyvae}
    \mathcal{L} = \mathbb{E}_{x_t, z_t} \left[\mathcal{D}_{\rm KL}(q(z_t \mid x_t) \parallel p(z_t)) + \lambda \cdot d(x_t, \hat{x}_t)\right].
\end{equation} 
If the posterior $q(z_t \mid x_t)$ is deterministic and discrete (e.g., when quantization is applied to $z_t$), optimizing~\eqref{eq:lossyvae} is equivalent to the rate-distortion optimization in lossy coding as proven in~\cite{balle2018variational}. The first term, Kullback–Leibler (KL) divergence $\mathcal{D}_{\rm KL}$, is turned into the (cross-) entropy of the latent variable $z_t$ to approximate the rate, $d(\cdot)$ is a distortion function, and $\lambda$ is the Lagrange multiplier that balances the rate and distortion tradeoff.

Among them, $p(z_t)$ serves as the probability mass function (PMF) to estimate the true distribution of symbols in $z_t$, allowing for lossless entropy coding to determine the actual bitrate needed for transmitting the frame. Existing learned image coding typically relied on using spatial or channel-wise auto-regressive contexts~\cite{minnen2018joint,Minnen2020ChannelWise} of the same frame to approximate $p(z_t)$ as close as possible. For videos, probabilistic predictive coding~\cite{habibian2019video,liu2020conditional} proposes to include temporal reference(s) as conditional contexts to parameterize priors for $z_t$:
\begin{align}
\label{eq:latent_condition}
p(z_1, ..., z_T) = \prod_{t=1}^T p(z_t \mid Z_{<t}),
\end{align}
where $Z_{<t}$ represents the set of temporal reference features preceding time $t$. By exploiting the temporal redundancy across frames in $Z_{<t}$ using an efficient prediction network, one can achieve more accurate probability estimation for latent variables of the current frame, thus reducing cross-entropy and thereby improving coding efficiency.

To enhance the capacity of the underlying probabilistic model for conditional probability estimation, VCT~\cite{mentzer2022vct} further introduced a block-level autoregressive model to augment additional spatial contexts:
\begin{align}
    p(z_t \mid Z_{<t}) = \prod_{i} p(z^i_t \mid z^{<i}_t, Z_{<t})
    \label{eq:cond_vct}
\end{align}
Here, $z^i_t$ corresponds to a latent token at position $i$ within a predefined block window of the current frame, while $z_t^{<i}$ denotes autoregressive neighbors that have been processed previously. As a result, compared to~\cite{liu2020conditional} that only concerns temporal (and hyper) reference mining for conditional contexts, VCT improves compression performance but comes with unbearable computational complexity that owes to the sequential processing of the autoregressive model with large-scale Transformer networks.

\subsection{Multiscale Probabilistic Modeling} \label{sec:3-hierarchical_predictive_coding}

Mining spatial or spatiotemporal reference features at a reduced, single-scale resolution in \eqref{eq:latent_condition} and \eqref{eq:cond_vct} is challenging to provide sufficient contexts for well-justified performance~\cite{liu2020conditional,mentzer2022vct}. This is because latent variables {$z_t$ are often presented at a resolution of $\frac{H}{16}\times\frac{W}{16}$, largely lacking fine-grained details and thus impeding the model performance. 

To overcome the limitation of single-scale probabilistic predictive coding shown in \eqref{eq:latent_condition} and \eqref{eq:cond_vct}, hierarchical VAEs~\cite{kingma2016improved, child2020very, vahdat2020nvae} were proposed and applied to a given image $x_t$ to have multiscale latent variables, denoted by $Z_t = \{z^1_t, ..., z^L_t\}$. Following the deductions discussed above, conditional probabilistic distribution of latent variables at $l$-th scale of image $x_t$ can be formulated {jointly} as:}
{\begin{align}
    p(Z_t) = \prod_{l=1}^Lp(z_t^l \mid Z_t^{<l}, M_{<t}^l), \label{eq:3-hierarchical_conditional} 
\end{align}
where $L$ is the total number of hierarchical scales, remaining the same for all frames in this study. Typically, $l = 1$ corresponds to the smallest scale,  while $l = L$ is for the largest spatial dimension.} $Z^{<l}_t$ includes latent variables preceding $l$-th scale, i.e., $Z^{<l}_t = \{z^1_t, ..., z^{l-1}_t\}$ from the same frame. In the meantime, {$M_{<t}^l = \{{m^{l}_{t-1},m^{l}_{t-2}},\cdots\}$ contains same-scale (e.g., $l$-th) features from temporal references preceding timestamp $t$. Using proper $M_{<t}^l$ is vital for compression efficiency, which is detailed in subsequent sections and experimentally evaluated in Sec.~\ref{sec:Ablation}. When omitting $M_{<t}^l$ in \eqref{eq:3-hierarchical_conditional}, each frame is encoded as independent image~\cite{duan2023lossy,duan2023qarv}. Typically, a proper prediction network is devised to fuse $Z^{<l}_t$ and $M_{<t}^l$ for contextual feature $c_t^l$ as in \eqref{eq:spatio-temporal-contexts}.

\subsection{Discussion} \label{sec:3-discussion}

Even though single-scale and multiscale probabilistic modeling, as respectively presented in \eqref{eq:cond_vct} and \eqref{eq:3-hierarchical_conditional}, have employed similar mechanisms to include conditional contexts to approximate the latent distribution, they are fundamentally different.

First, the hierarchical structure used in our serial explorations~\cite{duan2023lossy,lu2024deep} basically employs {\it conditional residual coding} scale by scale, where (causally available) spatial features from the lower scales are referenced to generate latent residual variables (through spatial abstraction with current-scale features) and aggregated with temporal reference features to predict probabilistic model parameters to entropy-encode the aforementioned latent residual. In contrast, VCT~\cite{mentzer2022vct}, a state-of-the-art single-scale predictive coding approach, only aggregates spatial neighbors with temporal reference(s) as conditional contexts to approximate the distribution of original latent variables but not latent residues.

\begin{figure}[t]
\centering
\includegraphics[width=0.95\linewidth]{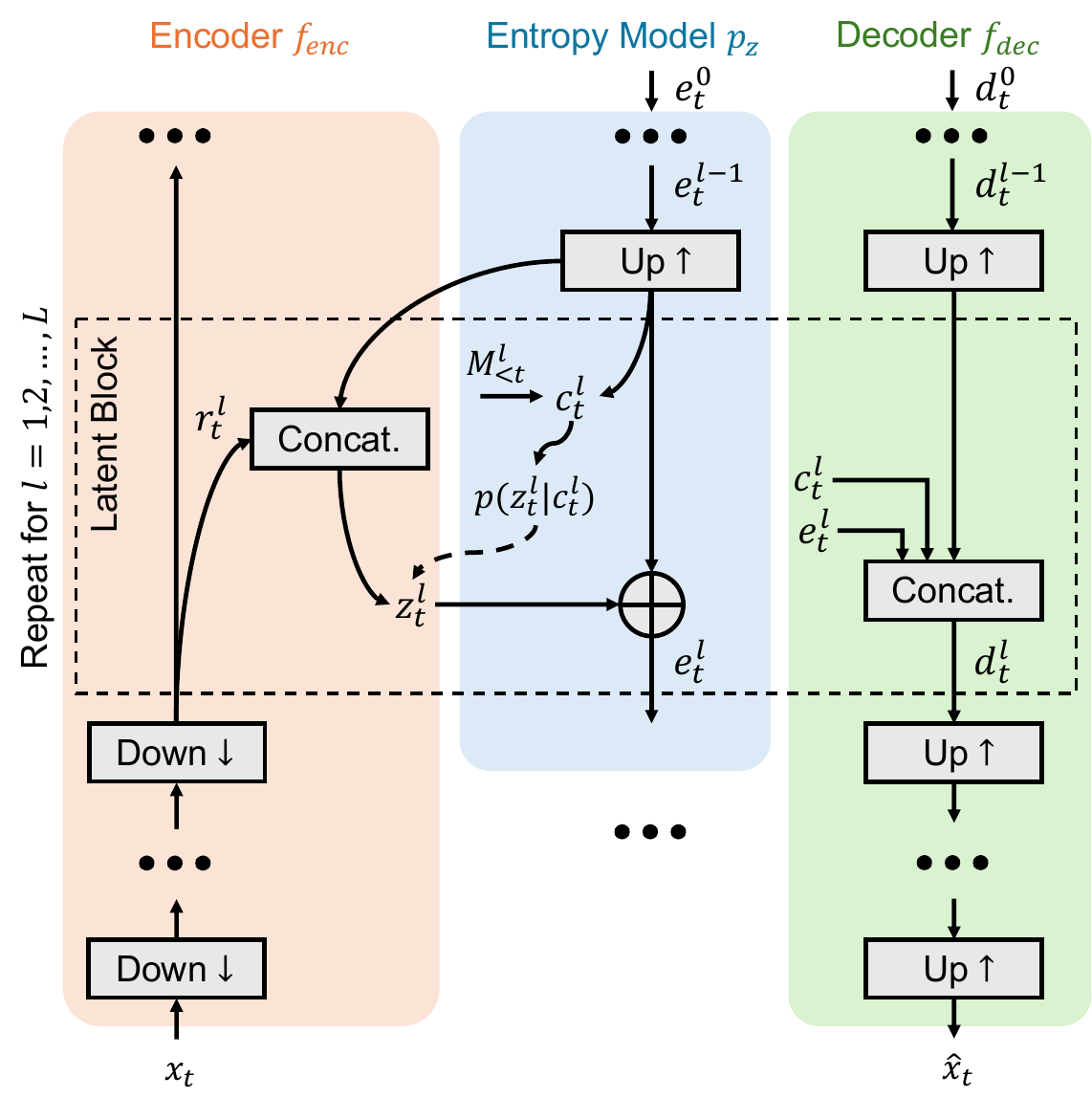}
\caption{{\bf A sketch of hierarchical predictive coding} framework used by DHVC. As scale-wise processing is repeated, we only exemplify the $l$-th scale.  ``Down $\downarrow$'' \& ``Up $\uparrow$'' stand for down- \& up-sampling, and ``Concat.'' represents feature-space concatenation. Neural networks are stacked to realize these operations.}
\label{fig:overview}
\end{figure}

Compressing the appropriate residues of given latent variables likely performs better than compressing latent variables themselves. As comparatively illustrated in Figs.~\ref{subfig:joint_latent},~\ref{subfig:vct_latent} and \ref{subfig:dvc_latent},~\ref{subfig:dcvcdc_latent},~\ref{subfig:hierarchical_latent}, the residual contains less information than the original signal and exhibits a more concentrated distribution, which can be better characterized using zero-mean Gaussian or Laplacian~\cite{balle2016density}. As shown in Fig.\ref{subfig:hierarchical_latent}, the second to fourth subplots of ``DHVC", corresponding to the second to fourth hierarchical scales, exhibit a similar concentrated distribution to those of ``DVC" and ``DCVC-DC" in Fig.~\ref{subfig:dvc_latent} and Fig.~\ref{subfig:dcvcdc_latent}, which also use residual coding. DCVC-DC uses conditional residual coding with generalized linear superimposition instead of the simple abstraction used in DVC. This results in a more concentrated and compressible latent distribution in Fig.~\ref{subfig:dcvcdc_latent}, partially explaining DCVC-DC's gains over DVC \cite{li2021deep}.

To facilitate the conditional residual coding,  we propose using ResNet VAE~\cite{kingma2016improved}, with which the encoding of $z^l_t$ depends on both $r^l_t$ ($l$-th scale representation or observation extracted from $x_t$) and $Z^{<l}_t$. The loss function for training ResNet VAEs can be extended from~\eqref{eq:lossyvae} for supervising multiscale optimization:
\begin{equation} 
\label{eq:lossyhvae}
    \mathcal{L} = \mathbb{E}_{x_t, Z_t} \left[\sum_{i=1}^L \mathcal{D}_{\rm KL}(q^l_t \parallel \, p^l_t)
    +
    \lambda \cdot d(x_t, \hat{x}_t)\right],
\end{equation}
where $q^l_t$ and $p^l_t$ are shorthand notations for the posterior and prior for the $l$-th scale latent residual variable, i.e., $q^l_t  = q(z^l_t \mid r^l_t, Z^{<l}_t),\mbox{~~and~~} p^l_t  = p(z^l_t \mid Z^{<l}_t, M^l_{<t}).$

Furthermore, multiscale probabilistic modeling in \eqref{eq:3-hierarchical_conditional} refines from a lower scale to a higher one, capturing coarse-to-fine characteristics for accurate data modeling. This approach surpasses single-scale predictive coding methods, offering superior performance compared to single-scale autoregressive models in distribution modeling \cite{duan2023qarv}.

\section{Hierarchical Predictive Coding} \label{sec:revisit_hierarchical_probabilistic_modeling}

This section details the proper use of spatial and temporal reference features for prediction in conditional coding.

\subsection{Hierarchical Architecture}~\label{sec:4-overview}

Figure~\ref{fig:overview} briefly depicts the hierarchical predictive coding architecture DHVC uses, involving a bottom-up path for the encoder $f_{enc}$ and two separate top-down paths for entropy coder using model $p_z$ and decoder $f_{dec}$. 

Given an input $x_t$, the bottom-up encoder $f_{enc}$ produces a hierarchy of observations (multiscale representations), e.g., $\{r_t^L, ..., r_t^2, r_t^1\}$, at different resolutions through successive downsampling $\mathcal{S_{D}}()$. Then, the encoding starts from the very first scale with $l=1$ and progressively refines to the highest scale with $l=L$.

For a given $l$-th scale, instead of compressing the latent observation $r_t^l$ directly, latent residual variables $z^l_t$ are first generated using $r^l_t$ and spatial reference accumulated from lower scales, e.g., $e^{l-1}_t$: 
\begin{align}
    z^l_t = \mathcal{G_R}(\mathcal{S_{U}}(e^{l-1}_t), r^{l}_t), \label{eq:spatial-abstraction}
\end{align} which we refer to as spatial abstraction. And then, $z^l_t$ is entropy-encoded using a conditional probability model $p_z$ having $ p(Z_t) = \prod_{l=1}^Lp(z_t^l \mid c_t^l)$. Its spatiotemporal contextual features $c_t^l$ are predicted using $e^{l-1}_t$ and $M^l_{<t}$ via 
\begin{align}
    &c_t^l = \mathcal{P_{CTX}}(\mathcal{S_{U}}(e^{l-1}_t), M^l_{<t}). \label{eq:spatio-temporal-contexts}
\end{align}
Here, $c_t^l$ is simply processed using 1$\times$1 convolutions to derive the entropy parameters mean $\hat{\mu}^l_t$ and scale $\hat{\sigma}_t^l$ of a Gaussian distribution to characterize the prior. Spatial reference $e_t^{l-1}$ starts from a learnable constant $e_t^0$ and aggregates all the information from $Z_t^{<l}$. The encoded feature $z_t^l$ is thus added with $\mathcal{S_U}(e_t^{l-1})$ to have $e_t^l$ for the next scale, e.g., 
\begin{align}
    e_t^l = \mathcal{S_U}(e_t^{l-1}) + z_t^l. \label{eq:spatial_ref}
\end{align} Such a recurrent refinement of $e_t^l$ from one scale to another allows us to best aggregate multiscale characteristics for better modeling. $M^l_{<t}$ comprises $l$-th scale temporal reference features from the past frames, including decoded features $D_{<t}^l$ and auxiliary features $a_{t-1}^l$ (if applicable). By default, we use two previous frames. 

Alongside the above entropy coding at $l$-th scale, the corresponding in-loop top-down decoder $f_{dec}$ aggregates decoded feature $d_t^{l-1}$ from preceding scale, spatiotemporal contextual feature $c_t^l$ and spatial reference feature $e_t^l$ for spatiotemporal compensation to generate $d_t^{l}$, restoring the final pixel-space reconstruction $\hat{x}_t$ or being accumulated into $D_{<t}^l$ as a portion of temporal reference $M^l_{<t}$ to help compress next frame:
\begin{align}
    d_t^{l} = \mathcal{G_D}( \mathcal{S_U}(d_t^{l-1}), c_t^l, e^l_t). \label{eq:decode_feature}
\end{align} 
Similar to the processing of spatial reference feature, $d_t^t$ also starts with a learnable constant $d_t^0$ and progressively refines as formulated in \eqref{eq:decode_feature}.

Functions mentioned above such as downsampling $\mathcal{S_{D}}()$, upsampling $\mathcal{S_U}( )$, latent residual generation $\mathcal{G_R}( )$, contextual prediction $ \mathcal{P_{CTX}}()$, and current-frame decoded feature generation $\mathcal{G_D}()$ are implemented using neural networks whose details are given in subsequent Sec.~\ref{sec:neural_network}.

{\bf Remark.} As in Fig.~\ref{fig:overview}, we intentionally separate the entropy coder and feature decoder paths, which is a significant departure from the design used in image coders \cite{duan2023lossy,duan2023qarv}. In video compression, a single model for characterizing $z_t^l$ isn't feasible. This is because intra coding uses only spatial reference, whereas inter coding uses both spatial and temporal references, resulting in significantly different $e^{l-1}_t$ and $z_t^l$. 

The separation proposed in DHVC limits latent residual generation to spatial references from preceding scales of the same frame (see \eqref{eq:spatial-abstraction}), allowing the decoder path to focus on generating decoded features for appropriate temporal contexts in conditional entropy coding. This separation not only unifies intra and inter coding under a single model but also supports continuous learning, extending the model's capacity when encountering unseen content \cite{duan2024towards}.

\begin{figure*}[t]
\centering
\subfigure[Training]{\includegraphics[width=0.354\linewidth]{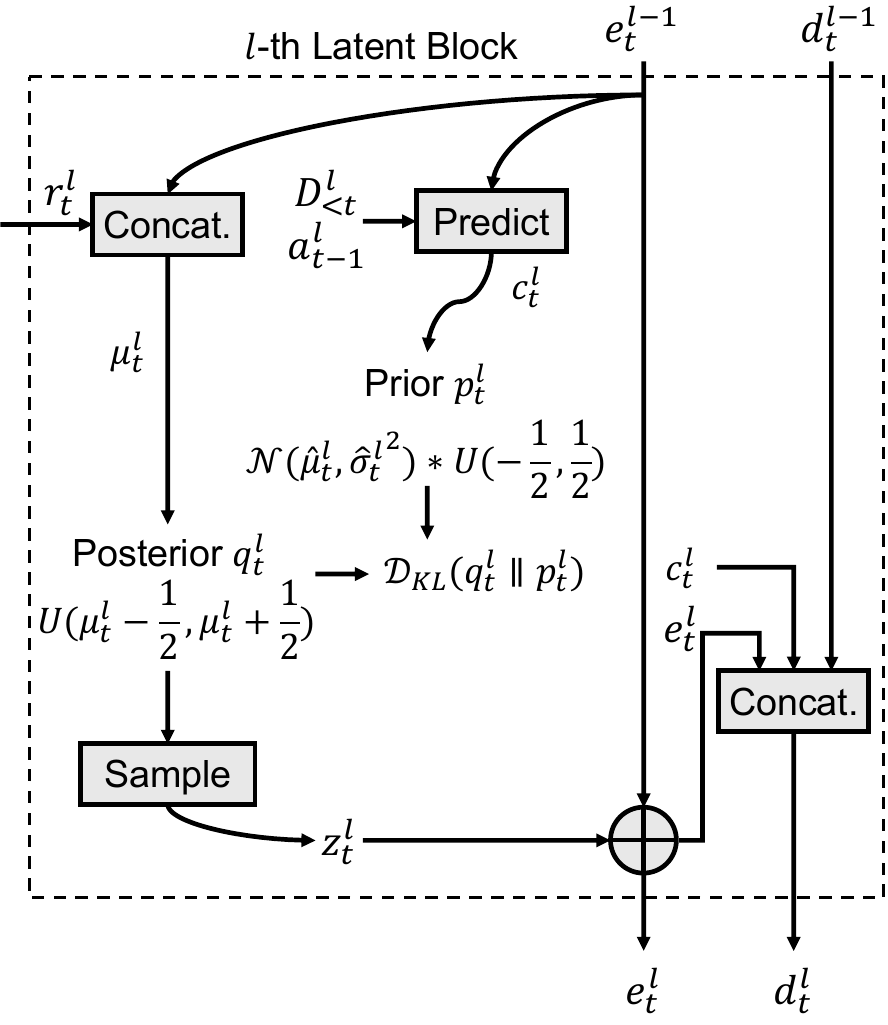}\label{fig:4-training}}
\hspace{0.5cm}
\subfigure[Testing: Compress]{\includegraphics[width=0.28\linewidth]{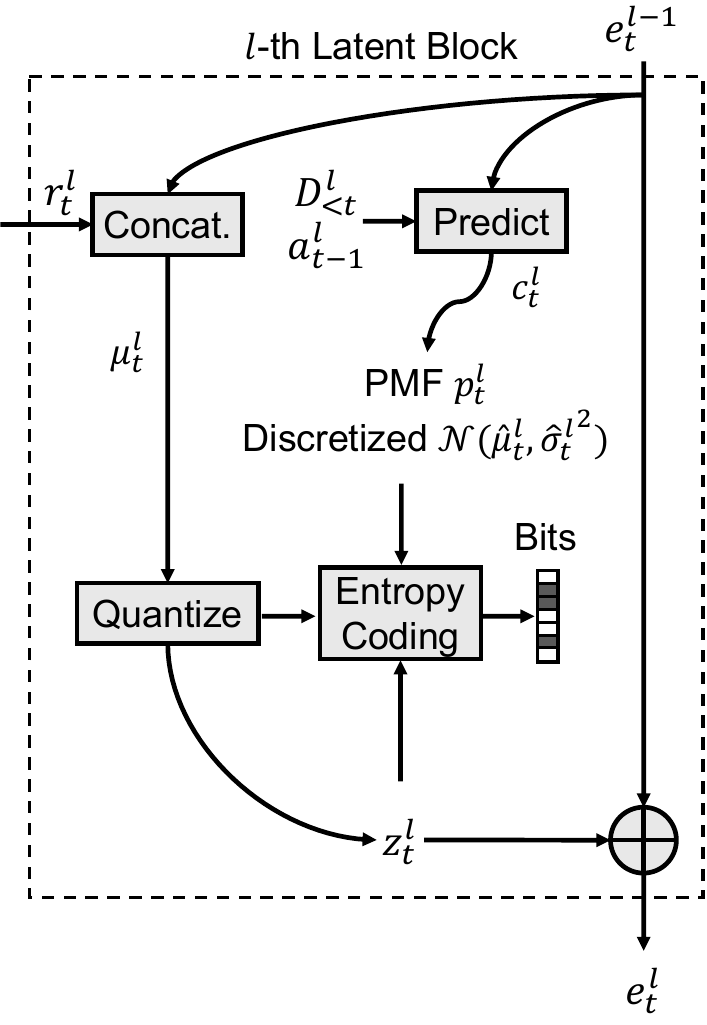}\label{fig:4-compress}}
\hspace{0.5cm}
\subfigure[Testing: Decompress]{\includegraphics[width=0.27\linewidth]{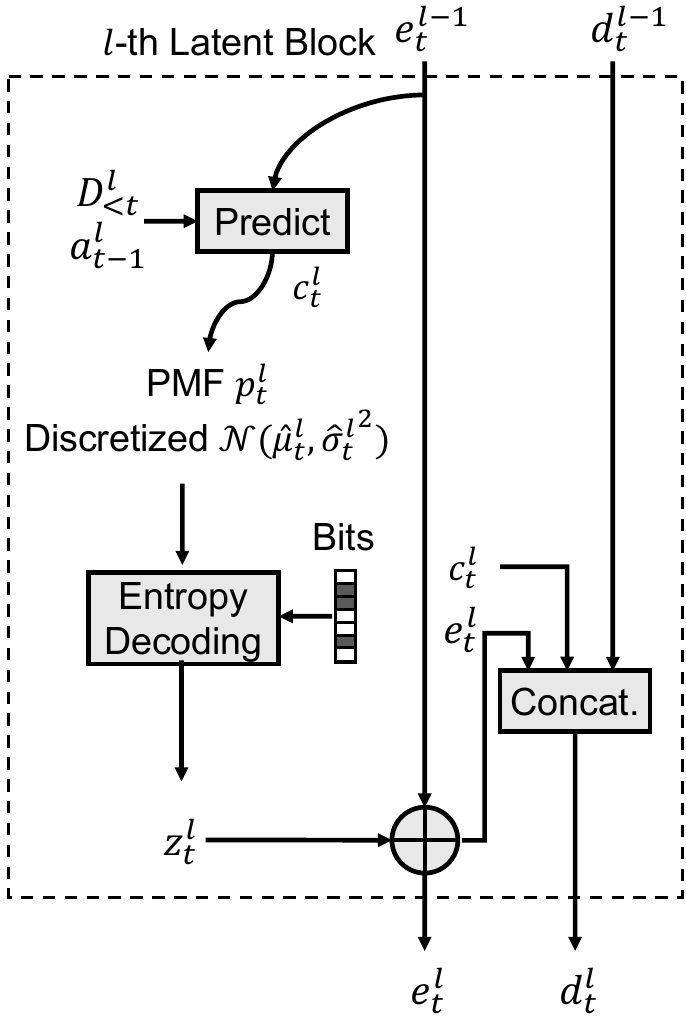}\label{fig:4-decompress}}
\caption{{\bf Latent Block} for conditional coding at each scale, which has thee modes: (a) training, (b) compressing, and (c) decompressing. In training, our model learns to minimize the KL divergence between all pairs of priors and posteriors, as well as the distortion between $\hat{x}_t$ and $x_t$. In compression inference, the uniform posteriors are replaced by uniform quantization, and the priors are turned into discrete PMFs (probability mass functions) for entropy coding.}
\label{fig:latent_block_task}
\end{figure*}

\subsection{Mining Temporal References as Conditional Contexts} \label{sec:4-spatial_temporal_predictive_learning}

Regarding the conditional coding above,  accurate prediction of spatiotemporal contextual features $c_t^l$ is a decisive factor controlling the compression performance. In \eqref{eq:spatio-temporal-contexts}, $c_t^l$ is determined by the proper construction of spatial and temporal references, e.g., $e^{l-1}_t$ and $M^l_{<t}$. As using spatial reference is already discussed in our image coding works~\cite{duan2023qarv}. This section focuses on constructing effective temporal references.

\subsubsection{Multiscale Decoded Features} \label{sec:temp_ref_decoded}

In DHVC 1.0~\cite{lu2024deep}, we simply accumulate latent residual variables $Z_{<t}^l$ to form temporal reference $M_{<t}^l$, which, for sure, is not sufficient as $Z_{<t}^l$ contains limited information after predictive coding.

Thus, we propose accumulating decoded features to form the corresponding temporal reference, yielding 
\begin{align}
    p(Z_t) = \prod_{l=1}^Lp(z_t^l \mid e_t^{l-1}, D_{<t}^l),
\end{align}
where $e_t^{l-1}$ represents spatial reference feature aggregated from the preceding scales at the current frame\footnote{Proper upsampling is devised to ensure the resolution consistency. Here, we omit it for simplicity.}, and $D_{<t}^l$ stand for temporal reference features accumulated from the preceding time at the same scale, e.g.,  $D_{<t}^l = \{d^l_{t-1}, d^l_{t-2}, \ldots\}$. As in \eqref{eq:decode_feature}, the decoded features comprise the information from encoded latent residual, spatial, and temporal references. 
In theory, the full information in the past frames can be utilized progressively scale by scale in feature space to form better contextual features for improving conditional entropy coding. 

\subsubsection{Auxiliary Refinement} \label{sec:temp_ref_aux}

Our hierarchical model has the potential to provide higher-quality predictions by capturing multiscale characteristics. However, the bidirectional dependencies between hierarchical latent variables—where higher-scale variables affect lower-scale ones and vice versa—can lead to optimization instability, particularly as the number of latent variables increases. While our deterministic hierarchical coding helps mitigate the issue of ``posterior collapse" seen in hierarchical VAEs~\cite{vahdat2020nvae,li2022out}, it still faces the challenge of information being concentrated at higher scales, resulting in inadequate representation capacity at lower scales. This problem becomes more pronounced in inter coding.

To enhance the latent features, particularly at lower scales and lower bitrates (with less information), we propose an auxiliary refinement method by aggregating additional features extracted from previously reconstructed pixel-space frame $\hat{x}_{t-1}$ to assist conditional distribution modeling, i.e.,
\begin{align}
    p(Z_t) = \prod_{l=1}^Lp(z_t^l \mid e_t^{l-1}, D_{<t}^l, a_{t-1}^l), \label{eq:spatial_temporal_aux}
\end{align}
where $a_{t-1}^l$ denotes the extracted feature at $l$-th scale. The derivation of $a_{t-1}^l$ uses simple ResBlock and downsampling (see Fig.~\ref{fig:network_overview}).
Starting from the reconstruction of the previous frame, we can effectively obtain information across different scales without being constrained by the hierarchical optimization issues mentioned above. This enhances the accuracy of hierarchical predictions, particularly at lower scales, and improves performance at higher scales.

Next, we will show how to perform multiscale probabilistic modeling to facilitate DHVC.

\subsection{Multiscale Probabilistic Modeling}~\label{sec:4-hierarchical_probabilistic_modeling}

{The probabilistic model in each latent block for training is depicted in Fig.~\ref{fig:4-training}. To support practical lossy compression using feasible entropy coding algorithms, we follow previous works~\cite{balle2018variational,duan2023lossy,duan2023qarv} and apply quantization-aware training using uniform posterior. 
We now give formal definitions below.}

\subsubsection{Posterior and Prior}

{The (approximate) posterior for the latent residual variable $z_t^l$ is uniformly distributed:
\begin{equation}
q(z_t^l \mid x_t, e_t^{l-1}) = \mathcal{U}(\mu_t^l - \frac{1}{2}, \mu_t^l + \frac{1}{2}),
\end{equation}
where $\mu_t^l$ is the output of the posterior branch $q_t^l$ in the latent block (see Fig.~\ref{fig:4-training}) by combining the observed feature $r_t^l$ and spatial reference $e_t^{l-1}$ from the previous scale. $z_t^l$ is sampled from $\mu_t^l$ with additive noise for entropy coding while straight-through quantization is used for aggregation with $e_t^{l-1}$.}

{Then, the prior distribution for each $z_t^l$ is defined as a Gaussian convolved with a uniform distribution:
\begin{equation}
\label{eq:method_prior}
\begin{aligned}
p(z_t^l \mid e_t^{l-1}, D_{<t}^l, a_{t-1}^l) = \mathcal{N}(\hat{\mu}_t^l, {\hat{\sigma}_t^l}\textsuperscript{\textsuperscript{2}}) * \mathcal{U}(- \frac{1}{2}, \frac{1}{2}),
\end{aligned}
\end{equation}
where $\mathcal{N}(\hat{\mu}_t^l, {\hat{\sigma}_t^l}\textsuperscript{\textsuperscript{2}})$ is Gaussian probability density function with the mean $\hat{\mu}^l_t$ and scale $\hat{\sigma}_t^l$ predicted in the prior branch $p_t^l$. Note that the parameters $\hat{\mu}_t^l$ and $\hat{\sigma}_t^l$ are dependent on both same-scale features from previous frames, e.g., $D_{<t}^l$ \& $a_{t-1}^l$ and same-frame lower-scale spatial reference, e.g., $e_t^{l-1}$.}

\subsubsection{Training Objective} {Typically, the hybrid motion and residual coding methods require multi-stage or simultaneous optimization of the optical flow, motion coding, and residual coding networks during the training phase. Differently, the training process of our model is as easy as optimizing a lossy image coding. 

As a result, the loss function $\mathcal{L}$ is extended from Eq.~\eqref{eq:lossyhvae} with the inclusion of temporal dependency:
\begin{align}
    \mathcal{L} 
    \!&=\! \mathbb{E}_{x_t,\Lambda,Z_t}\left[
    \sum_{l=1}^L \mathcal{D}_{KL}(q_t^l \parallel p_t^l)\!+\!\Lambda \!\cdot\! d(x_t,\hat{x}_t)\right]  \label{eq:4-rd_loss}\\ 
    \!&=\! \mathbb{E}_{x_t,\Lambda,Z_t}\left[
    \sum_{l=1}^L \log{\frac{q(z_t^l \mid Z_t^{<l},x_t,\Lambda)}{p(z_t^l \mid e_t^{l-1},D_{<t}^l,a_{t-1}^l,\Lambda)}}\!+\!\Lambda \!\cdot \!d(x_t,\hat{x}_t)\right] \nonumber \\
    \!&=\! \mathbb{E}_{x_t,\Lambda,Z_t}\left[
    \sum_{l=1}^L \log\frac{1}{{p(z_t^l \mid e_t^{l-1}},D_{<t}^l,a_{t-1}^l,\Lambda)}\!+\!\Lambda \!\cdot \!d(x_t,\hat{x}_t)\right].\! \nonumber 
\end{align} $x_t$ follows the training data distribution, $Z_t$ follows the joint distribution of $q_t^1 \cdots q_t^L$, and $\Lambda$ follows a distribution $p_{\Lambda}$ that controls the sampling strategy of the multiplier during training to enable variable rate compression.

\subsubsection{Compression and Decompression}

As for actual compression and decompression, the overall framework in Fig.~\ref{fig:overview} is unchanged, except the latent residual variables in Latent Block are quantized for entropy coding. Generally, we uniformly quantize $\mu_t^l$ from posterior instead of additive uniform noise or straight through estimator at training time. We also discretize the prior to form a discretized Gaussian probability mass function (PMF) $p_t^l$ for practical entropy coding.

\begin{figure*}[htbp]
\centering
\subfigure[Architecture Overview]{\includegraphics[width=0.9\linewidth]{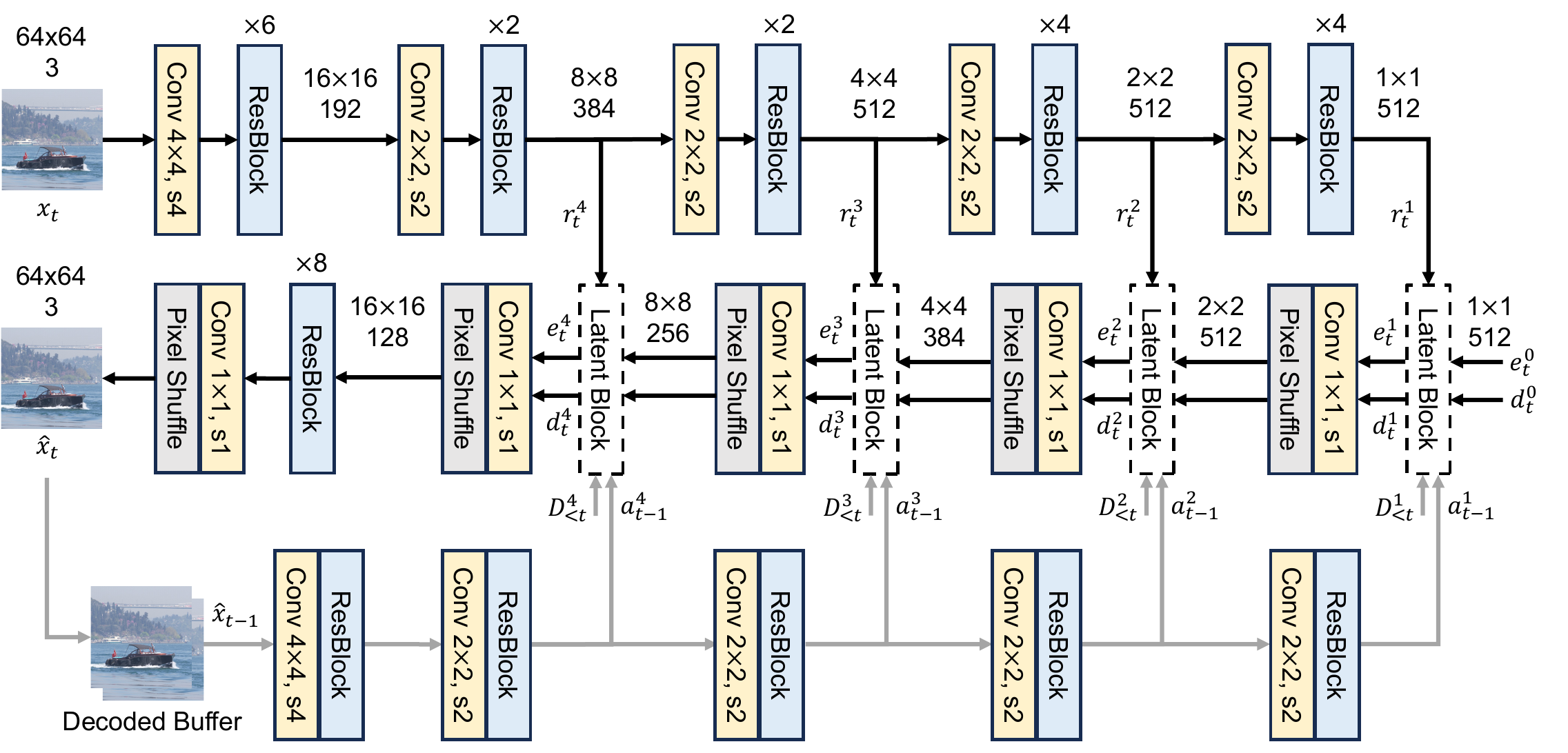}\label{fig:network_overview}} \\
\vspace{0.5cm}
\begin{minipage}{0.42\linewidth}
\centering
\subfigure[Latent Block]{\includegraphics[width=\linewidth]{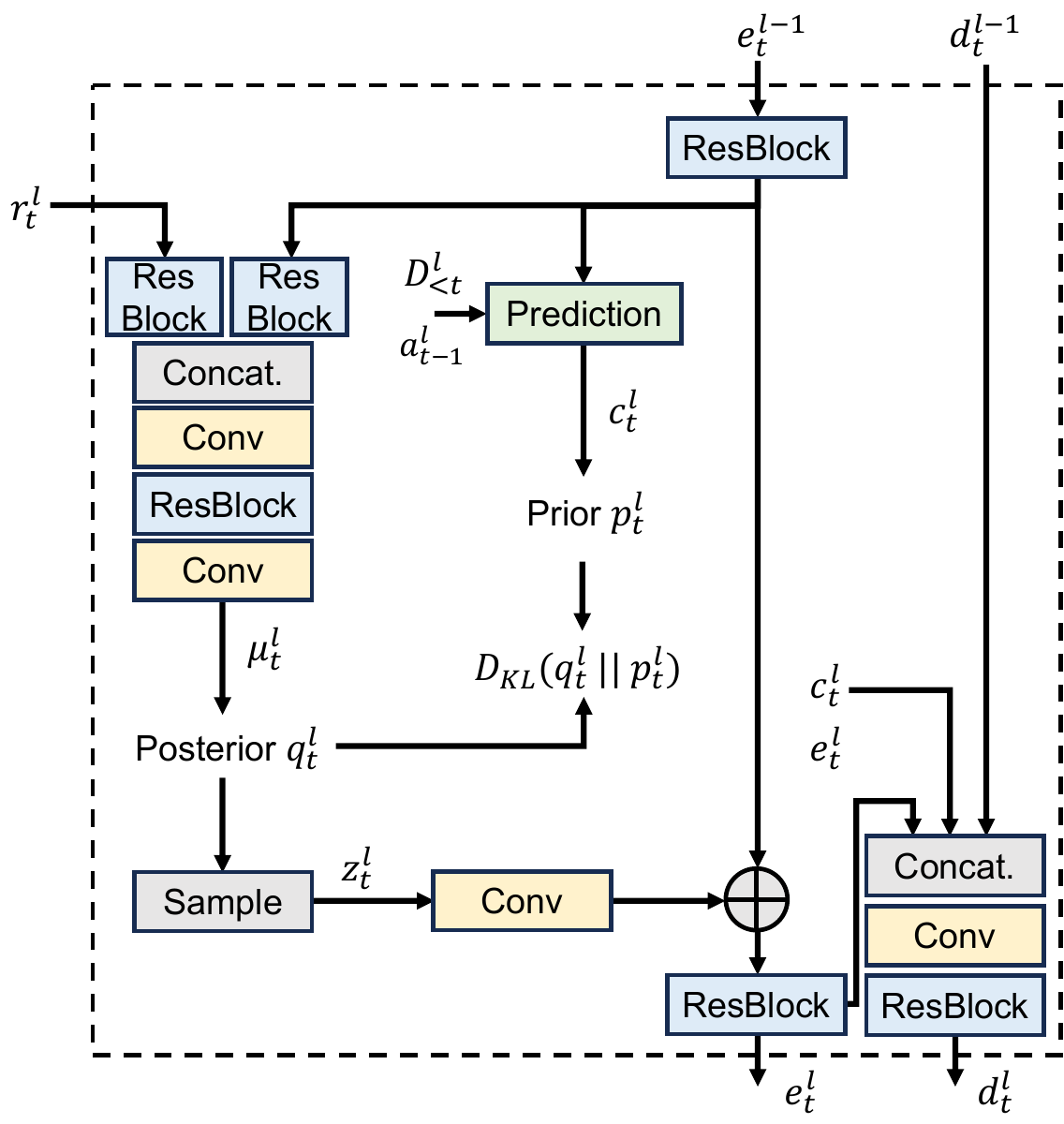}\label{fig:latent_block}}
\end{minipage}
\hspace{1.5cm}
\begin{minipage}{0.35\linewidth}
\centering
\subfigure[ResBlock]{\includegraphics[width=\linewidth]{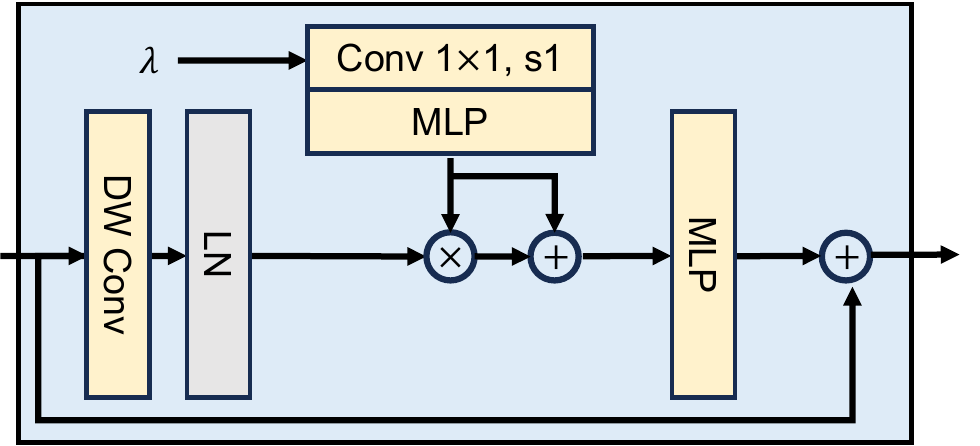}\label{fig:res_block}} \\
\subfigure[Contextual Prediction]{\includegraphics[width=\linewidth]{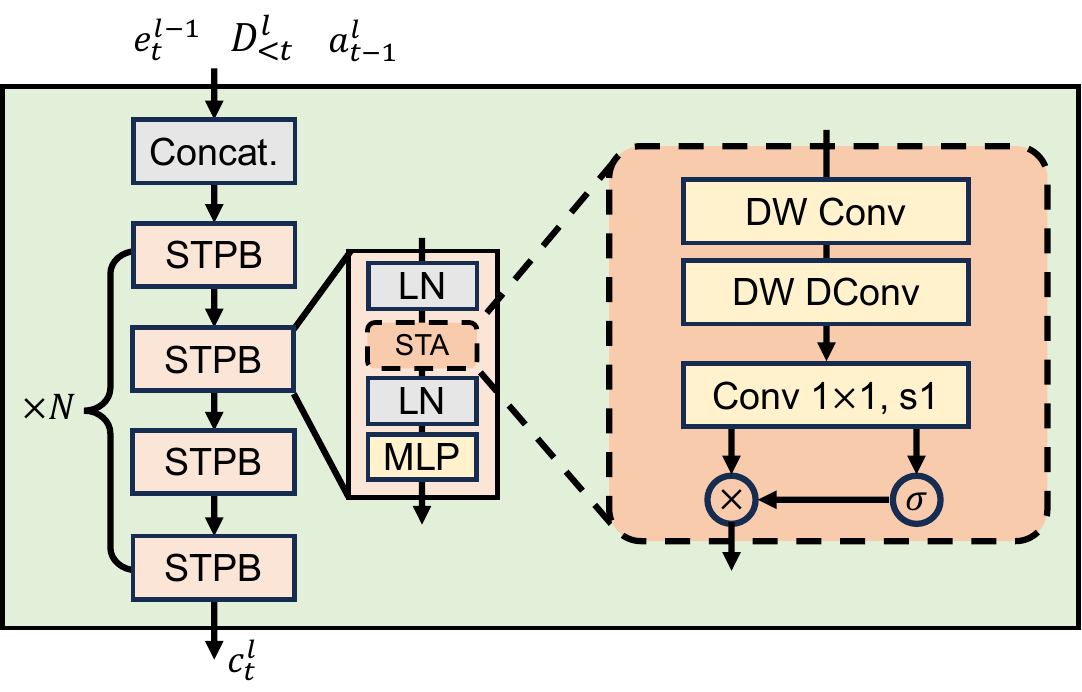}\label{fig:predict_network}}
\end{minipage}
\caption{{\bf DHVC's implementation} using neural network blocks.  Grey blocks denote nonparametric operations, and colored blocks indicate learnable neural network modules. Feature dimension adapts scale by scale accordingly. An input image of 64 $\times$ 64 $\times$ 3 (height $\times$ width $\times$ channel) is exemplified. Images with other sizes can be scaled proportionally.}
\label{fig:network}
\end{figure*}

The compression process is detailed in Fig.~\ref{fig:4-compress}. We follow the \cite{begaint2020compressai} by using the residual rounding for each $\mu_t^l$:
\begin{align}
    z_t^l = \hat{\mu}_t^l + \lfloor \mu_t^l - \hat{\mu}_t^l \rceil
\end{align}
where $\lfloor \cdot \rceil$ is the nearest integer rounding function. Then we can encode the quantized $z_t^l$ into bits using the PMF $p_t^l$. Each Latent Block produces a separate bitstream, so a compressed frame consists of $L$ bitstreams, corresponding to $L$-layer $z_t^1, ..., z_t^L$. Notably, since we have decoupled the entropy model and decoder into separate branches, we only need to generate the entropy features without waiting to complete decoding. This helps reduce the encoding time. But if we want to enable parallel pipelines according to the hierarchal scales, scale-wise decoding needs to be finished (see details in the supplemental material).

{Decompression (Fig.~\ref{fig:4-decompress}) is done in a similar way. We iteratively compute $p_t^l$ using  contextual feature $c_t^l$ predicted using $e_t^{l-1}$, $D^l_{<t}$ and $a^l_{t-1}$, and decode $z_t^l$ from the $l$-th bitstream at each Latent Block. Then the decoded feature $z_t^l$ is added to $e_t^{l-1}$ to produce $e_t^l$. By integrating $d_t^{l-1}$, $c_t^l$, and $e_t^l$, we can obtain the decoded feature $d_t^l$ for the current scale. After completing this process for all scales $l = 1,2,...,L$, the final pixel-space reconstruction $\hat{x}_t$ is produced.}

\section{Neural Network Implementation} \label{sec:neural_network}

This section details the neural network that facilitates the functionalities in DHVC.

\subsection{Overall Architecture} \label{sec:overall_arch_dhvc}

{Figure~\ref{fig:network_overview} first depicts the overall neural network implementation}. Given an input frame $x_t$, the encoder $f_{enc}$ applies down-sampling layers $\mathcal{S_D}()$ and ResBlocks, e.g., from left to right, to progressively extract a hierarchy of observations $\{r_t^L, ..., r_t^1\}$ at respective resolutions of $1/8$, $1/16$, $1/32$, and $1/64$ of the original input, both in height and width. Multiscale observations from $f_{enc}$ are sent to Latent Blocks that are inter-connected with the entropy coder $p_z$ and decoder $f_{dec}$ paths (e.g., from right to left) to produce a layered bitstream and multiscale decoded features for restoring pixel-space $\hat{x}_t$ or being accumulated as the temporal reference to help compress the next frame. Upsampling $\mathcal{S_{U}()}$ is devised appropriately. An auxiliary refinement path following the grey arrow is utilized to extract multiscale auxiliary features from $\hat{x}_{t-1}$ which are then augmented with spatial and temporal references to enhance the prediction of contexts as in \eqref{eq:spatial_temporal_aux}. Note that the number of feature channels in all processing paths increases as the scale decreases.


Neural networks used in DHVC are fully convolutional. The down-sampling $\mathcal{S_D}()$ is implemented using strided convolutions with stride size at 4 or 2 (e.g., ``s4'' or ``s2''), and up-sampling $\mathcal{S_U}()$ uses $1\times1$ convolutions followed by pixel shuffle. Basic feature processing unit -- ResBlock is formed using ConvNeXt unit~\cite{liu2022convnet}. The number of ResBlocks varies for different scales. To enable variable-rate compression, we simply embed Adaptive Layer Normalization (AdaLN)~\cite{duan2023qarv} in each ResBlock to modulate features using an additional multiplier factor $\lambda$ as depicted in Fig.~\ref{fig:res_block}.

\subsection{Latent Block}

Four Latent Blocks, e.g., one for each scale, are devised to fulfill the compression task shown in Fig.~\ref{fig:latent_block_task}. {This differs from our multi-block design within a single scale for image coding~\cite{duan2023lossy,duan2023qarv}, primarily due to the introduction of temporal references, which significantly reduces the need for repeated coding of latent residuals at the same scale.}

As for $l$-th Latent Block exemplified in Fig.~\ref{fig:latent_block}, the posterior branch stacks ResBlocks, a concatenation layer, and two convolutional layers with kernel size of $1\times1$ to combine $l$-th scale observation $r_t^l$ and upsampled spatial reference $\mathcal{S_U}(e_t^{l-1})$ to derive latent residual variables $z^l_t$ for compression. The channels for $z^l_t$ are set to  32, 96, 8, and 8 as $l =$ 1, 2, 3, and 4 respectively.  

The prior branch contains a contextual prediction network $\mathcal{P_{CTX}}()$ that purposely aggregates spatial, temporal, and/or auxiliary references to predict $c^l_t$ for characterizing entropy probabilistic model. Details are given below.

Along with the entropy coder path, entropy-decoded residual variables $z_t^l$ will be convoluted and added with $\mathcal{S_U}(e_t^{l-1})$ to construct $e_t^{l}$ as the spatial reference for next scale. In the meantime, along with the decoder path,   $e_t^{l}$ is concatenated with $c^l_t$ and $d^{l-1}_{t}$ to form current-scale decoded features $d^t_t$ through a convolutional layer and a ResBlock.

\subsection{Contextual Prediction Network} \label{sec:pred_net}

As in Fig.~\ref{fig:predict_network}, we concatenate $e_t^{l-1}$, $D_{<t}^l$ and $a_{t-1}^l$ and feed into $N$ ($N=4$ in this work) Spatial-Temporal Prediction Blocks (STPBs) to derive the contextual feature $c_t^l$. Each STPB consists of two Layer Normalization (LN) layers, a Spatio-Temporal Attention (STA) module, and a Multi-Layer Perceptron (MLP). The STA module includes a depth-wise convolution (DW Conv) focusing on local regions, a depth-wise dilated convolution (DW DConv) for capturing distant areas, and a $1\times1$ convolution for channel-wise attention. The aggregated information is then passed through a sigmoid function to reassign weights accordingly. Compared with the stacked convolutional blocks in DHVC 1.0~\cite{lu2024deep}, this contextual prediction network can better capture the temporal correlations with lightweight structure.  For intra coding, the temporal features $D_{<t}^l$ at each scale are set using learnable constants, while for inter coding, we use the decoded features from two previous frames, i.e., $d_{t-1}^l$ and $d_{t-2}^l$, to make up of $D_{<t}^l$. 

\section{Experimental Evaluations} \label{sec:experiment}

\subsection{Implementation Settings}

\subsubsection{Datasets}

{We use the popular Vimeo-90K~\cite{xue2019video} dataset to train our model, which consists of 64,612 video samples. Training batches comprise sequential frames that are randomly cropped to the size of $256 \times 256$. Commonly used test datasets, i.e., the UVG~\cite{mercat2020uvg}, MCL-JCV~\cite{wang2016mcl}, and HEVC Class B~\cite{bossen2013common} with resolution of 1080p, are used for evaluation. Original test sequences in YUV420 format are pre-processed following the suggestions in \cite{sheng2022temporal} to generate RGB frames as the input of learned models.}

\subsubsection{Training Details}

{We progressively train our model for fast convergence. First, the model is trained to encode a single frame independently for 2.5M iterations by setting the temporal features at each scale as learnable constants. Then, we train the aforementioned model for 3.5M steps using three successive frames, with temporal features hierarchically generated from previously decoded frames. Finally, another 100K steps are applied to fine-tune the model using seven successive frames, by which it better captures long-term temporal dependence~\cite{liu2020neural}. }

{We randomly sample $\lambda$  from $\Lambda$ = [256, 2048] or [4, 32] for MSE or MS-SSIM optimized models to cover wide rate ranges. Adam~\cite{kingma2014adam} is the optimizer with the learning rate at $10^{-4}$, which will decrease to $5\times10^{-6}$ at the final fine-tuning stage. Our model is trained using four NVIDIA RTX 3090 GPUs, and the batch size is fixed at eight.}

\subsection{Evaluation}

Evaluation experiments are all performed under the low-delay configuration. The first 96 frames of each test video are used for evaluation, and the group of pictures (GOP) is set at 32. For fair comparisons, these settings are the same as other methods. 
\begin{itemize}
    \item {\bf Traditional codecs}: we choose reference models HM 16.26\footnote{\url{https://vcgit.hhi.fraunhofer.de/jvet/HM}}, and VTM 18.0\footnote{\url{https://vcgit.hhi.fraunhofer.de/jvet/VVCSoftware_VTM}} as the benchmarks of respective 
    HEVC/H.265~\cite{sullivan2012overview}, and VVC/H.266~\cite{bross2021overview}. As the open-source implementation of HEVC/H.265, a.k.a., x265\footnote{\url{https://www.videolan.org/developers/x265.html}} is widely used, it is also included in the study. All of them use the default configuration. The detailed codec settings can be found in appendix.
    \item {\bf Learned codecs}: we compare with representative models using hybrid motion \& residual coding method like DVC-Pro~\cite{lu2020end}, MLVC~\cite{lin2020m}, RLVC~\cite{yang2020learning}, DCVC~\cite{li2021deep}, and DCVC-DC~\cite{Li_2023_CVPR}, using probabilistic predictive coding approach like VCT~\cite{mentzer2022vct} and our preliminary exploration DHVC 1.0~\cite{lu2024deep}. {For fair comparison, all of them are retrained using the same dataset as ours and following the training strategies outlined in corresponding papers.}
\end{itemize}

\begin{figure*}[htbp]
\centering
\subfigure[]{\includegraphics[width=0.325\linewidth]{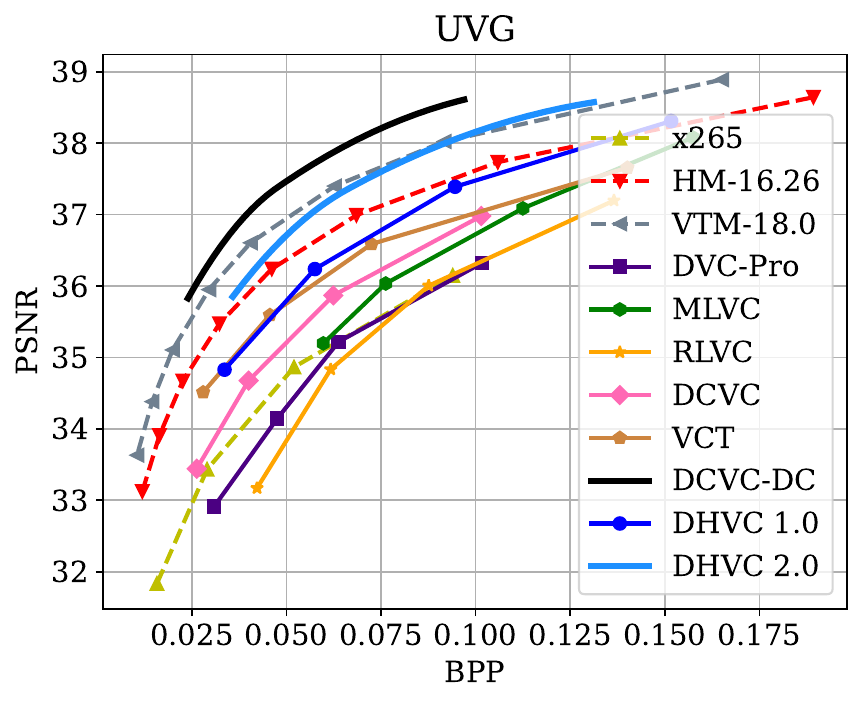}}
\subfigure[]{\includegraphics[width=0.325\linewidth]{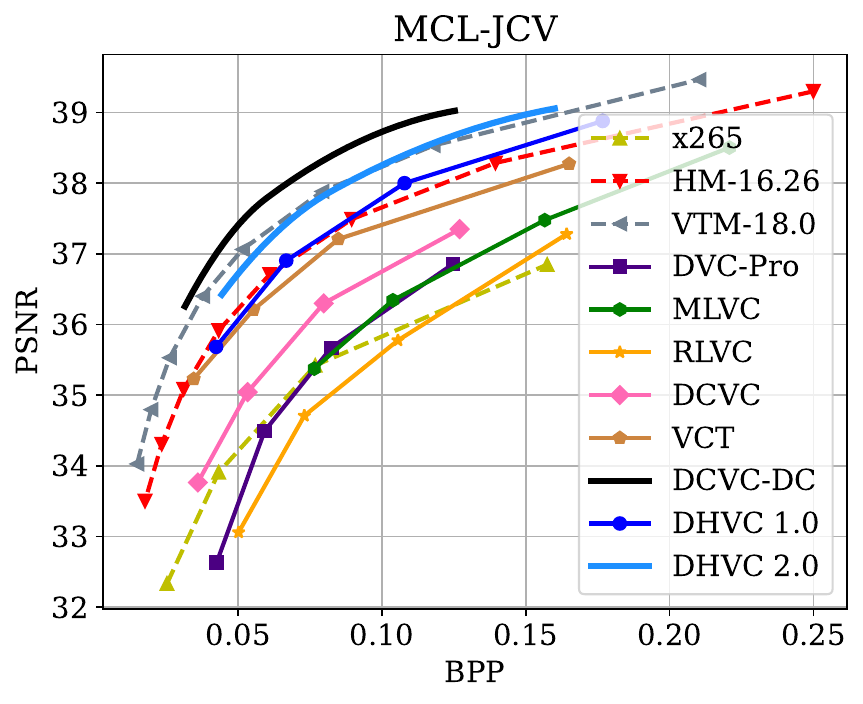}} 
\subfigure[]{\includegraphics[width=0.325\linewidth]{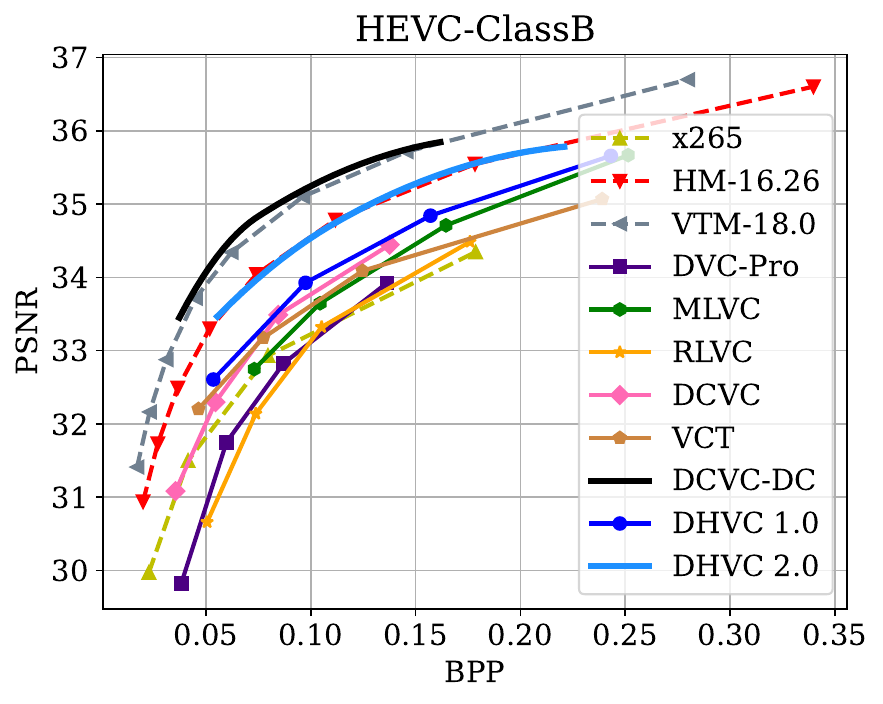}} \\
\subfigure[]{\includegraphics[width=0.325\linewidth]{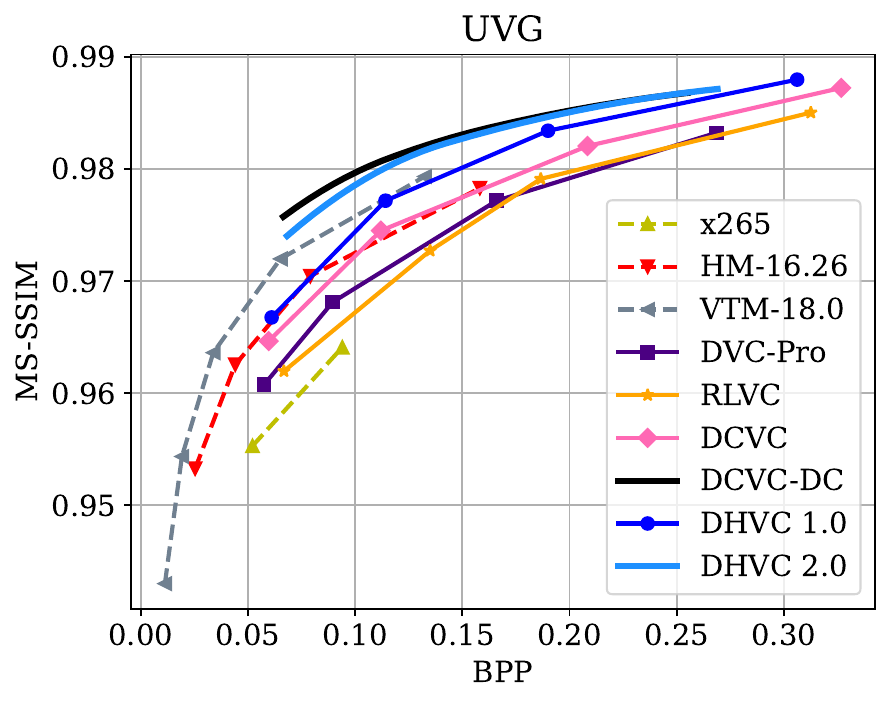}} 
\subfigure[]{\includegraphics[width=0.325\linewidth]{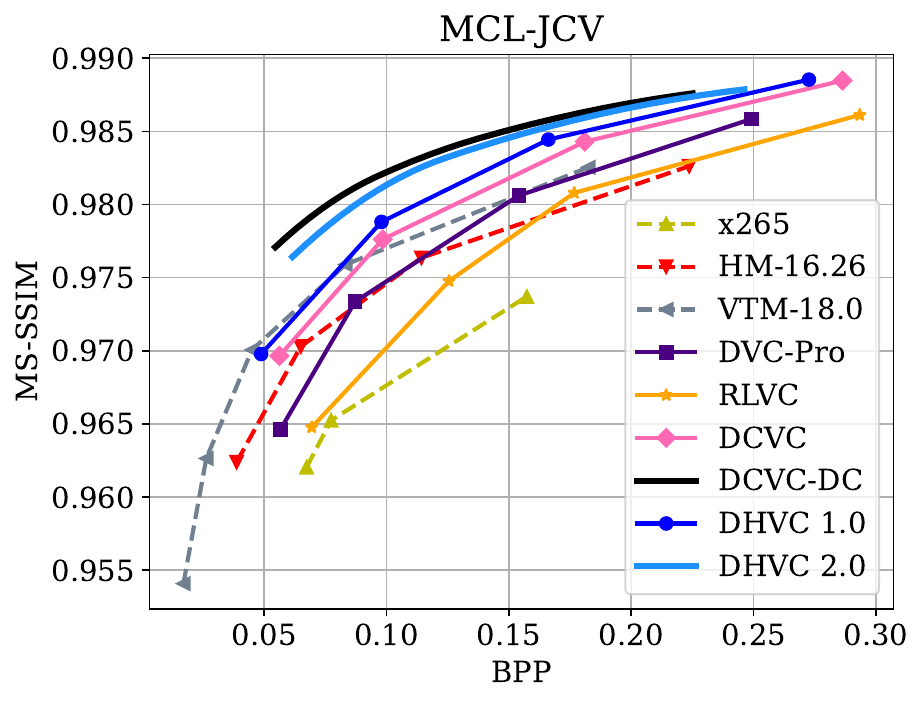}}
\subfigure[]{\includegraphics[width=0.325\linewidth]{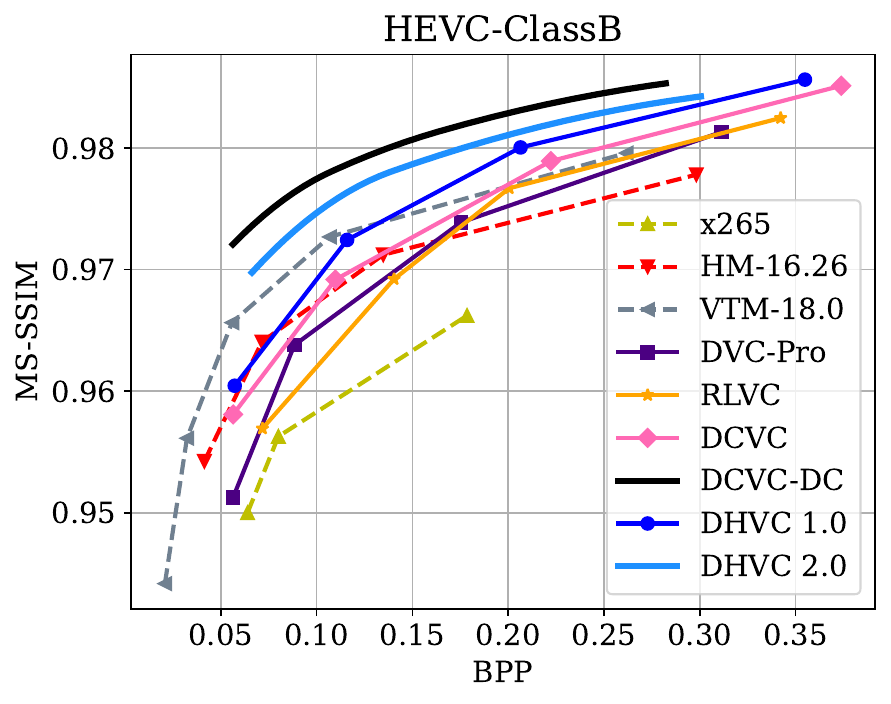}}
\caption{{\bf Compression efficiency} comparisons using rate-distortion (R-D) curves. Results in (a) to (c) are optimized with MSE, while results in (d) to (f) are optimized with MS-SSIM. Dashed lines indicate traditional codecs; Solid lines with \& without markers represent fixed- and variable-rate models, respectively.}
\label{fig:rd_curve}
\end{figure*}

\begin{table*}[t]
\caption{Performance-Complexity Evaluation of  Learning-based Video Codecs}
\label{tab:effficiency}
\centering
\begin{tabular}{lcccccccc}
\toprule
\multirow{2}{*}{\bf Method} 
& \multicolumn{4}{c}{\bf Complexity} & \multicolumn{2}{c}{\bf Latency (s)} & \multicolumn{2}{c}{\bf BD-rate (\%) w.r.t. HM 16.26} \\ 
&{\bf Models} & {\bf Params. (M)} & {\bf KMACs/pixel} & {\bf Memory (G)} & {\bf Enc.} & {\bf Dec.} & {\bf PSNR} & {\bf MS-SSIM} \\
\midrule
\textit{Hybrid models} & & & & & & & \\
DVC-Pro & 4+4 & 69.21 
& 855.06 & 17.31 & 0.49 & 0.24 & 112.95\% & 30.41\% \\
DCVC & 4+4 & 77.64 
& 1126.40 & 11.71 & 13.78 & 46.48 & 57.67\% & 4.75\% \\
DCVC-DC & 1+1 & \underline{50.78}
& 1416.49 & 13.29 & 0.65 & 0.52 & \underline{-41.28\%} & \underline{-46.03\%} \\
\midrule
\textit{Unified models} & & & & & & & \\
VCT & 4 & 751.28 & 3042.20 & 10.51 & 1.64 & 1.58 & 33.59\% & - \\
DHVC 1.0 & 4 & 449.84 & 433.81 & 4.27 & 0.25 & 0.21 & 19.52\% & -11.17\% \\
DHVC 2.0 & \underline{1} & 107.97 & 400.71 & 4.87 & 0.23 & 0.14 & -15.97\% & -37.55\% \\
DHVC 2.0 (Parallel) & \underline{1} & 91.07 & \underline{349.44} & \underline{2.82} & \underline{0.05} & \underline{0.03} & -9.87\% & -27.47\% \\
\bottomrule
\end{tabular}
\end{table*}

\subsubsection{Compression Performance}

Figure~\ref{fig:rd_curve} qualitatively depicts various methods' R-D (rate-distortion) curves upon different datasets. As seen, DHVC 2.0 outperforms most methods in MSE-optimized comparisons, including both rules-based and learned approaches, and even achieves performance close to VVC on the UVG and MCL-JCV sequences. Such a pronounced efficiency of DHVC 2.0 primarily stems from its novel hierarchical predictive coding design. 

As  DHVC 2.0 is in its infancy, which is more or less close to the DCVC phase~\cite{li2021deep}, we are optimistic about its potential, although DHVC 2.0 still lags slightly behind the best-performing DCVC-DC. For example, we can incorporate frame-level rate allocation and other techniques developed to improve the DCVC model in the past years~\cite{li2022hybrid,sheng2022temporal,Li_2023_CVPR} to fulfill the purpose.

In the MS-SSIM optimized results, our compression performance is on par with DCVC-DC on the UVG and MCL-JCV datasets and consistently surpasses VVC across all test sequences. This further demonstrates the strong representational capability of our hierarchical architecture.

Corresponding quantitative measures of BD-rate against the HEVC anchor are given in Table~\ref{tab:effficiency} (see last two columns).

\begin{figure}[t]
\centering
\includegraphics[width=0.97\linewidth]{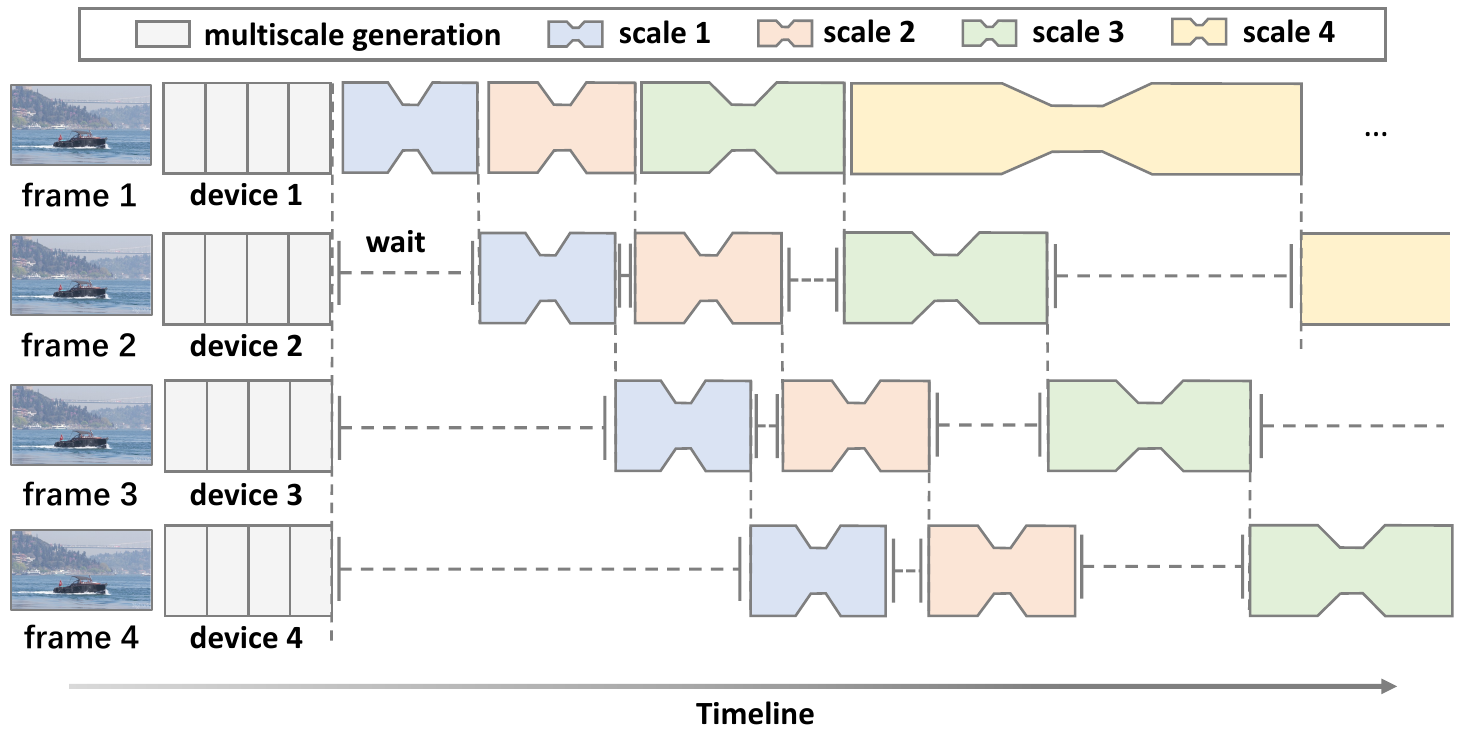}
\caption{{\bf Parallel pipeline} in DHVC 2.0 (Parallel). The ``multiscale generation'' represents the generation of hierarchical observation of each frame, and ``scale $l$'' specifically represents the encoding of latent residual variables in $l$-th scale. As a higher scale is with a larger resolution (i.e., less downsampling), we purposely use a larger box to indicate more processing time to process more elements. Other operations, such as memory copy, are omitted for simplicity.}
\label{fig:parallel}
\end{figure}

\subsubsection{Complexity Efficiency}~\label{sec:complexity_efficiency}
Performance-complexity evaluations of using different learned approaches to process 1080p videos are summarized in Table~\ref{tab:effficiency}.
{All learned codecs are trained and tested on NVIDIA RTX 3090 GPU (24G RAM).}
Commonly-used metrics include
\begin{itemize}
    \item {\bf Models}: number of trained models to support intra \& inter coding modes across various bitrate points\footnote{Four models are exemplified for fixed-rate models to support four bitrate points. $N$ models are required for $N$ bitrate points. };
    \item {\bf Params. (M)}: number of model parameters in millions;
    \item {\bf KMACs/pixel}: average thousand multiply-accumulate operations per pixel;
    \item {\bf Memory}: peak memory in Gigabytes in the inference;
    \item {\bf Latency}:  seconds used for encoding/decoding a frame.
\end{itemize}

DHVC 2.0 is the first solution to use a single model to support both intra- and inter-coding modes with variable-rate options, giving it a significant advantage in real-life deployment. 

Except for the model size, the proposed DHVC, regardless of version 1.0 or 2.0, shows clear advantages for other metrics, reporting the least requirements of respective KMACs/pixel, peak memory, and encoding/decoding time. This also suggests that the model size is unrelated to the complexity of running codecs in practice. The relatively sizeable model DHVC uses mainly applies to using default ConvNeXt units to form the ResBlocks and Latent Blocks, which can be simplified later for practical implementation. In contrast, DHVC's lightweight complexity is attributed to its feature-space hierarchical probabilistic prediction using simple convolutions.

Concerning the encoding and decoding time that particularly matters from the application perspective, the spatial autoregressive model used in DCVC and VCT\footnote{As VCT limits the autoregressive processing in a local 4$\times$4 block, it provides faster encoding and decoding than DCVC that applies pixel-level autoregression.} for entropy coding exhibits  more than 1 second per frame, which is impractical for real-life services.  Although DCVC-DC employs a fast parallel context model, its encoding/decoding speed is less than 2 FPS due to the use of complicated hybrid coding, which is still far from the real-time threshold. In contrast, in DHVC, entropy contextual autoregression and hybrid motion \& residual coding are completely bypassed with faster speed, demonstrating encouraging potential for real-time processing.

Besides utilizing multiscale characteristics for improving compression performance, the proposed hierarchical structure in DHVC makes it friendly for parallel processing. Figure~\ref{fig:parallel} gives a quick example of DHVC 2.0 (Parallel) by dispatching the encoding of four consecutive input frames on four GPU devices. Devising more advanced techniques, such as hyperthreading on the same device, to facilitate fine-grained parallel pipelines is also possible. However, it is not the focus of this work and is deferred as our future study.  

Considering the referencing dependency across hierarchical scales, encoding the $l$-th scale of the $t$-th frame starts as soon as the encoding of the $l$-th scale of the $(t-1)$-th frame completes. A waiting break is often encountered when the temporal reference is not ready (see dash lines in between) for a scale to be encoded. The same rules apply to the decoding.

As the parallel pipeline runs at a sub-frame scale, e.g., we are not able to have a fully decoded temporal reference if we pursue the parallelism, so we are currently removing the auxiliary refinement path in the implementation of DHVC 2.0 (Parallel). Additionally, half-precision computation is applied, e.g., FP16 or float16 instead of full-precision float32 (FP32), with no performance degradation (to the FP32 model) but less memory footprint and faster computation.

As shown in Table~\ref{tab:effficiency}, DHVC 2.0 (Parallel) achieves a 1080p encoding of 20 FPS (0.05s) and decoding of 33 FPS (0.03s), making real-time neural video coding practically sound. The performance loss is due to the removal of auxiliary refinement for parallel processing. Nevertheless, DHVC 2.0 (Parallel) surpasses the HEVC reference model HM 16.26 with about 10\% BD-rate gain.  In comparison, other acclaimed real-time neural video codecs~\cite{le2022mobilecodec,van_Rozendaal_2024_WACV} are still largely inferior to HM 16.26. The supplemental material will give more details regarding the parallel acceleration of the encoding and decoding process.

\begin{figure}[t]
\centering
\centering
\includegraphics[width=0.8\linewidth]{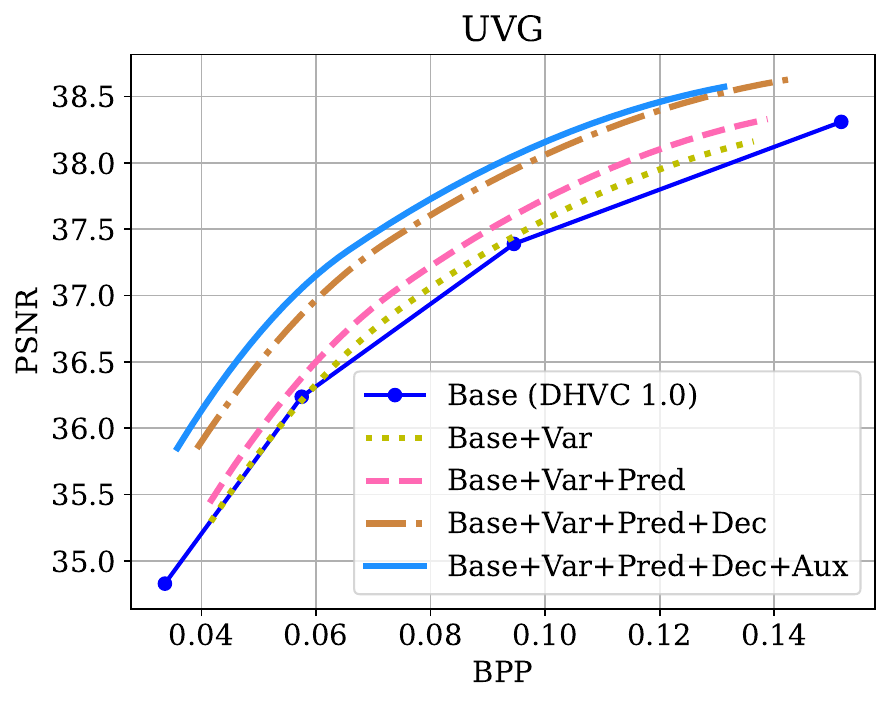}
\caption{{\bf Modular contribution} in DHVC 2.0. Tests are evaluated using UVG sequences but similar patterns keep for other videos.}
\label{fig:ablation}
\end{figure}

\begin{table}[t]
\caption{Impact of the number of frames used to generate temporal reference features. Using two frames is the default.}
\centering
\begin{tabular}{lcccc}
\toprule
Temp Num. & 1 & 2 & 3 & 5 \\ 
\midrule
BD-rate (\%) & 12.39\% & - & 6.21\% & -3.77\% \\
\bottomrule
\end{tabular}
\label{tbl:ablation_temporal}
\end{table}

\section{Deep Dive}
Detailed discussions are given below to help further understand the capacity of DHVC.

\subsection{Modular Contribution} \label{sec:Ablation}
DHVC 2.0 has substantially enhanced DHVC 1.0 by introducing various improvements, including variable-rate coding (Var), better contextual prediction network (Pred), temporal reference using decoded features (Dec), and auxiliary refinement (Aux).  Letting fixed-rate model DHVC 1.0~\cite{lu2024deep} as the base, Figure~\ref{fig:ablation} clearly illustrates the performance progress of augmenting aforementioned tools, for example,
\begin{itemize}
    \item ``Base + Var'': slightly better performance but only using one model for all cases (Sec.~\ref{sec:overall_arch_dhvc}); 
    \item ``Base + Var + Pred'': notable gain observed due to using a novel network instead of simple ResNet stacking to better characterize and embed spatiotemporal correlations (Sec.~\ref{sec:pred_net});
    \item ``Base + Var + Pred + Dec'': significant performance boost when accumulating decoded features $d_t^l$ with richer information instead of latent residual $z_t^l$ as temporal references for better contextual prediction (Sec. \ref{sec:temp_ref_decoded});
    \item ``Base + Var + Pred + Dec + Aux'':  noticeable gain attributed to using extra features from the pixel-space reconstructions to partially alleviate reference feature collapse (Sec. \ref{sec:temp_ref_aux}).
\end{itemize}

Recalling that properly utilizing temporal reference is vital for DHVC's performance, {we thus study the impact of the number of previous frames used to generate temporal reference features. By default, we accumulate multiscale decoded features from two previous frames\footnote{This work only considers the low-delay video coding with forward frames as references. Bidirectional reference can be easily extended.}. Table~\ref{tbl:ablation_temporal} reveals that if we only use the information from one previous frame,  an absolute 12.39\% BD-rate loss is reported as compared with the default setting. When three previous frames are used, coding performance degrades instead, likely due to the instability in fusing temporal references. Even when the number of reference frames increases to five, only a slight 3.77\% gain is reported. Considering the balance of performance and complexity, we choose to use two previous frames for generating temporal reference features.

Next, we show DHVC's superiority in effectively capturing various motion patterns, its efficiency for network transmission, and its resilience to packet loss.

\begin{figure}[t]
\centering
\includegraphics[width=\linewidth]{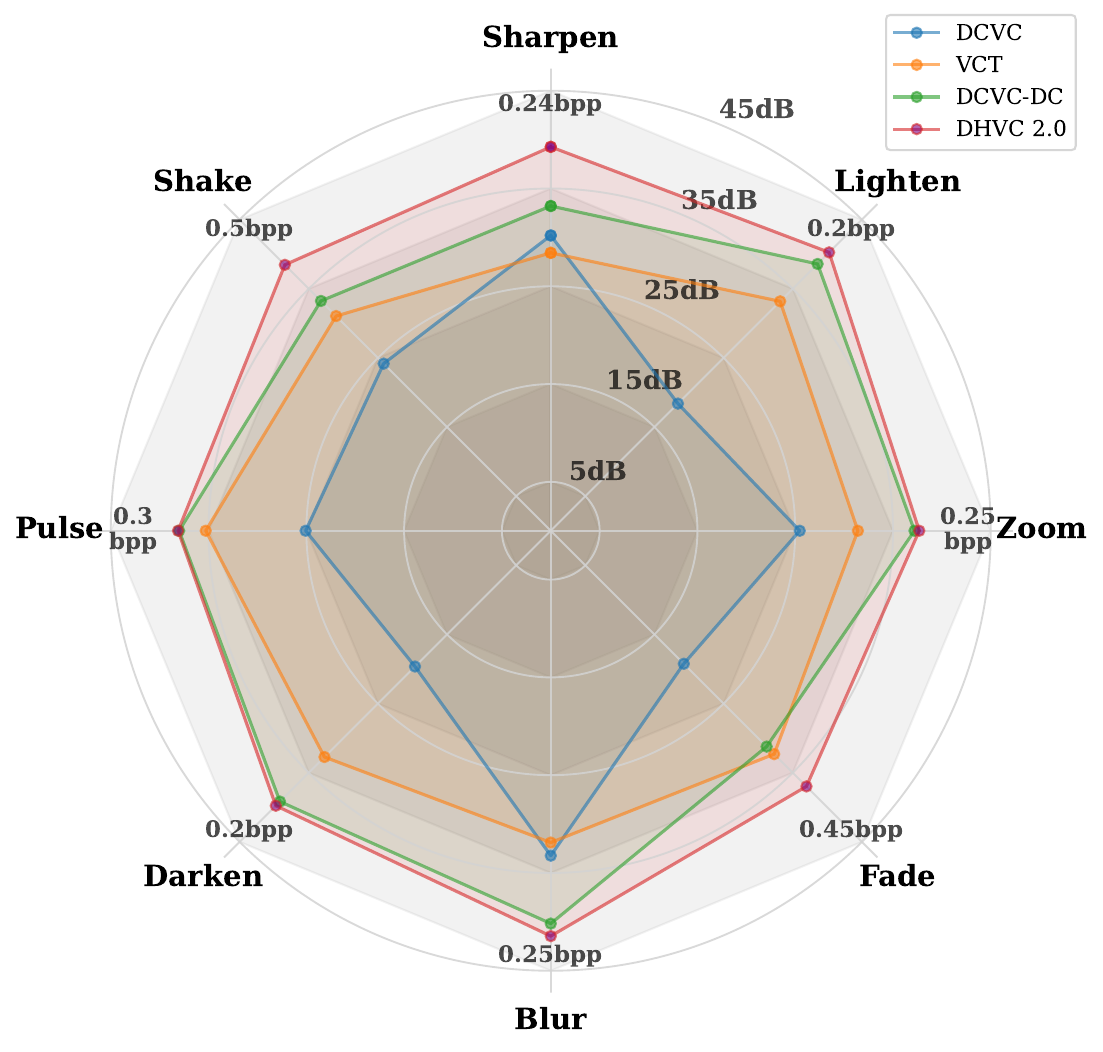}
\caption{{\bf Impact of motion patterns} on compression performance using various synthetic data. All comparisons are measured using PSNR values under the same bitrate (marked next to different motion patterns).}
\label{fig:rd_synthetic_data}
\end{figure}

\begin{figure*}[htbp]
\centering
\includegraphics[width=\linewidth]{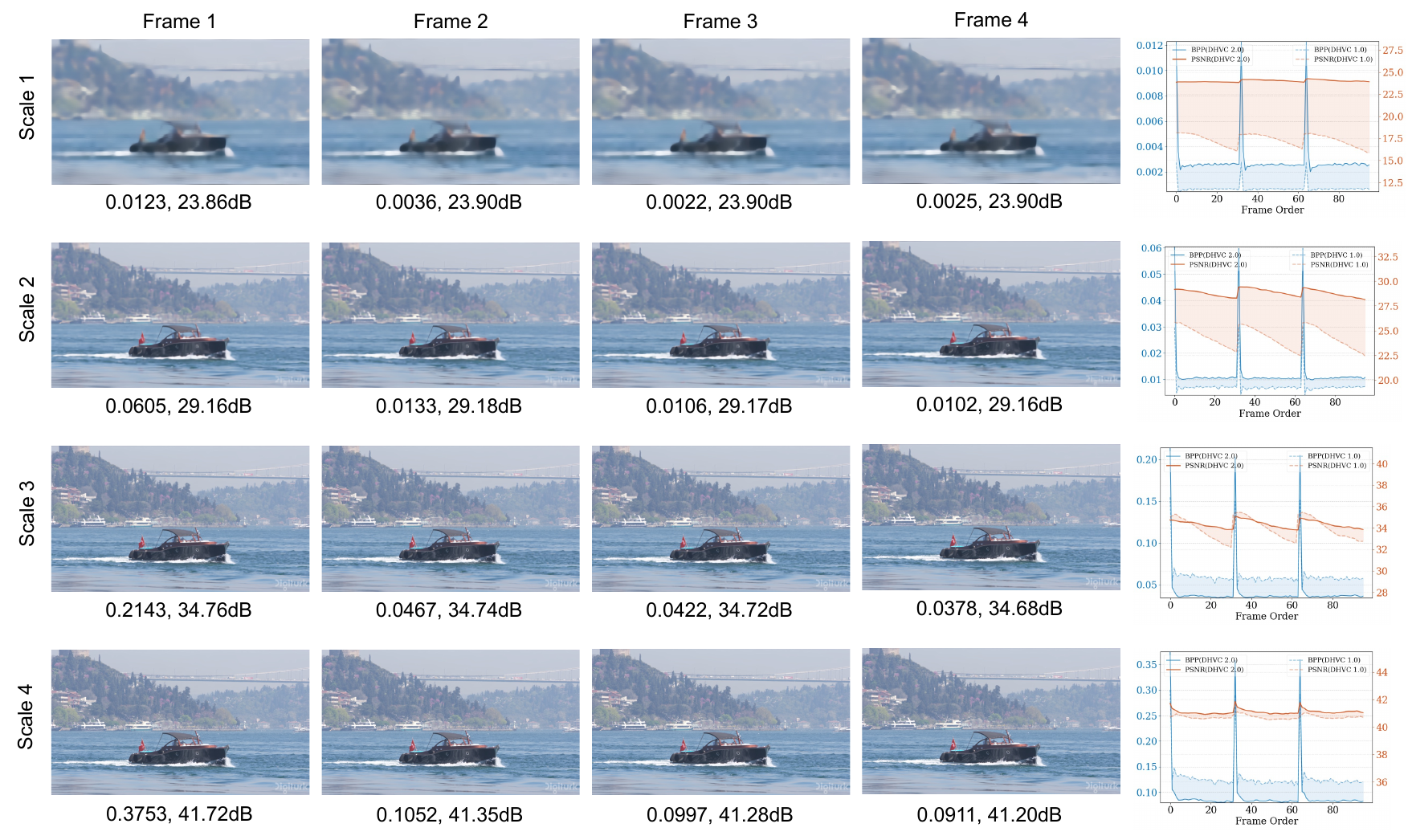}
\caption{{\bf Progressive decoding} by visualizing reconstruction at each scale across frames. The last column shows the sequential bitrate (BPP) and quality (PSNR) at different scales. Please zoom in for more details.}
\label{fig:progressive}
\end{figure*}

\subsection{Generalization to Various Motion Patterns}

Motion generalization is critical for video coding models to handle different motion patterns. To study its impact, we applied motion synthesis to the CLIC2020 test dataset~\cite{toderici2020clic}, generating video sequences with 32 frame. For each sequence, a specific types of motion is augmented alongside the frame order $i \in [0,32)$:
\begin{itemize}
    \item {\bf Sharpen}: Progressive sharpening is applied frame by frame with a kernel size of $\frac{i}{32}+1$.
    \item {\bf Lighten}: Brightness is gradually increased at a rate of $\frac{i}{32}$ per frame.
    \item {\bf Zoom}: A combination of left-to-right panning with a fixed step size of $100\times\frac{i}{32}$ and scaling at a rate of 0.98 is applied frame by frame.
    \item {\bf Fade}: Frame-by-frame fading is applied at a rate of $\frac{i}{32}$.
    \item {\bf Blur}: A progressive blurring effect is applied frame by frame with a kernel size of $2 \times \left\lfloor \frac{5i}{64} \right\rfloor + 1$.
    \item {\bf Darken}: Brightness is gradually reduced at a rate of $1-\frac{i}{32}$ per frame.
    \item {\bf Pulse}: A brightness fluctuation with a period of 2 frames is applied within each sequence.
    \item {\bf Shake}: Random translations with a maximum amplitude of 20 pixels in any direction are applied to simulate shaking.
\end{itemize}

Visualizations of various motion patterns are provided in the supplemental materials. Figure~\ref{fig:rd_synthetic_data} illustrates how these motion patterns affect compression efficiency by fixing the bitrate in bits per pixel (bpp) and comparing the PSNR. As shown, DHVC 2.0 surpasses all other methods across all synthetic datasets, highlighting the impressive representation capability of hierarchical predictive coding. Regardless of the motion pattern or the speed of scene changes, our method remains consistently effective. In real-world scenarios, multiple motion modes often occur simultaneously, along with brightness fluctuations and noise. Compared to optical flow-based motion prediction methods like DCVC and DCVC-DC, our hierarchical predictive coding exhibits strong generalization. This underscores the potential of our method for practical video coding applications.

\subsection{Progressive Decoding}\label{sec:progressive_decoding}

DHVC's hierarchical structure inherently supports progressive coding as scale-wise processing can run independently instead of waiting for the completion of a frame (see discussion regarding parallel pipelines in Sec.~\ref{sec:complexity_efficiency}), which enables rapid delivery of a portion of the bitstream and the decoding of only a few scales for a coarse reconstruction.

For a 1080p video in Fig.~\ref{fig:progressive}, accessing the bitstream at just the first scale, like ``scale 1", can already provide a clear preview of the content. This functionality, rarely supported by existing methods, allows streaming a scene preview using minimal bandwidth, which is particularly advantageous for remote decision-making in emergencies. With additional packets of higher scales received, we can observe reconstruction quality improvement row by row from upper to bottom (e.g., scale 1 to scale 4).

{Till frame $t$, assuming we have successfully decoded bitstream packets up to the first $l-1$ scales and bitstream packets are not available for $l$-th and further scales,
\begin{enumerate}
    \item  Decoded $e_t^{l-1}$ and $d_t^{l-1}$, as well as accumulated $D_{<t}^{l}$\footnote{As we don't have the packets available for $l$-th and further scales, we set a redefined constants, i.e., $d_0^l$, $d_0^{l+1},\ldots$, $d_0^L$, for the first frame and recurrently accumulate it to form $D_{<t}^{l}$ using predicted $e^l_t$ and $c_t^l$ and decoded $d_t^{l-1}$.} are ready to predict $c_t^l$ using \eqref{eq:spatio-temporal-contexts} and thus entropy parameters $\hat{\mu}_t^l$ and scale $\hat{\sigma}_t^l$ using $c_t^l$;
    \item  We then directly use predicted $\hat{\mu}_t^l$ as a replacement of $z_t^l$ and add it to $e_t^{l-1}$ for $e_t^l$ as in \eqref{eq:spatial_ref}.  $e_t^l$ is subsequently augmented with predicted $c_t^l$ and $d^{l-1}_t$ to have $d_t^l$, as formulated in \eqref{eq:decode_feature};
\end{enumerate}
Steps 1) and 2) are repeated beyond $l$-th scale until the final reconstruction of the pixel-space RGB sample is reached.}

We also provide the rate and PSNR variations of the entire sequence in the rightmost column of Fig.~\ref{fig:progressive}. Compared to version 1.0, DHVC 2.0 demonstrates smoother temporal quality across all scales, offering better progressive decoding results. This proves the superiority of using multiscale decoded features for predicting conditional contexts. Meanwhile, DHVC 2.0 consumes more bitrates at lower scales, significantly improving the reconstruction quality at these scales (for instance, achieving a gain of over 5dB at the first scale). At higher levels (third and fourth), while maintaining better PSNR, the bitrate consumption is greatly lower than that in DHVC 1.0. This indicates that better compression at lower scales can reduce the encoding burden at higher scales, thereby enhancing overall compression performance.

Additionally, this approach broadens our understanding: it allows for content decoding in networked applications, even with packet loss when only partial packets are received. In congested connections, we can even choose to drop latent packets associated with higher scales for smooth streaming. The next section will provide more details on the presence of packet loss.

\begin{figure}[t]
\centering
\includegraphics[width=\linewidth]{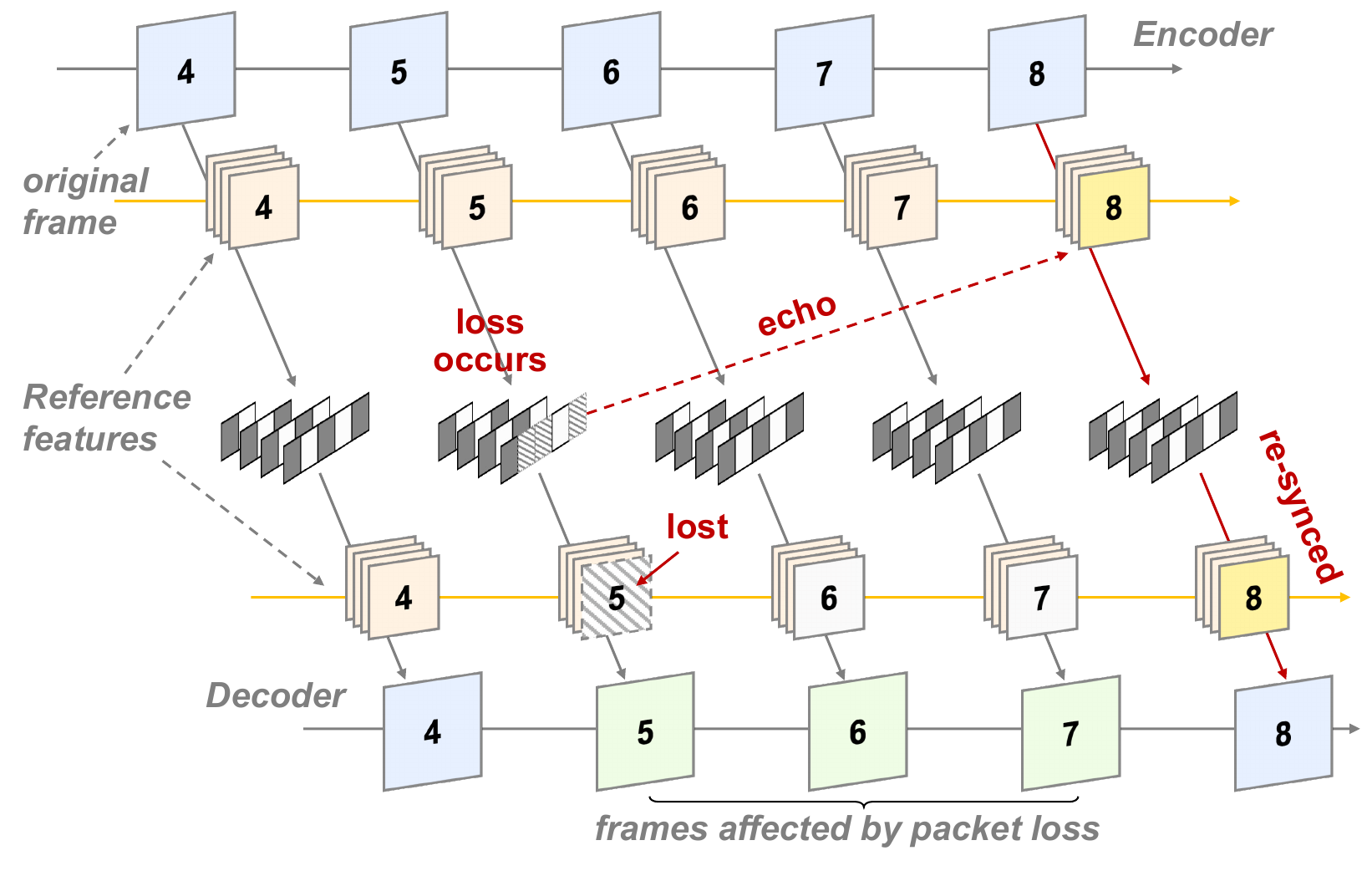}
\caption{{\bf Packet loss resilience} using decoder-to-encoder echo indication.}
\label{fig:loss_resiliance}
\end{figure}

\subsection{Packet Loss Resilience}

{This section reports DHVC's resilience to the presence of packet loss in transmission. As exemplified in Fig.~\ref{fig:loss_resiliance}, when the bitstream packet is lost in the fourth scale of the fifth frame, our model can still decode and reconstruct the frame using the bitstreams from the first three scales, which is similar to the progressive decoding described above. However, we can achieve better results when decoding the fourth scale for upcoming frames, given packets corresponding to the fourth scale from previous frames are successfully decoded to produce temporal references. In contrast, progressive decoding assumes the loss of all fourth-scale information and simply accumulates temporal references from predefined constants since the first frame. 

Given the lost packet at the fourth scale of the fifth frame, even if fourth-scale packets of subsequent frames are received correctly, the quality of decoded frames beyond the fifth frame will be impaired as we can only approximate the reconstruction without successful decoding. How many future frames after the fifth frame will be impacted depends on the successful delivery of the ``echo'' message to the encoder that indicates the scale of the lost packets.  In~\ref{fig:loss_resiliance}, the echo message is received at the eighth frame. As a result, frames 6 and 7 are impacted with reduced reconstruction quality (e.g., lower PSNR) shown in Fig.~\ref{fig:packet_loss_a} (left), and frame 8 enforces the encoding of fourth-scale latent variables using spatial references only to eliminate temporal error propagation going forward (noted as re-synced) but with a slight increase of bitrate consumption.  Usually, the longer the period of echo delivery, the greater the quality degradation and the higher the bitrate required at the re-synchronization point, as shown in Fig.~\ref{fig:packet_loss_a} (left vs. right). 

\begin{figure}[t]
\centering
\subfigure[]{\includegraphics[width=1.0\linewidth]{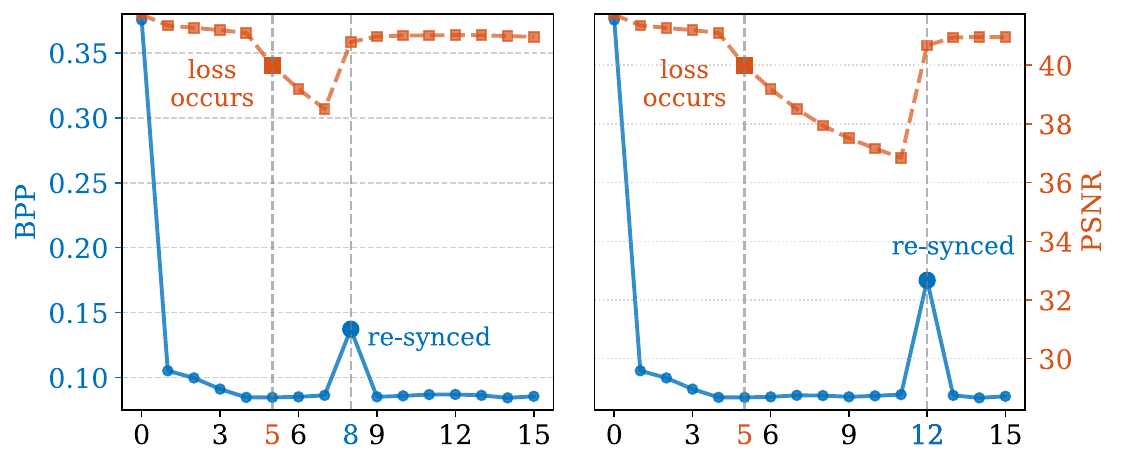}\label{fig:packet_loss_a}} \\
\subfigure[]{\includegraphics[width=1.0\linewidth]{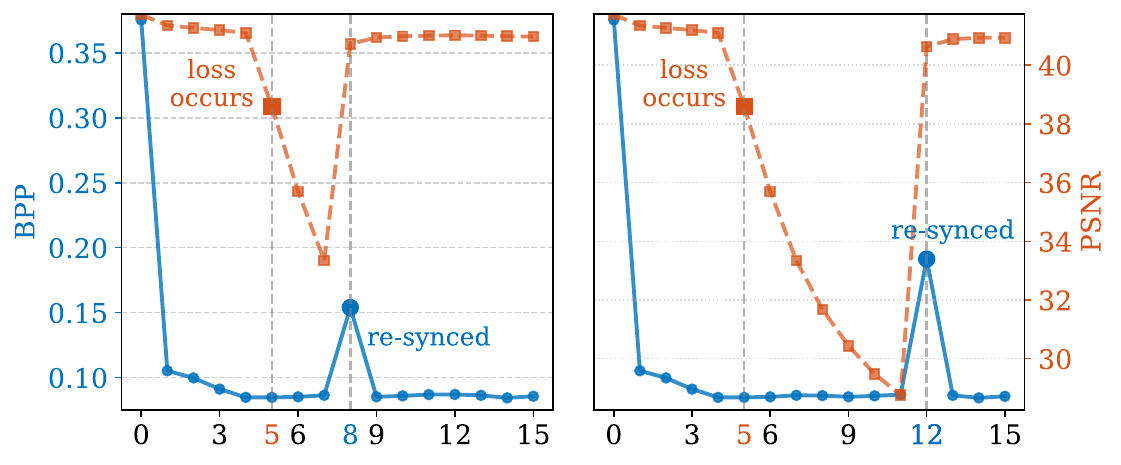}\label{fig:packet_loss_b}}
\caption{{\bf Packet loss} at (a) fourth scale only; (b) second scale. The resilient recovery depends on the period used to deliver the echo message to the encoder for re-synchronization, e.g., left vs. right plots.}
\label{fig:packet_loss}
\end{figure}

We also test more severe packet loss scenarios, such as losses occurring at the second scale in Fig.~\ref{fig:packet_loss_b}. In this case, only the first scale can be decoded correctly, leading to more pronounced quality degradation. However, compared to existing approaches, which either require additional bitrate by Forward Error Correction (FEC)~\cite{mizuochi2006recent}, re-encoding the intra frame, or performing interpolation at the decoder with minimal gains~\cite{suh1997error}, our hierarchical coding method inherently presents loss resilience, upon which a simple decoder-to-encoder echo indication is sufficient. 

It can be observed that temporal reference features play a crucial role in the reconstruction process, ensuring that even if a particular scale's features are lost in the current frame, the reconstruction quality remains satisfactory. It is important to note that the first scale's bitstream must not experience packet loss due to the hierarchical dependency. In our design, since the first scale occupies a minimal portion of the bitstream, its packet loss probability is relatively low, and even with strong FEC protection, the overhead is marginal. More details and packet loss cases can be found in the appendix.

\section{Conclusion}
The proposed DHVC demonstrates that learned video compression can achieve top-tier performance with a reasonable complexity cost, making it suitable for real-time applications. This marks a significant advancement over previous studies, as industry leaders have generally been skeptical about employing learned codecs in real-time services. These unprecedented benefits stem from the innovative hierarchical predictive coding framework, which transforms each frame into multiscale representations for scale-wise processing. When compressing latent residual variables at a given scale, the framework flexibly uses spatial references from the preceding scale of the same frame and temporal references from the same scale of previous frames to accurately define probabilistic contexts for conditional coding. This hierarchical structure inherently supports parallel processing through scale-wise pipeline arrangement and enables progressive decoding and packet loss resilience. These capabilities are rarely seen in existing studies but are highly desirable for practical applications.

Future studies to improve our DHVC include but not limited to:
\begin{itemize}
    \item Bi-directional temporal referencing for delay-insensitive applications like Video-on-Demand;
    \item Effective bit allocation across frames and across scales within a frame to optimize the overall quality;
    \item Fixed-point inference and platform oriented optimization to further accelerate the processing speed and reduce the power consumption.
\end{itemize} 
This work will be made to the public at \url{https://github.com/NJUVISION/DHVC} to encourage reproducible research. All interested parties are welcome to contribute
to its further improvement.

\bibliographystyle{IEEEtran}
\bibliography{reference}

\begin{IEEEbiography}[{\includegraphics[width=1in,height=1.25in,clip,keepaspectratio]{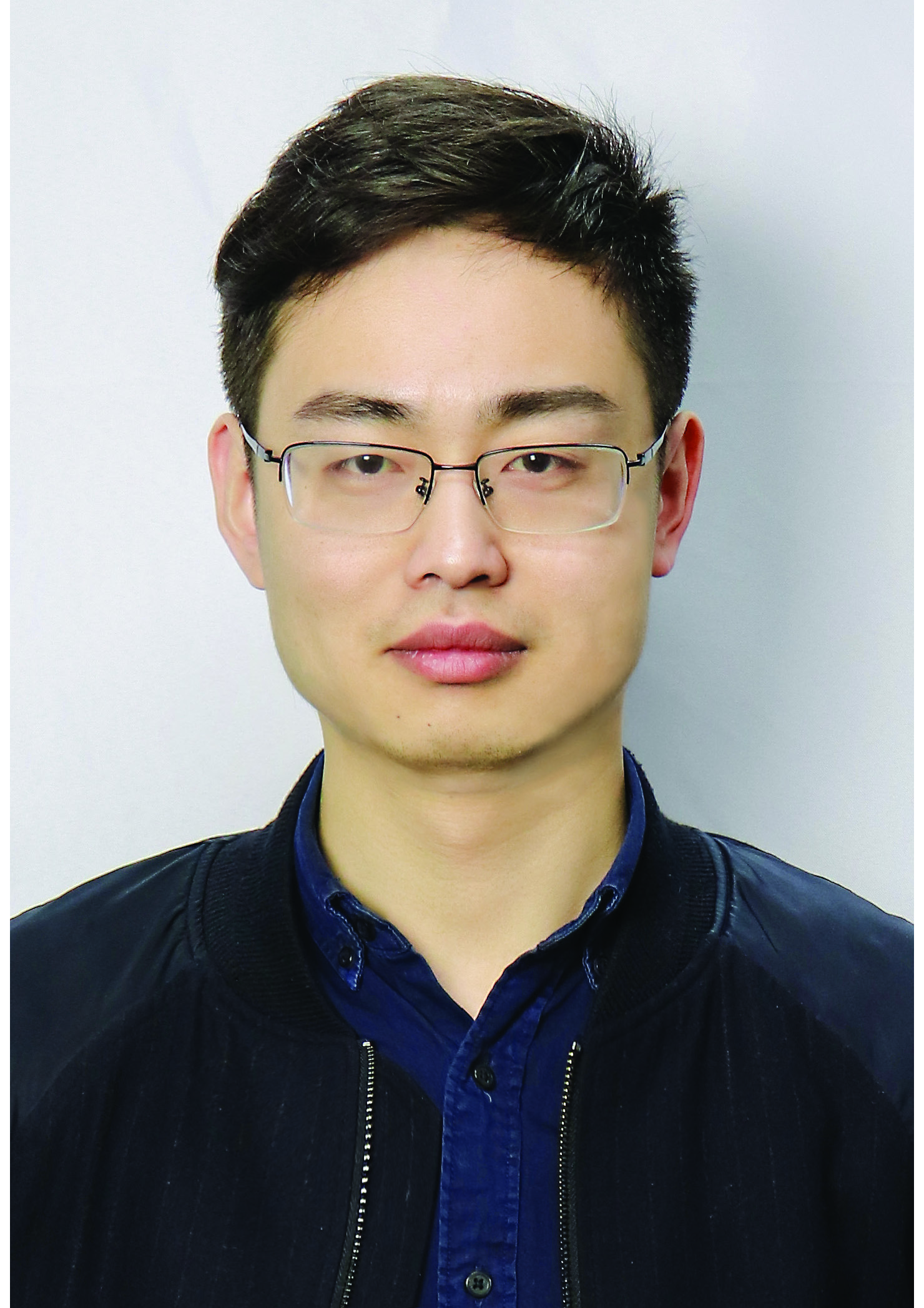}
}]{Ming Lu} is an Associate Researcher in the School of Electronic Science and Engineering, Nanjing University, Jiangsu, China. He received his B.S. and Ph.D. degrees in the School of Electronic Science and Engineering from Nanjing University, Jiangsu, China, in 2016 and 2023 respectively. His current research focuses on deep learning-based image/video coding. He is a co-recipient of the 2018 ACM SIGCOMM Student Research Competition Finalist, the 2020 IEEE MMSP Image Compression Grand Challenge Best Performing Solution, the 2023 IEEE WACV Best Algorithms Paper Award, and the 2023 IFTC Best Paper Award.
\end{IEEEbiography}

\begin{IEEEbiography}[{\includegraphics[width=1in,height=1.25in,clip,keepaspectratio]{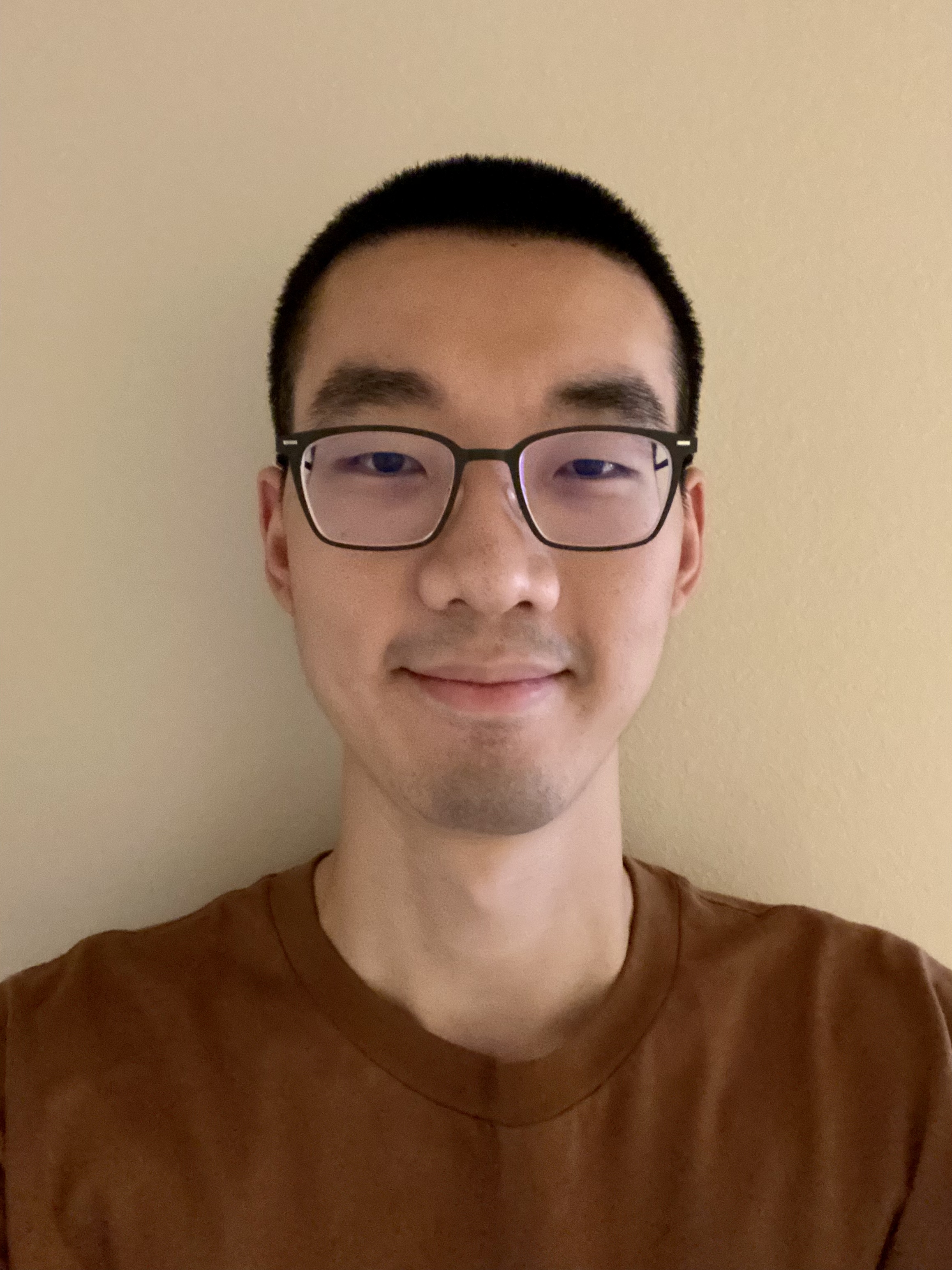}}]{Zhihao Duan} received a B.E. degree in electrical engineering from Shanghai Jiao Tong University, Shanghai, China, in 2018 and an M.S. degree in electrical and computer engineering from Boston University, Boston, MA, USA, in 2020. He is now a Ph.D. candidate at the Video and Image Processing Laboratory at Purdue University, West Lafayette, IN, USA. His research interests lie in the intersection of data compression and machine learning. Recently, he received the 2022 IEEE PCS Best Paper Finalist and the 2023 IEEE WACV Best Algorithms Paper Award.
\end{IEEEbiography}

\begin{IEEEbiography}[{\includegraphics[width=1in,height=1.25in,clip,keepaspectratio]{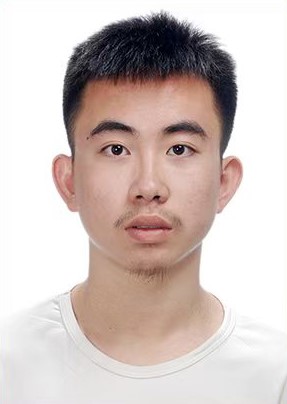}}]{Wuyang Cong} received the B.S. degree in electronic science and technology, from Central South University, Changsha, China, in 2023. He is pursuing the M.S. degree in the School of Electronic Science and Engineering, Nanjing University, Nanjing, China. His current research focus on learning-based video coding.
\end{IEEEbiography}

\begin{IEEEbiography}[{\includegraphics[width=1in,height=1.25in,clip,keepaspectratio]{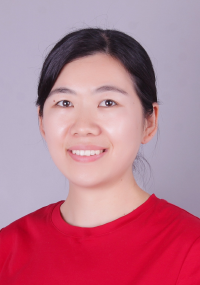}}]{Dandan Ding} received her B. Eng. degree (Honor.) in Communication Engineering from Zhejiang University, China, in 2006. From 2007 to 2008, she was an exchange student in Microelectronic Systems Laboratory (GR-LSM) of EPFL, Switzerland. After returning to Zhejiang University, she earned her Ph.D. degree in June 2011, and served first as a postdoc researcher from July 2011 through June 2013, then as a research associate in Zhejiang University till 2015. Since 2016, she has been with the School of Information Science and Technology, Hangzhou Normal University as a faculty member with tenure track. Her research interests include artificial intelligence based image/video processing, video coding algorithm design and optimization, and point cloud processing.
\end{IEEEbiography}

\begin{IEEEbiography}[{\includegraphics[width=1in,height=1.25in,clip,keepaspectratio]{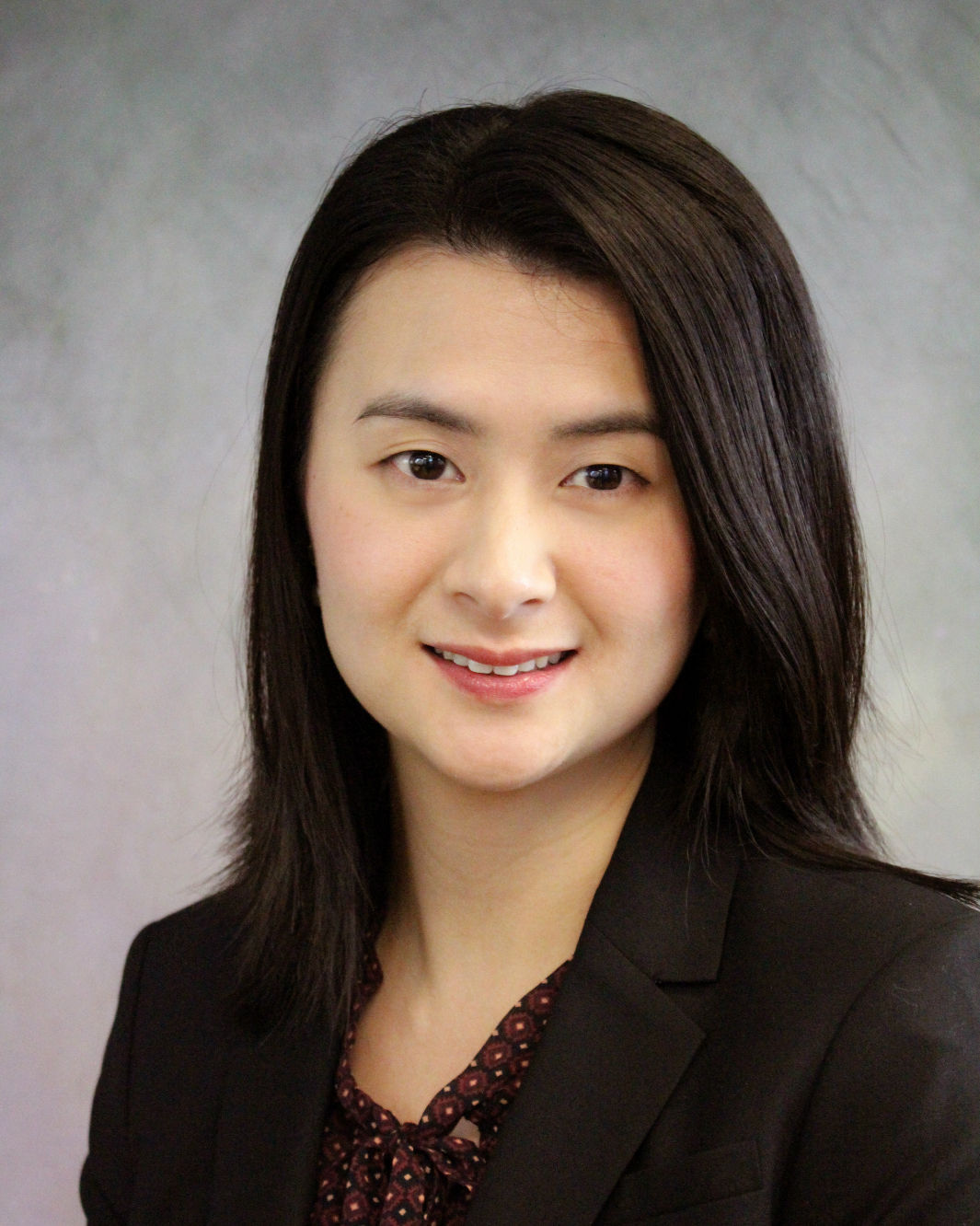}}]{Fengqing Zhu} is an Associate Professor of Electrical and Computer Engineering at Purdue University, West Lafayette, Indiana. She received the B.S.E.E. (with highest distinction), M.S. and Ph.D. degrees in Electrical and Computer Engineering from Purdue University. From 2012 to 2015, she was a Staff Researcher with Futurewei Technologies, Inc., Santa Clara, CA, USA, where she received the Certification of Recognition for Core Technology Contribution in 2012. Her research interests include smart health with a focus on image based dietary assessment and wearable sensor data analysis, visual coding for machines, and application-driven visual data analytics. She is the recipient of an NSF CISE Research Initiation Initiative (CRII) award in 2017, a Google Faculty Research Award in 2019, and an ESI and trainee poster award for the NIH Precision Nutrition workshop in 2021.
\end{IEEEbiography}

\begin{IEEEbiography}[{\includegraphics[width=1in,height=1.25in,clip,keepaspectratio]{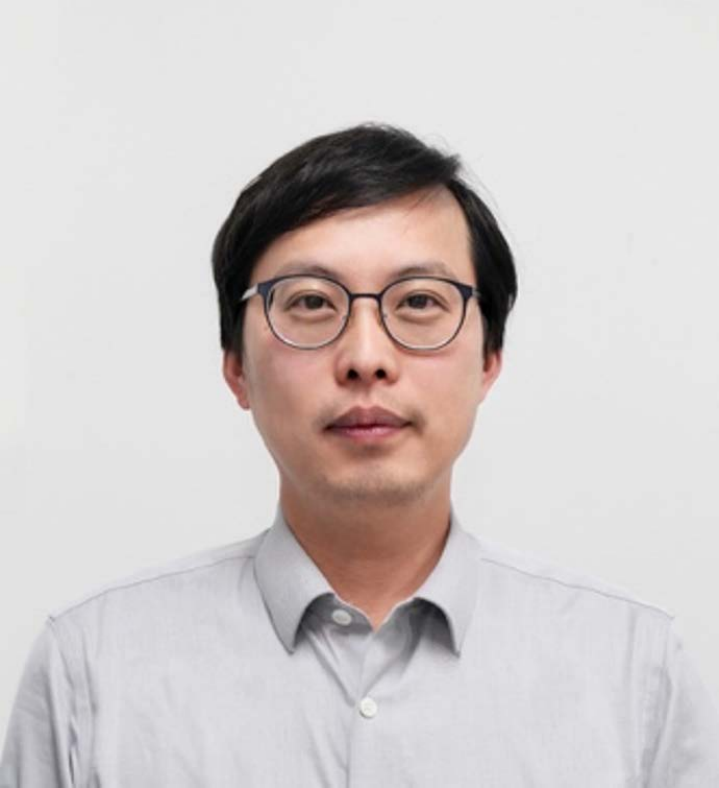}}]{Zhan Ma} is a Professor in the School of Electronic Science and Engineering, Nanjing University, Jiangsu, 210093, China. He received his Ph.D. from New York University, New York, in 2011 and his B.S. and M.S. from the Huazhong University of Science and Technology, Wuhan, China, in 2004 and 2006 respectively. From 2011 to 2014, he has been with Samsung Research America, Dallas, TX, and  Futurewei Technologies, Inc., Santa Clara, CA, respectively. His research focuses include learned image/video coding and computational imaging. He was awarded the 2018 PCM Best Paper Finalist, the 2019 IEEE Broadcast Technology Society Best Paper Award,  the 2020 IEEE MMSP Grand Challenge Best Image Coding Solution, the 2023 IEEE WACV Best Algorithms Paper Award, and the 2023 IEEE Circuits and Systems Society Outstanding Young Author Award.
\end{IEEEbiography}



\clearpage
\appendix

This supplemental material provides additional implementation details and experimental results for a more thorough description and validation of our method. 


\section*{Settings for Traditional Codecs}

The configuration parameters of traditional codecs affect the performance of the baseline model. In this work, we maintain the settings that yield the best performance in low-latency scenarios, allowing us to better evaluate the compression performance of our model.

The command lines of x265 used in our paper are:

\begin{lstlisting}[language=Bash]
ffmpeg
-y
-pix_fmt yuv420p
-s {width}x{height}
-framerate {frame rate}
-i {input file name}
-vframes 96
-c:v libx265
-preset veryslow
-tune zerolatency
-x265-params
``qp={qp}:keyint=32:csv-log-level=1:
csv={csv path}:verbose=1:psnr=1''
{bitstream file name}
\end{lstlisting}

The command lines of HM 16.26 used in our paper are:
\begin{lstlisting}[language=Bash]
TAppEncoder
-c encoder_lowdelay_main_rext.cfg
--InputFile={input file name}
--SourceWidth={width}
--SourceHeight={height}
--InputBitDepth=8
--OutputBitDepth=8
--OutputBitDepthC=8
--InputChromaFormat=444
--FrameRate={frame rate}
--FramesToBeEncoded=96
--IntraPeriod=32
--DecodingRefreshType=2
--QP={qp}
--Level=6.2
--BitstreamFile={bitstream file name}
\end{lstlisting}

The command lines of VTM 18.0 used in our paper are:
\begin{lstlisting}[language=Bash]
EncoderAppStatic
-c encoder_lowdelay_vtm.cfg
--InputFile={input file name}
--SourceWidth={width}
--SourceHeight={height}
--InputBitDepth=8
--OutputBitDepth=8
--OutputBitDepthC=8
--InputChromaFormat=444
--FrameRate={frame rate}
--FramesToBeEncoded=96
--IntraPeriod=32
--DecodingRefreshType=2
--QP={qp}
--Level=6.2
--BitstreamFile={bitstream file name}
\end{lstlisting}

\begin{figure*}[t]
\centering
\subfigure[]{\includegraphics[width=0.33\linewidth]{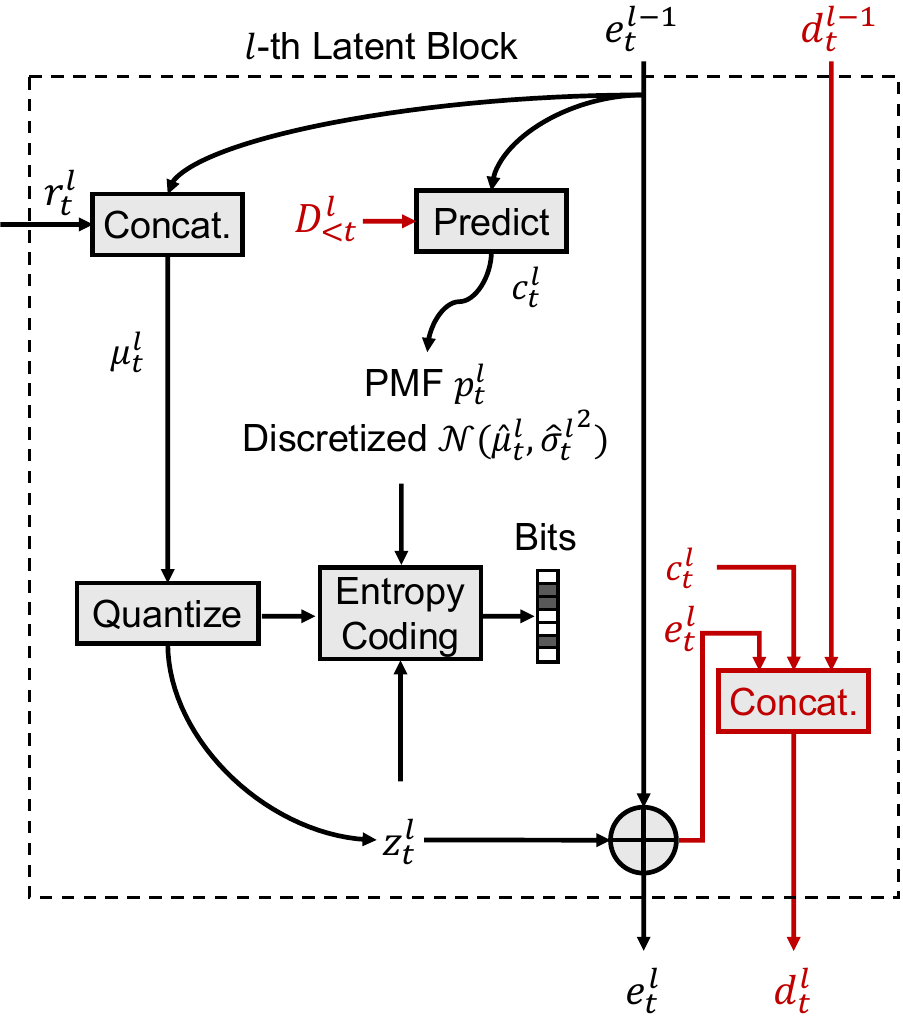}\label{fig:compress_parallel}}
\hspace{1cm}
\subfigure[]{\includegraphics[width=0.25\linewidth]{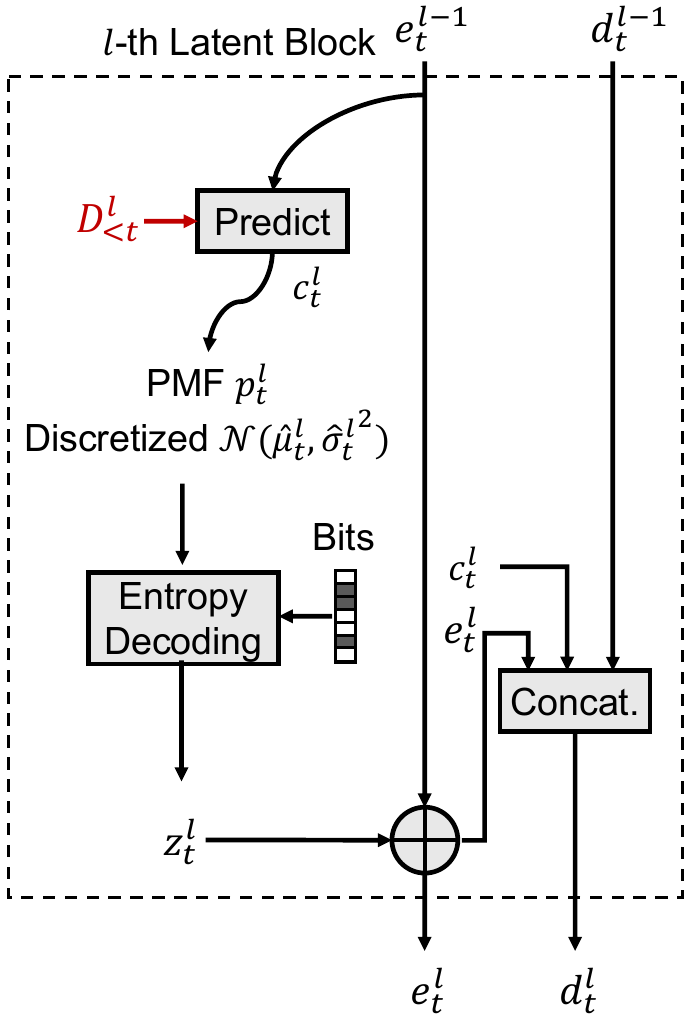}\label{fig:decompress_parallel}}
\hspace{1cm}
\subfigure[]{\includegraphics[width=0.25\linewidth]{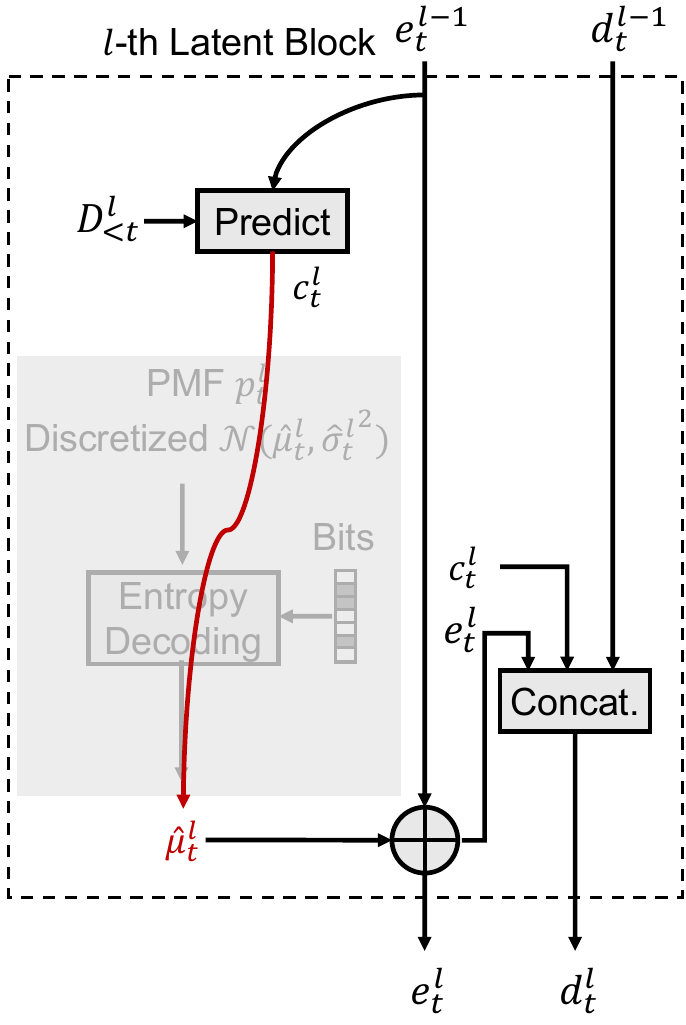}\label{fig:decompress_progressive}}
\caption{DHVC 2.0 (Parallel) (a) compression, (b) decompression, (c) and decompression w/o entropy decoding modes.}
\label{fig:appendix-fast-coding}
\end{figure*}

\section*{Parallel Acceleration}

In Sec.~\ref{sec:complexity_efficiency}, we introduce the parallel pipelines to accelerate the encoding and decoding model, referred to as DHVC 2.0 (Parallel). As described in the main paper, to enable parallel processing across scales, we remove the auxiliary refinement path. Thus, the overall compression and decompression process differs from that shown in Fig.~\ref{fig:4-compress} and Fig.~\ref{fig:4-decompress}. We present the actual compression and decompression process of DHVC 2.0 (Parallel) in Fig.~\ref{fig:compress_parallel} and Fig.~\ref{fig:decompress_parallel} with modifications highlighted in red.

For compression, only the decoded features $D_{<t}^l$ are used as temporal information for context prediction. Right after obtaining spatial reference features 
$e_t^l$ at the current $l$-th scale, we need to fuse it with the contextual features $c_t^l$, and lower-scale decoded feature $d_t^{l-1}$ to generate $d_t^l$. This allows us to obtain the decoded features to immediately serve as temporal references for encoding subsequent frames.

For decompression, the only difference is that we have removed the auxiliary information $a_{t-1}^l$ in Fig.~\ref{fig:4-decompress} during the context prediction process. 

Additionally, we have compared the processing speed of DHVC 2.0 (Parallel) on four devices or a single device, as shown in Fig.~\ref{fig:parallel_compare}. Through parallel pipeline processing shown in Fig.~\ref{fig:parallel}, we achieve speedups of $2.3 \times$ for encoding and $2.7 \times$ for decoding.

\begin{figure}[t]
\centering
\includegraphics[width=0.8\linewidth]{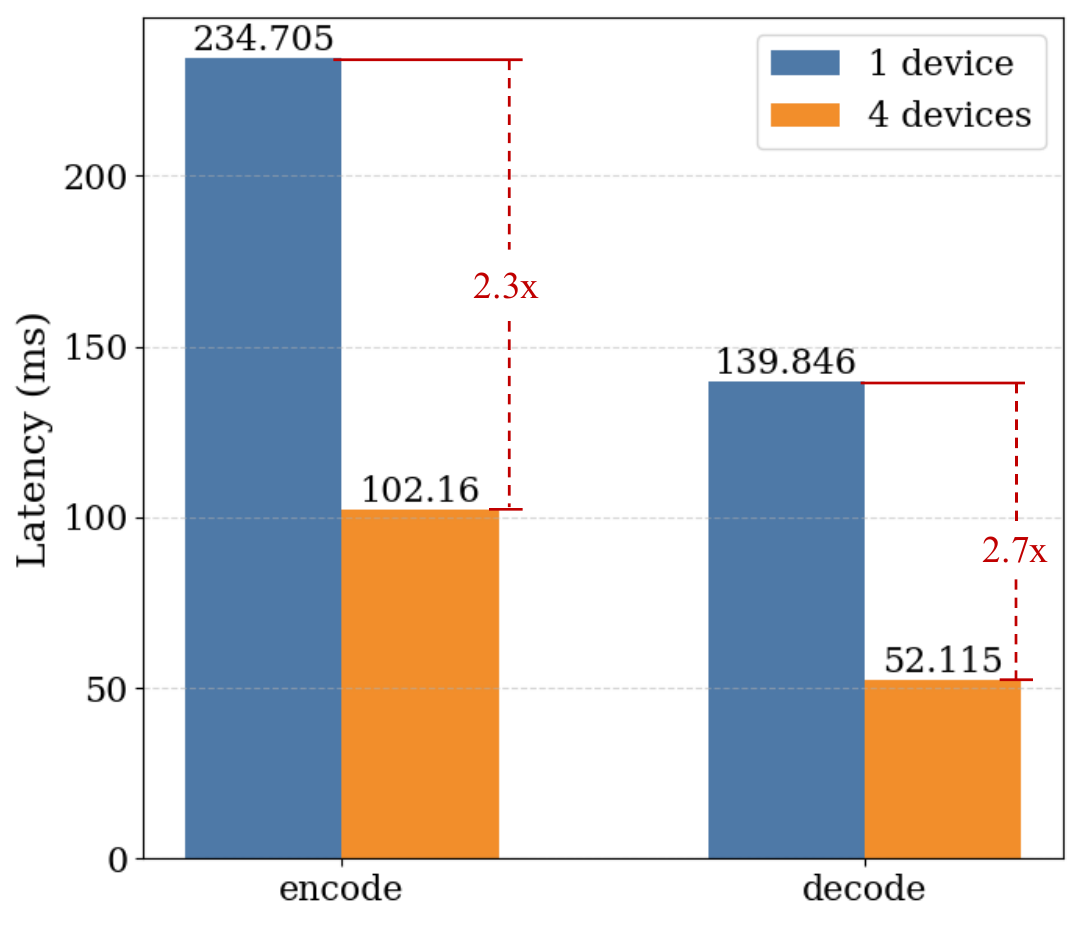}
\caption{Parallel acceleration for encoding and decoding.}
\label{fig:parallel_compare}
\end{figure}

\section*{Motion Patterns}

Figure~\ref{fig:appendix-motion_pattern1}, Figure~\ref{fig:appendix-motion_pattern2}, Figure~\ref{fig:appendix-motion_pattern3}, and Figure~\ref{fig:appendix-motion_pattern4} respectively depict the motion patterns we used in this work. Please view each subplot from left to right, top to bottom to fully understand the motion characteristics. Unlike the regular translational motion in standard test sequences, e.g., UVG dataset with 120 FPS, we believe these synthesized motion patterns are common in real-world scenarios, which provide a valuable assessment of our motion generalization.

\begin{figure*}[t]
    \centering
    \subfigure[Blur]{\includegraphics[width=\linewidth]{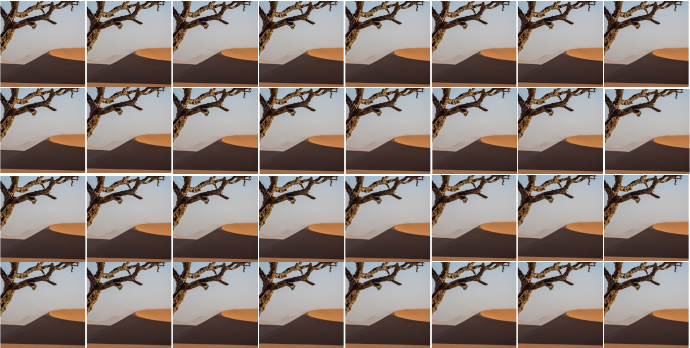}} \\
    \subfigure[Fade]{\includegraphics[width=\linewidth]{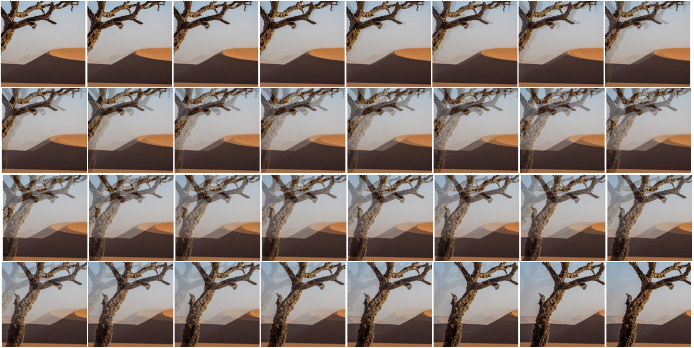}}
    \caption{Illustration of motion patterns in our experiments for motion generalization evaluation (Part 1). Please zoom in for detailed visualization.}
    \label{fig:appendix-motion_pattern1}
\end{figure*}

\begin{figure*}[t]
    \centering
    \subfigure[Darken]{\includegraphics[width=\linewidth]{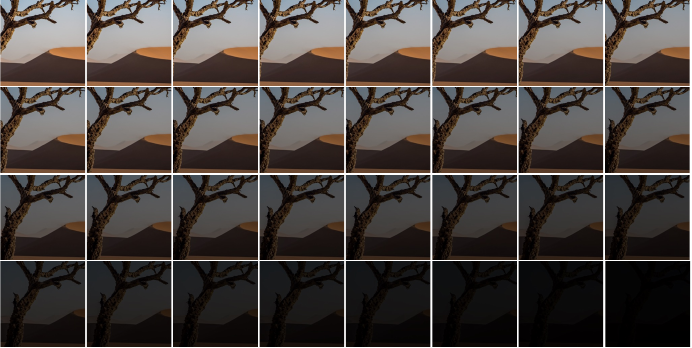}} \\
    \subfigure[Lighten]{\includegraphics[width=\linewidth]{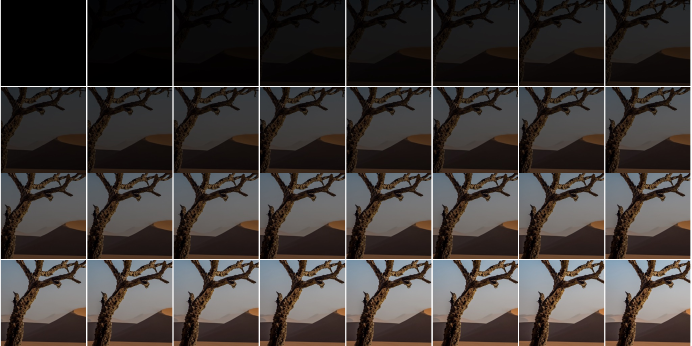}}
    \caption{Illustration of motion patterns in our experiments for motion generalization evaluation (Part 2). Please zoom in for detailed visualization.}
    \label{fig:appendix-motion_pattern2}
\end{figure*}

\begin{figure*}[t]
    \centering
    \subfigure[Zoom]{\includegraphics[width=\linewidth]{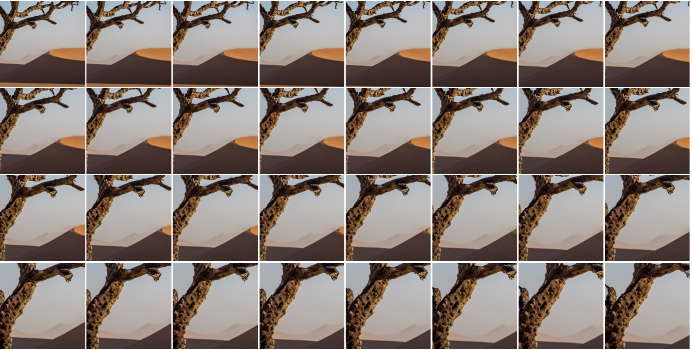}} \\
    \subfigure[Pulse]{\includegraphics[width=\linewidth]{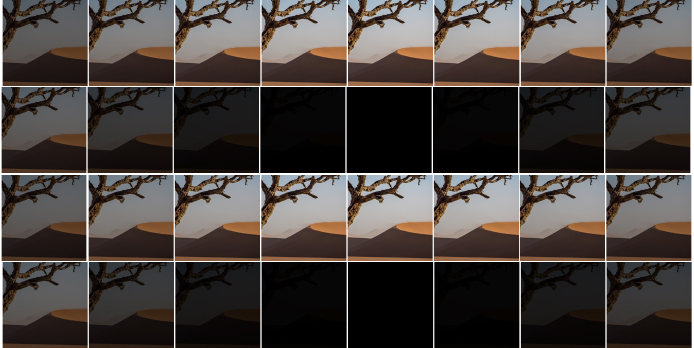}}
    \caption{Illustration of motion patterns in our experiments for motion generalization evaluation (Part 3). Please zoom in for detailed visualization.}
    \label{fig:appendix-motion_pattern3}
\end{figure*}

\begin{figure*}[t]
    \centering
    \subfigure[Shake]{\includegraphics[width=\linewidth]{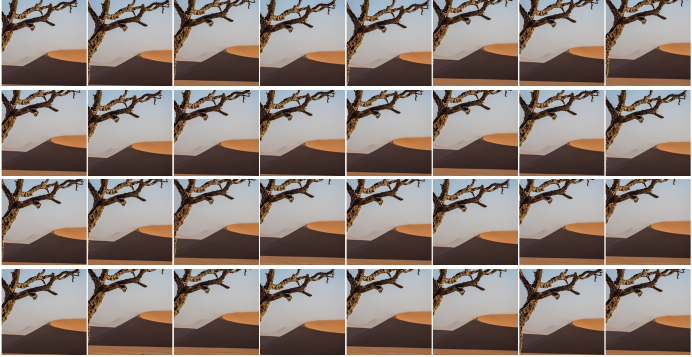}} \\
    \subfigure[Sharpen]{\includegraphics[width=\linewidth]{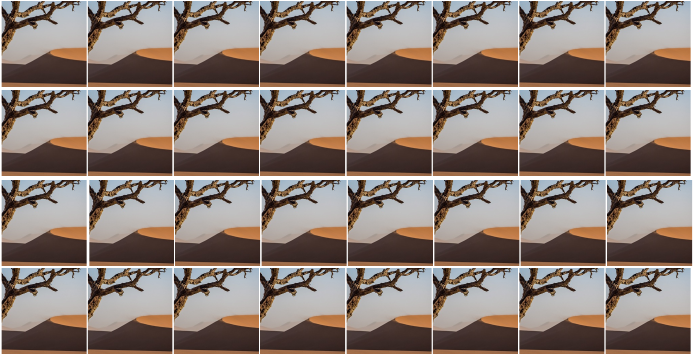}}
    \caption{Illustration of motion patterns in our experiments for motion generalization evaluation (Part 4). Please zoom in for detailed visualization.}
    \label{fig:appendix-motion_pattern4}
\end{figure*}

\section*{Progressive Decoding}

We visualize the pipeline of our progressive decoding without the bitstream of the current Latent Block in Fig.~\ref{fig:decompress_progressive}, which has already been discussed in Sec.~\ref{sec:progressive_decoding}. Compared to the normal decoding process, as shown in Fig.~\ref{fig:4-decompress}, the main difference is that since the $l$-th bitstream is not available, we skip entropy decoding and use the predicted mean $\hat{\mu}_t^l$ from $c_t^l$ to replace the original $z_t^l$ to be decoded, adding it to $e_t^{l-1}$ to obtain $e_t^l$. The entire process does not involve any probability-based encoding or decoding. Therefore, even if there is a mismatch with the encoding side in added information for $e_t^l$, it will only reduce performance due to the loss of encoded latent residual, without causing any decoding errors.

Two video sequences, i.e., ``videoSRC16'' in the MCL-JCV dataset and ``BasketballDrive'' in the HEVC Class B dataset, are used to showcase the progressive decoding of our method, as visualized in Fig.~\ref{fig:appendix-progressive_a} and Fig.~\ref{fig:appendix-progressive_b}. It clearly shows that as more scales (layers) of latent variables are received and decoded, more bitrates are consumed, and the reconstruction quality of the corresponding frames gradually improves. Even if we receive partial bitstreams and can only decode up to ``scale 1'', the reconstruction quality is still relatively stable. This also explains that our approach is resistant to the presence of packet loss.

Recall that the DHVC model produces $L=4$ bitstreams for each frame. Each bitstream corresponds to a different latent variable, and the lengths of the bitstreams vary for different frames. To better understand how the rate distributes over latent variables and how inter-coding impacts such distributions, we visualize them in Fig.~\ref{fig:rate_distribution} using the UVG dataset as an example.

Several interesting facts can be observed from Fig.~\ref{fig:bpp_distribution}. First, we see that the intra-frame is at a higher rate because of the lack of temporal references. When looking at the rates for individual scales, one can observe that the rate increases as the scale enlarges, and the overall rate consumption is dominated by compressing latent variables at higher scales. We also depict in Fig.~\ref{fig:percentage_distribution} the percentage (instead of absolute values) of scale-wise rate with respect to the total rate of each frame. The observation is consistent with Fig.~\ref{fig:bpp_distribution}. The majority of the bit rate is consumed by latent variables at higher scales, regardless of intra-coded or inter-coded frame. This also confirms that even in the case of packet loss, the likelihood of packet loss occurring at higher scales is significant. As a result, the additional cost of defining an FEC mechanism to protect the first scale is minimal.

\begin{figure*}[t]
\centering
\subfigure[]{\includegraphics[width=\linewidth]{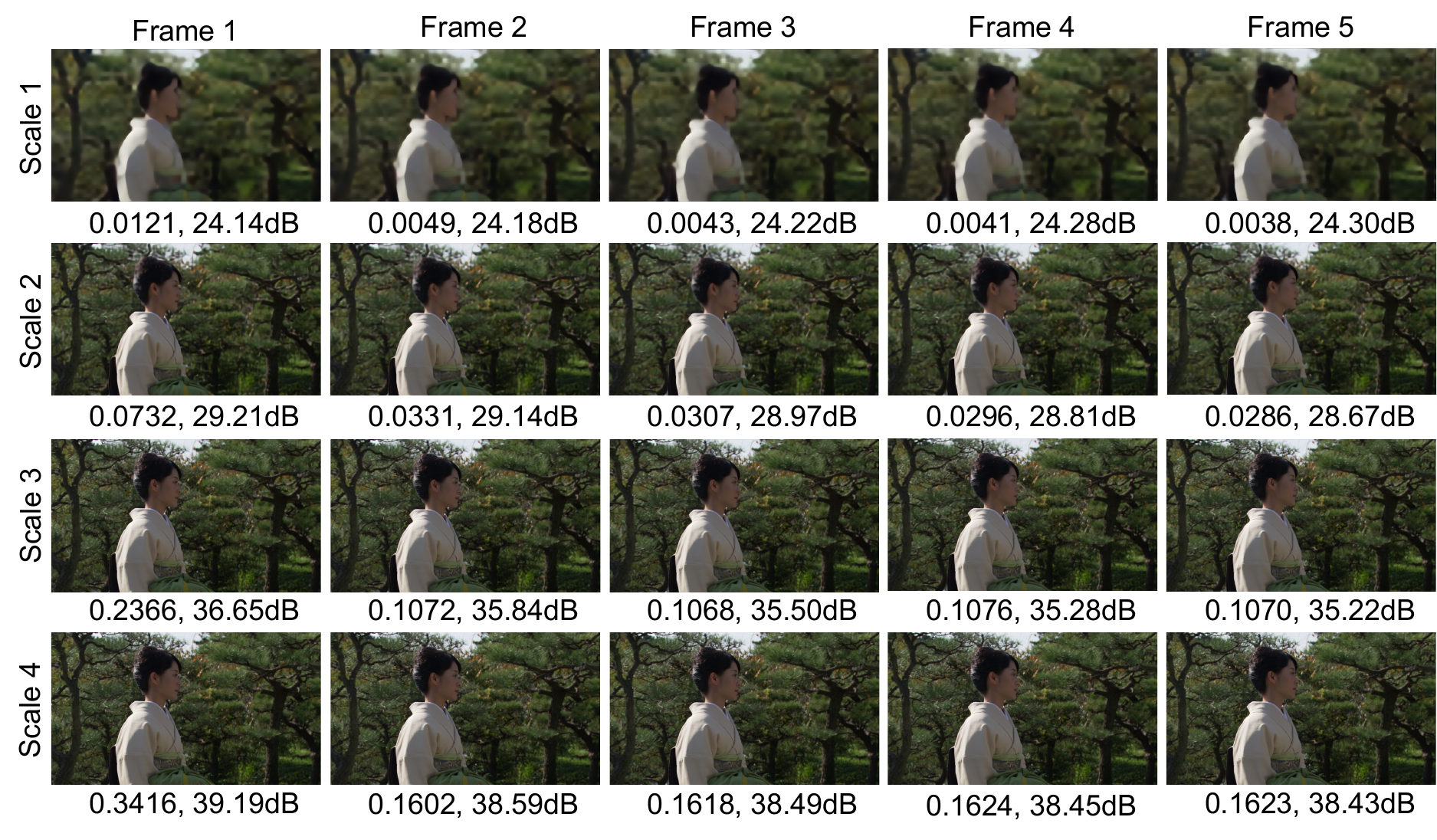}\label{fig:appendix-progressive_a}} \\
\subfigure[]{\includegraphics[width=\linewidth]{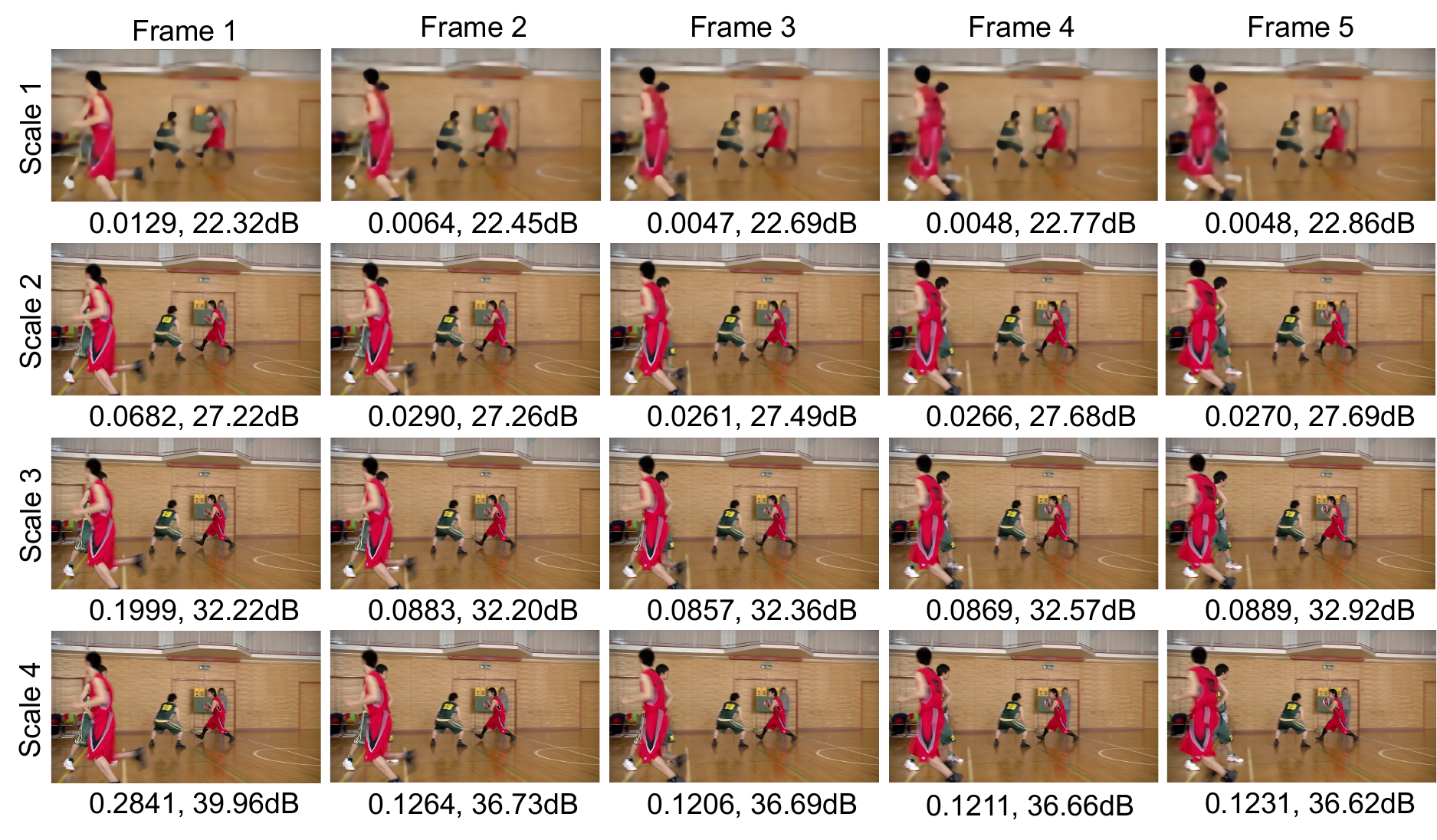}\label{fig:appendix-progressive_b}}
\caption{{Progressive decoding} by visualizing reconstruction at each scale across frames. Please zoom in for more details.}
\label{fig:appendix-progressive}
\end{figure*}

\begin{figure*}[htbp]
\centering
\subfigure[]{\includegraphics[width=0.8\linewidth]{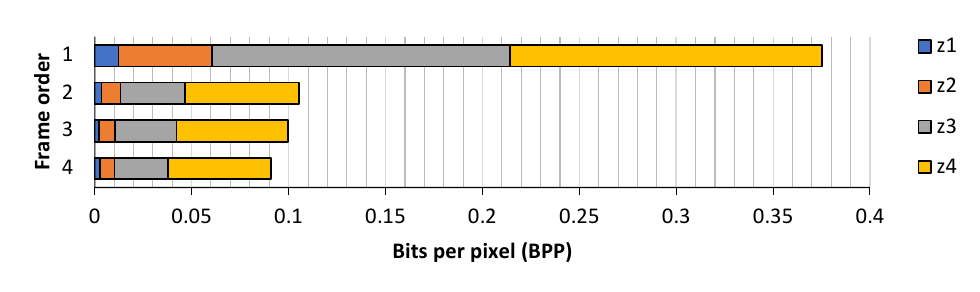} \label{fig:bpp_distribution}} \\
\subfigure[]{\includegraphics[width=0.8\linewidth]{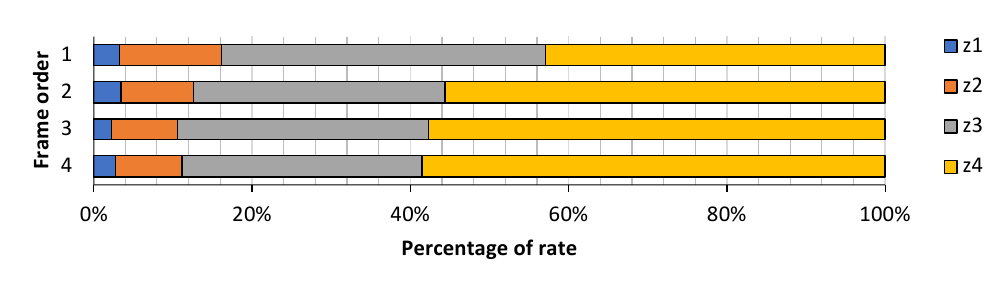} \label{fig:percentage_distribution}}
\caption{Rate distribution over latent variables for successive frames within a video sequence. The x-axis is BPP in (a) and percentage in (b). In both figures, rows correspond to different frame orders. In each row, different colors represent latent variables at different scales: $z1$ has the smallest spatial dimension ($64\times$ downsampled from the input frame), and $z4$ has the largest spatial dimension ($8\times$ downsampled from the input frame).}
\label{fig:rate_distribution}
\end{figure*}

\section*{Packet Loss Resilience}

If a single packet at a given scale $l$ is lost, we will discard the entire bitstream of that scale and skip the entropy decoding. This process is consistent with the progressive decoding workflow as illustrated in Fig.~\ref{fig:decompress_progressive}. The only difference is that although the bitstream of $l$-th scale for the current frame is lost, the information from previous frames is intact. Therefore, the decoded features $D_{<t}^l$ used as references can be used to generate better $\hat{\mu}_t^l$ and thus better quality when compared with progressive decoding. 

Additionally, we consider consecutive packet loss in real-world scenarios and attempt to combine our loss resilience mechanism to achieve better transmission quality. Specifically, we conduct a test with one GOP consisting of 32 frames and assume that packet loss randomly occurs between frames 4 and 28. The loss rate is set from 20\%, 40\%, 60\%, to 80\% respectively as illustrated in Fig.~\ref{fig:packet_loss_supp}. We take Fig.~\ref{fig:packet_loss_supp_80}, which exhibits the highest packet loss rate, as an example. The loss of more scales will result in subsequent frames being decoded with fewer available scales, i.e., frame 10 can only be decoded using the first scale's bitstream as last 3 scales' bitstreams of preceding frames have been lost. At this point, we send the synchronized echo to the encoder and attempt to restore quality through additional bitstream consumption. However, the transmitted re-synced frame (i.e., frame ``14'') experiences packet loss again, resulting in only the first two scales being decodable. Consequently, the reconstruction quality still can not recover to its original level, which is an inevitable consequence of consecutive packet losses. Despite this, it at least prevents further degradation in quality and provides a potential for recovery. As the network conditions improve, the quality can be stretched back to a better state. In scenarios with lower packet loss rates (e.g., 20\%), our method demonstrates excellent loss resilience performance. Combining the FEC mechanism with improved error control algorithms, we anticipate uncovering better loss resilience methods in our future work.

\begin{figure*}[t]
\centering
\subfigure[]{\includegraphics[width=0.49\linewidth]{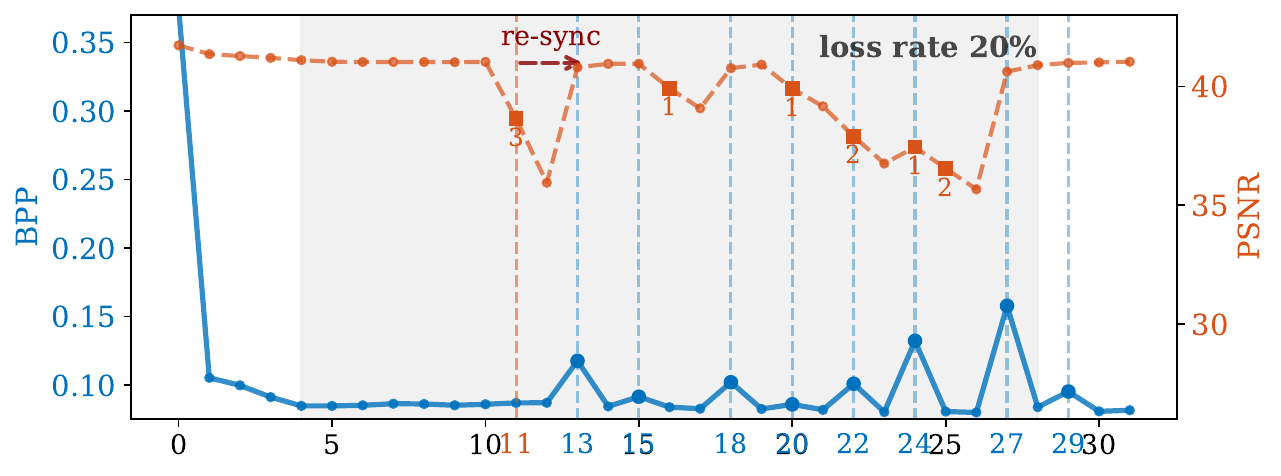}} 
\subfigure[]{\includegraphics[width=0.49\linewidth]{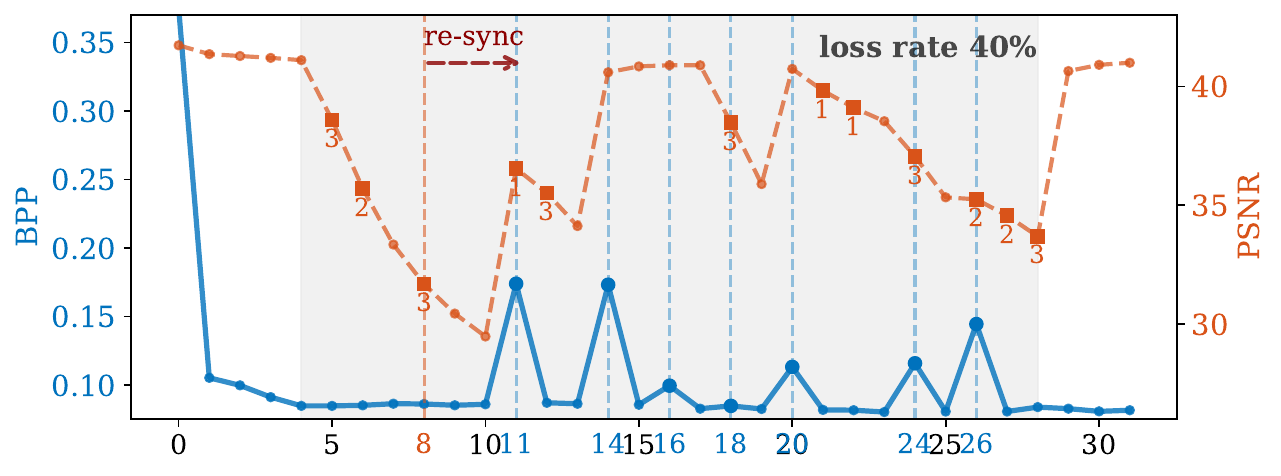}} \\
\subfigure[]{\includegraphics[width=0.49\linewidth]{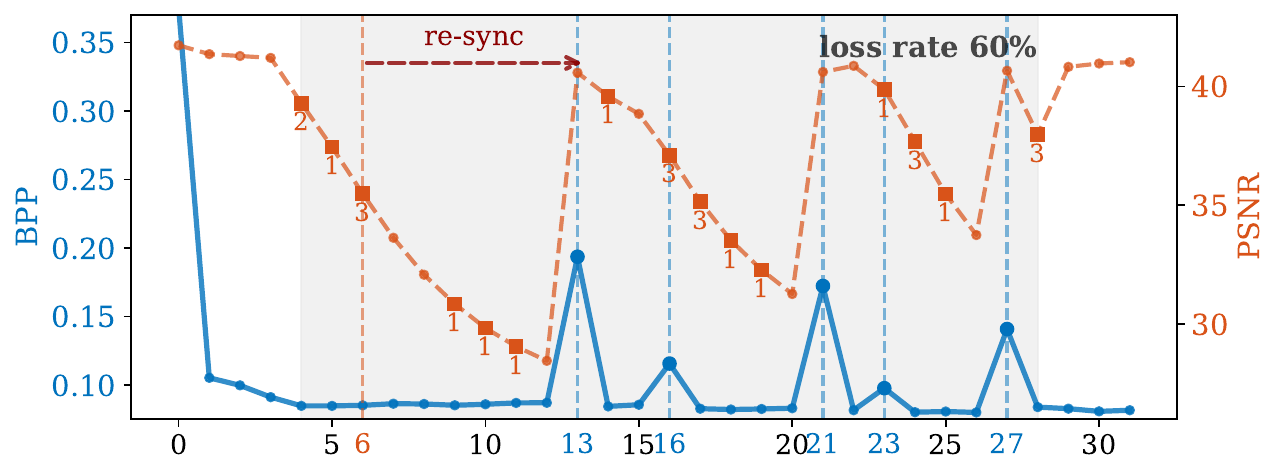}} 
\subfigure[]{\includegraphics[width=0.49\linewidth]{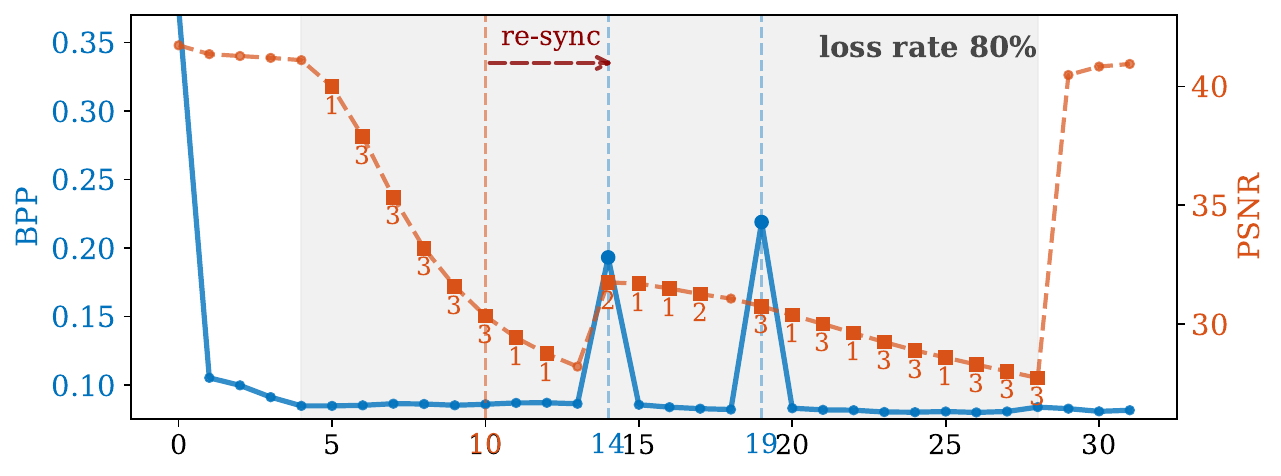}\label{fig:packet_loss_supp_80}} 
\caption{Packet loss simulation under different loss rates. Even for a loss rate of 80\%, our model can still offer smooth playback and recover the lost packets.}
\label{fig:packet_loss_supp}
\end{figure*}





\ifCLASSOPTIONcaptionsoff
  \newpage
\fi

\end{document}